\documentclass[11pt]{revtex4}
\usepackage{mathtools}
\usepackage{hyperref}
\usepackage{pdflscape}
\usepackage{graphics}
\usepackage{subfigure}
\usepackage{epsfig}
\usepackage{amsmath,amsfonts}
\usepackage{amssymb}
\usepackage{color}
\usepackage{ulem}
\usepackage{amsmath}
\allowdisplaybreaks

\begin{document}
\title{New gravitational wave polarization modes in the torsionless spacetime}

\author{Yu-Qi Dong$^{a,b}$}
\email{dongyq2023@lzu.edu.cn}

\author{Xiao-Bin Lai$^{a,b}$}
\email{laixb2024@lzu.edu.cn}

\author{Yu-Zhi Fan$^{a,b}$}
\email{fanyzh2024@lzu.edu.cn}

\author{Yu-Xiao Liu$^{a,b}$}
\email{liuyx@lzu.edu.cn (corresponding author)}

\affiliation{$^{a}$ Lanzhou Center for Theoretical Physics, Key Laboratory of Theoretical Physics  of Gansu Province, Key Laboratory of Quantum Theory and Applications of MoE, Gansu Provincial Research Center for Basic Disciplines of Quantum Physics, Lanzhou University, Lanzhou 730000, China\\
	$^{b}$ Institute of Theoretical Physics \& Research Center of Gravitation, Lanzhou University, Lanzhou 730000, China}

\begin{abstract}
\textbf{Abstract:} In this study, we investigate the polarization properties of gravitational waves within a torsionless spacetime framework, as described by the Palatini formalism. Our analysis uncovers the presence of two novel polarization modes, referred to as shear modes, which extend beyond the traditional set of six modes in a four-dimensional Riemannian spacetime.  These shear modes, uniquely driven by vector degrees of freedom associated with non-metricity, are classified as vector modes, and their detection provides a unique opportunity to explore the fundamental structure of spacetime and to test gravity theories. These modes extend the standard gravitational wave polarization paradigm and provide novel observational signatures for gravitational wave detectors.
\end{abstract}
	
\maketitle

\section{Introduction}
\label{sec: intro} 
The successful detection of gravitational waves \cite{Abbott1,Abbott2,Abbott3,Abbott4,Abbott5} has provided a powerful tool for testing various modified gravity theories. One key feature of gravitational waves is their polarization, which can offer insights into the nature of gravity. In general relativity, gravitational waves exhibit only two polarization modes. However, a general analysis suggests that \cite{Eardley}, in four-dimensional spacetime, if we require free particles to follow the usual geodesic equation, the theory allows for a maximum of six independent polarization modes. These additional polarization modes are often predicted by modified gravity theories. Different theories make different predictions about the polarization modes of gravitational waves and their corresponding mode dispersion relations. Therefore, by conducting polarization tests on observed gravitational waves, we can test various candidate modified gravity theories. Especially, once a new polarization mode of gravitational waves is definitively discovered, it will be incontrovertible that general relativity will require modification.

To rigorously test modified gravity theories through gravitational wave polarization detection, it is essential to analyze the polarization modes of gravitational waves and their corresponding dispersion relations in these theories. Numerous studies have explored the polarization modes of gravitational waves in specific modified gravity theories \cite{f(R,Orfeu Bertolami,f(R2,Horndeski0,TeVeS,Horava,STVG,fT,dCS and EdGB,Y.Dong,S. Bahamonde,J.Lu,L.Shao,S. Nojiri,Y.Dong3,Shaoqi Hou,Xiao-Bin Lai,Y. Fan,Miguel Barroso Varela,S Capozziello}. Moreover, in recent work, we have developed a systematic parametric framework \cite{Y.Dong4,Y.Dong5} that allows for a unified investigation of gravitational wave effects across all theories conforming to specific fundamental physical principles. This framework simplifies the process of determining the gravitational wave polarization properties of a specific modified gravity theory by simply inputting the relevant parameters. Using this approach, we have completed an analysis of the polarization modes and dispersion relations for gravitational waves in the most general pure metric theory \cite{Y.Dong4}, scalar-tensor theory \cite{Y.Dong4}, and second-order vector-tensor theory \cite{Y.Dong5}. In terms of experimental observation, ground-based gravitational wave detectors \cite{J. Aasi,F. Acernese,T. Akutsu,M. Punturo,D. Reitze,Lijing Shao}, space-based gravitational wave detectors \cite{P. Amaro-Seoane,Z. Luo,J. W. Mei,A. Torres-Orjuela,H. B. Jin,Y. Y. Wang}, and pulsar timing arrays \cite{H. Xu,F. Jenet,B. W. Stappers,R. N. Manchester,B. C. Joshi}, which cover different frequency bands of gravitational waves, are actively observing or planning to observe additional polarization modes. Some analysis of pulsar timing array data seems to support the existence of additional polarization modes \cite{Qing-Guo Huang1,Qing-Guo Huang2}, but this preliminary evidence requires further confirmation.

Lovelock's theorem \cite{David Lovelock1,David Lovelock2} states that in a four-dimensional spacetime, if a gravity theory is described solely by the metric and has second-order field equations, only general relativity satisfies these conditions. Therefore, adding extra degrees of freedom is the primary method for modifying gravity. From the viewpoint that classical gravity is fully expressed geometrically, general relativity is formulated within the framework of Riemannian geometry, where all geometric effects are described by the metric \cite{MTW}. The connection, which defines parallel transport, is a priori expressed as a metric function in the form of the Levi-Civita connection. However, in more general geometries, the connection can be defined independently of the metric. Thus, by assuming that the connection is independent of the metric and treating both as independent variational quantities, we naturally introduce a way to modify gravity. This approach may provide additional degrees of freedom through the connection. The modified gravity theory constructed within a non-Riemannian geometry is called metric-affine gravity, where, in addition to curvature, spacetime typically exhibits torsion and non-metricity \cite{Lavinia Heisenberg2}. Metric-affine gravity has garnered significant attention due to the important geometric role of these extra degrees of freedom, and its potential to offer new geometric insights into gravity theory \cite{Friedrich W. Hehl1,Vincenzo Vitagliano}. Recent years have witnessed a notable increase in research interest in this area, especially in the context of the geometrical trinity and its extensions \cite{Lavinia Heisenberg1,Yi-Fu Cai,Sebastian Bahamonde,Salvatore Capozziello,Lavinia Heisenberg2}.

In this paper, we focus on general metric-affine gravity in a torsionless spacetime, where the action is formulated using the Palatini formalism. We refer to a gravity theory in this spacetime as Palatini theory to distinguish it from a theory in a spacetime with torsion. We aim to study the polarization properties of gravitational waves in Palatini theory. In particular, we aim to investigate whether entirely new polarization modes—distinct from the six known conventional types—can arise in a torsionless non-Riemannian spacetime.
Polarization modes of gravitational waves are defined by the relative motion of two test particles \cite{Eardley}. Therefore, if non-metricity modifies the equations governing the relative motion of test particles, it may give rise to new polarization modes of gravitational waves. These additional modes could provide valuable insights into gravitational-wave astronomy, offer new avenues for testing gravitational theories, and deepen our understanding of the fundamental nature of spacetime.


Based on the above considerations, this paper is organized as follows: In Sec. \ref{sec: 2}, we construct a series of gauge invariants by combining perturbations within the linearized Palatini theory. These invariants are often used to analyze polarization modes of gravitational waves and help us examine the physical meaning more directly.
In Sec. \ref{sec: 3}, we derive the equation of motion for free particles in a torsionless spacetime. Here, a free particle is defined as one that interacts only with the metric and the connection. In fact, previous studies \cite{Dirk Puetzfeld1, Dirk Puetzfeld2, Damianos Iosifidis} have provided the equation of motion for free particles in the context of general metric-affine gravity. The equation we derive here is obtained by considering the previous results in the special case of torsionless spacetime. In Sec. \ref{sec: 4}, we derive the equation of relative motion for neighboring free particles and examine the correction to gravitational wave polarization modes due to the non-metricity in the theory. In particular, non-metricity induces two previously unidentified polarization modes of gravitational waves, which we refer to as shear modes. In Sec. \ref{sec: 5}, we further demonstrate the existence of the novel shear modes in certain theories by providing a concrete example, confirming that they arise within the context allowed by the field equations.
Sec. \ref{sec: 6} provides the conclusion. Additionally, in Appendices \ref{app: C} and \ref{app: D}, we establish a comprehensive theoretical framework within the most general second-order Palatini theory to systematically analyze the polarization modes of gravitational waves. This provides strong support for utilizing gravitational wave detectors to test and refine gravitational theories.

We adopt the convention $c=G=1$ and use the metric signature $(-,+,+,+)$. The indices $(\mu, \nu, \lambda, \rho)$ correspond to four-dimensional spacetime indices, ranging from $0$ to $3$, while the indices $(i, j, k, l)$ represent three-dimensional spatial indices, ranging from $1$ to $3$, corresponding to the $(+x, +y, +z)$ directions, respectively. $\Gamma^{\lambda}_{~\mu\nu}$ denotes a general connection, while $\widehat{\Gamma}^{\lambda}_{~\mu\nu}$ refers to the Levi-Civita connection. Similarly, for any quantity $A$ that depends on the connection, $\widehat{A}=A\arrowvert_{\Gamma^{\lambda}_{~\mu\nu}=\widehat{\Gamma}^{\lambda}_{~\mu\nu}}$.

\section{Gauge invariants in the linearized Palatini theory}
\label{sec: 2}

Gravitational waves are extremely weak, so it is sufficient to study their polarization modes using the linear approximation of theoretical perturbations on a flat spacetime background, known as linearized gravity theory \cite{MTW,Michele Maggiore}. In this framework, the general covariance of the theory manifests as gauge invariance, where the gauge transformation corresponds to an infinitesimal coordinate transformation. 

Due to the linear nature of the gauge transformation, the linearized field equations remain invariant under it. Moreover, the flatness of the background and the linearity of the gauge transformations allow us to derive several gauge invariants by combining perturbations, enabling us to express the linearized field equations in terms of these invariants (As long as we note that in momentum space, the computation of field equations and gauge transformations reduces to linear algebra, this result can be easily seen). This process naturally eliminates the redundant degrees of freedom introduced by arbitrary gauge choices (coordinate choices). In addition, similarly to Maxwell's electromagnetic theory, if the field equations of a theory are fully described by gauge invariants, all observable quantities in the theory should also be gauge invariant. Thus, identifying all the gauge invariants formed by combining perturbations is a convenient strategy. This approach simplifies the subsequent analysis and offers a clearer characterization of the observables in gravitational waves.

Based on the above discussion, we construct the gauge invariants of Palatini theory in torsionless spacetime by combining perturbations in this section. The action of Palatini theory can generally be expressed as
\begin{eqnarray}
	\label{Palatini action}
	S=S_{g}\left[g_{\mu\nu},\Gamma^{\lambda}_{~\mu\nu}\right]
	+S_{m}.
\end{eqnarray}
Here, $S_{g}$ is the action of gravity, while $S_{m}$ is the action of the matter field, which will be discussed in the next section. It can be seen that in Palatini theory, gravity is described by both the metric $g_{\mu\nu}$ and the connection $\Gamma^{\lambda}_{~\mu\nu}$. Since we are considering a torsionless spacetime, the connection satisfies the prior relationship
\begin{eqnarray}
	\label{torsionless relationship}
	\Gamma^{\lambda}_{~\mu\nu}
	=
	\Gamma^{\lambda}_{~\nu\mu}.
\end{eqnarray}

When analyzing gravitational wave solutions, we need to consider only the vacuum situation $S_{m}=0$. The flat spacetime background is 
\begin{eqnarray}
	\label{flat spacetime background}
	g_{\mu\nu}=\eta_{\mu\nu},\quad
	\Gamma^{\lambda}_{~\mu\nu}=0.
\end{eqnarray}
At this background value, the theory maintains Lorentz and CPT symmetries. Then, we can take the background perturbations as
\begin{eqnarray}
	\label{flat spacetime background perturbations}
	g_{\mu\nu}=\eta_{\mu\nu}+h_{\mu\nu},\quad
	\Gamma^{\lambda}_{~\mu\nu}=-\Sigma^{\lambda}_{~\mu\nu}.
\end{eqnarray}
In the context of linearized gravity theory, considering the infinitesimal coordinate transformation
\begin{eqnarray}
	\label{infinitesimal coordinate transformation}
	x^\mu \rightarrow x^\mu+\xi^{\mu}(x), 
\end{eqnarray} 
where $\xi^{\mu}$ is an arbitrary function, the gauge transformation of the perturbations $h_{\mu\nu}$ and $\Sigma^{\lambda}_{~\mu\nu}$ is
\begin{eqnarray}
	\label{gauge transformation}
	h_{\mu\nu} \!\rightarrow\! h_{\mu\nu}-\partial_{\mu}\xi_{\nu}-\partial_{\nu}\xi_{\mu}, \quad
	\Sigma^{\lambda}_{~\mu\nu} \!\rightarrow\!
	\Sigma^{\lambda}_{~\mu\nu}+\partial_{\mu}\partial_{\nu}\xi^{\lambda}.
\end{eqnarray} 

Now we construct the gauge invariants obtained by combining $h_{\mu\nu}$ and $\Sigma^{\lambda}_{~\mu\nu}$. It is convenient to consider the following decomposition \cite{James M. Bardeen,Y.Dong2,Eanna E Flanagan,Y.Dong3} for our subsequent analysis:
\begin{eqnarray}
	\label{decompose perturbations}
	h_{00}&=&h_{00}, \nonumber\\
	h_{0i}&=&\partial_{i}\gamma+\beta_{i},\nonumber\\
	h_{ij}&=&h^{TT}_{ij}+\partial_{i}\epsilon_{j}+\partial_{j}\epsilon_{i}
	+\frac{1}{3}\delta_{ij}H+(\partial_{i}\partial_{j}-\frac{1}{3}\delta_{ij}\Delta)\zeta,
	\nonumber\\
	\Sigma^{0}_{~00}&=&\Sigma^{0}_{~00}, \nonumber \\
	\Sigma^{i}_{~00}&=&\partial_{i}F+G_{i},\nonumber \\
	\Sigma^{0}_{~0i}&=&\partial_{i}M+N_{i},\nonumber \\
	\Sigma^{0}_{~ij}&=&S_{ij}+\partial_{i}C_{j}+\partial_{j}C_{i}+\partial_{i}\partial_{j}B+\delta_{ij}A,\nonumber \\
	\Sigma^{i}_{~0j}&=&U_{ij}+V_{ij}+\partial_{i}\bar{C}_{j}+\partial_{j}\bar{D}_{i}+\partial_{i}\partial_{j}\bar{B}+\delta_{ij}\bar{A},\nonumber \\
	\Sigma^{i}_{~jk}&=&B^{i}_{~jk}+\partial_{i}C_{jk}+\partial_{j}D_{ik}+\partial_{k}D_{ij}+\partial_{j}E_{ik}+\partial_{k}E_{ij}\nonumber \\
	&+&\partial_{i}\partial_{j}f_{k}+\partial_{i}\partial_{k}f_{j}+\partial_{j}\partial_{k}g_{i}+\delta_{jk}h_{i}+\delta_{ij}q_{k}+\delta_{ik}q_{j}\nonumber\\
	&+&\partial_{i}\partial_{j}\partial_{k}l+\delta_{jk}\partial_{i}m+\delta_{ij}\partial_{k}n+\delta_{ik}\partial_{j}n,
\end{eqnarray}
where,
\begin{eqnarray}
	&\partial_{i}\beta^{i}=\partial_{i}\epsilon^{i}=\partial_{i}G^i=\partial_{i}N^i=\partial_{i}C^i=\partial_{i}{\bar{C}}^i=\partial_{i}{\bar{D}}^i=\partial_{i}f^i=\partial_{i}g^i=\partial_{i}h^i=\partial_{i}q^i=0,\nonumber
\\
	&h^{TTi}_{~~~~i}=S^{i}_{~i}=U^{i}_{~i}=V^{i}_{~i}=C^{i}_{~i}=D^{i}_{~i}=E^{i}_{~i}=0,\nonumber
	\\
	&\partial_{i}h^{TTi}_{~~~~j}=\partial_{i}S^{i}_{~j}=\partial_{i}U^{i}_{~j}=\partial_{i}V^{i}_{~j}=\partial_{i}C^{i}_{~j}=\partial_{i}D^{i}_{~j}=\partial_{i}E^{i}_{~j}=0,\nonumber
	\\
	&h^{TT}_{ij}=h^{TT}_{ji},~S_{ij}=S_{ji},~ U_{ij}=U_{ji},~ C_{ij}=C_{ji},\nonumber
	\\
	&D_{ij}=D_{ji},~ V_{ij}=-V_{ji},~ E_{ij}=-E_{ji},\nonumber
\\
	&B^{i}_{~jk}=	B^{i}_{~kj},~ B^{i}_{~ik}=B^{i}_{~jk}\delta^{jk}=0,~ \partial_{i}B^{i}_{~jk}=\partial^{j}B^{i}_{~jk}=0.
\end{eqnarray}
In the above equations, we have not distinguished between the upper and lower spatial indices, but this does not lead to any confusion. This is because, in linearized gravity theory, indices are raised and lowered by $\eta_{\mu\nu}$, while spatial indices are raised and lowered by
$\delta_{ij}$.

The decomposition method given in Eq. (\ref{decompose perturbations}) exists and is unique. The proof of this can be attributed to the existence and uniqueness of solutions to the Poisson's equation $\Delta u=f$ \cite{Weinberg}. Among them, considering that physical quantities generally tend to zero at infinity, we also require $f$ and $u$ to tend to zero at infinity, which imposes boundary conditions on the Poisson's equation. A solution to the Poisson's equation can be constructed using the Green's function, which also proves the existence of the solution. The uniqueness of the solution can be established using the maximum principle for harmonic functions \cite{Evans}.

After performing the same decomposition on $\xi^{\mu}$:
\begin{eqnarray}
	\label{decompose perturbations of xi}
	\xi^{0}=\xi^{0},\quad
	\xi_{i}=\xi^{T}_{~i}+\partial_{i}\xi,
\end{eqnarray}
where $\partial^{i}\xi^{T}_{~i}=0$, we can easily obtain the gauge transformation of the decomposed quantities in Eq. (\ref{decompose perturbations}). Furthermore, we find that the following combinations are invariant under the gauge transformation (\ref{gauge transformation}):

\begin{itemize}
	\item For transverse traceless third-order spatial tensor (here and in the subsequent text, spatial tensors refer to quantities that transform according to tensor laws under spatial rotations):
	\begin{eqnarray}
		\label{gauge invariant third-order spatial tensors}
		B^{i}_{~jk}.
	\end{eqnarray}
	
	\item For transverse traceless second-order spatial tensors:
	\begin{eqnarray}
		\label{gauge invariant second-order spatial tensors}
		h^{TT}_{ij},~S_{ij},~ U_{ij},~V_{ij},~ C_{ij},~D_{ij},~ E_{ij}.
	\end{eqnarray}
	
	\item For transverse spatial vectors:
	\begin{eqnarray}
		\label{gauge invariant spatial vectors}
		&N_{i},~C_{i},~\bar{C}_{i},~f_{i},~h_{i},~q_{i},
		\nonumber\\
		&\Xi_{i}\coloneqq \beta_{i}-\partial_{0}\epsilon_{i},
		\nonumber\\
		&\Omega_{i}\coloneqq G_{i}-\partial_{0}\bar{D}_{i},~
		\bar{\Omega}_{i}\coloneqq \bar{D}_{i}-\partial_{0}g_{i},~
		K_{i}\coloneqq\beta_{i}+\bar{D}_{i}.
	\end{eqnarray}
	
	\item For spatial scalars:
	\begin{eqnarray}
		\label{gauge invariant spatial scalars}
		&A,~\bar{A},~m,~n,
		\nonumber\\
		&\Theta\coloneqq \frac{1}{3}\left(H-\Delta\zeta\right),~
		\phi\coloneqq -\frac{1}{2}h_{00}+\partial_{0}\gamma-\frac{1}{2}\partial_{0}^{2}\zeta,
		\nonumber\\
	    &\Psi\coloneqq \Sigma^{0}_{~00}-\partial_{0}M,~
	    \bar{\Psi}\coloneqq M-\partial_{0}B,
	    \nonumber\\
	    &\Pi\coloneqq F-\partial_{0}\bar{B},~
	    \bar{\Pi}\coloneqq \bar{B}-\partial_{0}{l},
	    \nonumber\\
	    &K \coloneqq -h_{00}+2M,~
	    L \coloneqq H+2\Delta l.
	\end{eqnarray}
\end{itemize}

In decomposition (\ref{decompose perturbations}), there are 1 third-order transverse traceless spatial tensor, 7 second-order transverse traceless spatial tensors, 11 transverse spatial vectors, and 14 spatial scalars. The number of independent gauge invariants we combine is as follows: 1 third-order transverse traceless spatial tensor, 7 second-order transverse traceless spatial tensors, 10 transverse spatial vectors, and 12 spatial scalars. This is consistent with the fact that Eq. (\ref{decompose perturbations of xi}) contains an arbitrary transverse spatial vector and two arbitrary spatial scalars, which allow the gauge transformation to eliminate one transverse spatial vector degree of freedom and two spatial scalar degrees of freedom. Now, we have completed the construction of gauge invariant perturbations.

\section{The equation of motion for free particles in a torsionless Spacetime}
\label{sec: 3}

In this section, we examine the motion of free particles in a general torsionless spacetime. In the context of general relativity, Papapetrou first demonstrated that the geodesic equation governing the motion of free particles can be derived from the covariant conservation of the energy-momentum tensor of the matter field \cite{A. Papapetrou}. This conservation equation can be derived using the general covariance of the action of the matter field. This provides us with a method to derive the equations of motion for free particles based on the general covariance of the matter field action. Reference \cite{Damianos Iosifidis} investigated the general metric-affine spacetime case following this approach, and here, we apply the corresponding results to the case of torsionless spacetime. The latter exhibits an interesting property compared to that of general metric-affine spacetime.

\subsection{The equation of motion in general relativity}

We begin by reviewing the case of general relativity. In this context, the action of the matter field is a functional of both the metric $g_{\mu\nu}$ and the matter fields $\Phi^{a}$:
\begin{eqnarray}
	\label{Sm in GR}
	S_{m}\left[g_{\mu\nu},\Phi^{a}\right].
\end{eqnarray}
Here, we use the indicator $a$ to label multiple matter fields $\Phi^{a}$, so the above action can represent the general case. By varying the action (\ref{Sm in GR}) with respect to the metric, we can define the energy-momentum tensor of the matter fields
\begin{eqnarray}
	\label{energy-momentum tensor of the matter field}
	T_{\mu\nu}\coloneqq -\frac{2}{\sqrt{-g}}\frac{\delta S_{m}}{\delta g^{\mu\nu}}.
\end{eqnarray}
For the point-particle situation we are considering, we can write $T^{\mu\nu}$ in the following form \cite{Weinberg2}:
\begin{eqnarray}
	\label{Tmunu in point particle}
	T^{\mu\nu}\left(t,\textbf{x}\right)=\frac{1}{\sqrt{-g}}\frac{M^{\mu\nu}}{u^{0}}\delta^{3}\left(\textbf{x}-\textbf{X}\left(t\right)\right).
\end{eqnarray}
Here, $M^{\mu\nu}$ is a tensor, $\textbf{X}\left(t\right)$ is the particle's trajectory and $u^{0}$ is the temporal component of the particle's four-velocity $u^{\mu}\coloneqq dX^{\mu}/d\tau$ ($\tau$ is the proper time of the particle.).

From general covariance, it can be concluded that $T^{\mu\nu}$ satisfies the covariant conservation equation \cite{landau}
\begin{eqnarray}
	\label{covariant conservation equation of T in GR}
	\widehat{\nabla}_{\mu}T^{\mu\nu}=0.
\end{eqnarray}
Reference \cite{A. Papapetrou} states that, by applying the above equation in conjunction with Eq. (\ref{Tmunu in point particle}), one can conclude that a free particle in general relativity follows the geodesic equation
\begin{eqnarray}
	\label{geodesic equation}
	\frac{\widehat{D}u^{\mu}}{d\tau}=0.
\end{eqnarray}
A detailed derivation is presented in Appendix \ref{app: A}.

It is worth noting that as long as the action of the matter fields satisfies the form of Eq. (\ref{Sm in GR}), the equation of motion for free particles is the same as in general relativity.

\subsection{The equation of motion in a torsionless spacetime}
\label{The equation of motion in a torsionless spacetime}
Now, we study the motion of free particles in a torsionless spacetime. In this case, there is no reason to believe that the action of the matter fields lacks a connection. Therefore, in the general Palatini theory, the action of the matter fields has the form
\begin{eqnarray}
	\label{Sm in Palatini}
	S_{m}\left[g_{\mu\nu}, \Gamma^{\lambda}_{~\mu\nu}, \Phi^{a}\right].
\end{eqnarray}
{In the current theoretical framework of matter fields (i.e., the Standard Model), there is no direct coupling between matter fields and the connection. Nevertheless, one can hypothetically introduce such a coupling without contradicting existing solar-system experiments. As an illustrative example, consider a scalar matter field $\phi$ governed by the action
\begin{eqnarray}
	\label{Sm in phi}
	S_{m}=\int d^4{x}\sqrt{-g}
	\left[
	-\frac{1}{2}\partial_{\mu}\phi\!~\!\partial^{\mu}\phi-\frac{1}{2}m^{2}\phi^{2}-\xi R \phi^2
	\right].
\end{eqnarray}
Here, $R \coloneqq g^{\nu\sigma}R^{\mu}_{~\nu\mu\sigma}$, and $R^{\mu}_{~\nu\lambda\sigma}$ is the curvature tensor. In the solar system, where $R$ is extremely small, the parameter $\xi$ can be tuned so that the dynamics of $\phi$ are indistinguishable from those of a Klein-Gordon field with mass $m$. In contrast, in strong gravitational fields, $\phi$ exhibits behavior that deviates from the standard Klein-Gordon field. Moreover, as noted in Ref. \cite{Damianos Iosifidis6}, introducing a coupling between matter fields and the connection provides new avenues for exploring questions related to cosmic evolution.}

Just as in the case of general relativity, by varying the action (\ref{Sm in Palatini}) with respect to the metric, we can define $T^{\mu\nu}$, which is referred to here as the metric energy-momentum tensor:
\begin{eqnarray}
	\label{metric energy-momentum tensor of the matter field}
	T_{\mu\nu}\coloneqq -\frac{2}{\sqrt{-g}}\frac{\delta S_{m}}{\delta g^{\mu\nu}}.
\end{eqnarray}
Similarly, by varying the action (\ref{Sm in Palatini}) with respect to the connection $\Gamma^{\lambda}_{\mu\nu}$, we can define the hypermomentum tensor:
\begin{eqnarray}
	\label{hypermomentum tensor of the matter field}
	\mathcal{H}_{\lambda}^{~\mu\nu} \coloneqq -\frac{2}{\sqrt{-g}}\frac{\delta S_{m}}{\delta \Gamma^{\lambda}_{~\mu\nu}}.
\end{eqnarray}
When spacetime is torsionless, due to the symmetry of the $(\mu,\nu)$ indices in the connection, $\mathcal{H}_{\lambda}^{~\mu\nu}$ is symmetric with respect to the indices $(\mu,\nu)$. Both $T_{\mu\nu}$ and $\mathcal{H}_{\lambda}^{~\mu\nu}$ serve as sources in the gravitational field equations. By utilizing general covariance, it can be inferred that the covariant conservation equation in this case is satisfied \cite{Damianos Iosifidis2}
\begin{eqnarray}
	\label{covariant conservation equation in Palatini}
	\sqrt{-g}\left(2\widehat{\nabla}_{\mu}T^{\mu}_{~\alpha}-\mathcal{H}^{\lambda\mu\nu}R_{\lambda\mu\nu\alpha}\right)
	+\nabla_{\mu}\nabla_{\nu}\left(\sqrt{-g}\mathcal{H}_{\alpha}^{~\mu\nu}\right)=0.
\end{eqnarray}
It can be seen that when $\mathcal{H}_{\lambda}^{~\mu\nu}=0$, Eq. (\ref{covariant conservation equation in Palatini}) reduces to Eq. (\ref{covariant conservation equation of T in GR}). Therefore, a free particle without hypermomentum still satisfies Eq. (\ref{geodesic equation}).

When $\mathcal{H}_{\lambda}^{~\mu\nu} \neq 0$, it is convenient to introduce a canonical energy-momentum tensor. By expressing $S_{m}$ in the form of the tetrad $e_{\mu}^{~a}$ and the spin connection $\omega_{\mu ab}$ \cite{Nikodem J. Poplawski}, the canonical energy-momentum tensor can be defined as \cite{Damianos Iosifidis2}
\begin{eqnarray}
	\label{canonical energy-momentum tensor of the matter field}
	\Sigma^{\mu}_{~\lambda}\coloneqq \frac{1}{\sqrt{-g}}\frac{\delta S_{m}}{\delta e_{\mu}^{~a}}e_{\lambda}^{~a}.
\end{eqnarray}
It should be noted that $\Sigma^{\mu\nu}$ is generally not symmetric. $\Sigma^{\mu}_{~\lambda}$ is not independent, it can be represented by $T^{\mu\nu}$ and $\mathcal{H}_{\lambda}^{~\mu\nu}$ \cite{Damianos Iosifidis2}:
\begin{eqnarray}
	\label{relationship canonical energy-momentum tensor}
	\Sigma^{\mu}_{~\lambda}
	=T^{\mu}_{~\lambda}+\frac{1}{2\sqrt{-g}}\nabla_{\nu}\left(\sqrt{-g}\mathcal{H}_{\lambda}^{~\mu\nu}\right).
\end{eqnarray}
When taking the solution $g_{\mu\nu}=\eta_{\mu\nu}$ and $\Gamma^{\lambda}_{~\mu\nu}=0$ in a flat spacetime, where the theory should reduce to special relativity, Eq. (\ref{covariant conservation equation in Palatini}) shows that $\Sigma^{\mu}_{~\nu}$ is conserved in this case: $\partial_{\mu}\Sigma^{\mu}_{~\nu}=0$. In this sense, $\Sigma^{\mu}_{~\lambda}$ can be reasonably assumed to be an extension of the typical conserved energy-momentum tensor in special relativity for general spacetime, rather than $T^{\mu}_{~\nu}$. 

After introducing $\Sigma^{\mu}_{~\lambda}$, Eqs. (\ref{covariant conservation equation in Palatini}) and (\ref{relationship canonical energy-momentum tensor}) can be equivalently represented as \cite{Damianos Iosifidis2,Damianos Iosifidis,Damianos Iosifidis3,Damianos Iosifidis4}
\begin{eqnarray}
	\label{covariant conservation equation in Palatini equivalently hypermomentum}
	\widehat{\nabla}_{\nu}\mathcal{H}_{\lambda}^{~\mu\nu}\!&=&\!
	2\left(\Sigma^{\mu}_{~\lambda}-T^{\mu}_{~\lambda}\right)
	-N^{\mu}_{~\alpha\beta}\mathcal{H}_{\lambda}^{~\alpha\beta}
	+N^{\alpha}_{~\lambda\beta}\mathcal{H}_{\alpha}^{~\mu\beta},
	\\
	\label{covariant conservation equation in Palatini equivalently canonical energy-momentum tensor}
	\widehat{\nabla}_{\mu}\Sigma^{\mu}_{~\alpha}\!&=&\!
	\frac{1}{4}\widehat{R}_{\lambda\mu\nu\alpha}
	\left(\mathcal{H}^{\lambda\mu\nu}-\mathcal{H}^{\mu\lambda\nu}\right)
	+\frac{1}{2}\widehat{\nabla}_{\lambda}\left(\mathcal{H}^{\mu\nu\lambda}N_{\mu\nu\alpha}\right)
	-\frac{1}{2}\mathcal{H}_{\lambda}^{~\mu\nu}\widehat{\nabla}_{\alpha}N^{\lambda}_{~\mu\nu}.
\end{eqnarray}
Here, 
\begin{eqnarray}
	\label{N lambda munu}
     N^{\lambda}_{~\mu\nu}\coloneqq \Gamma^{\lambda}_{~\mu\nu}-\widehat{\Gamma}^{\lambda}_{~\mu\nu}
\end{eqnarray}
represents the deviation of the connection from the Levi-Civita connection. Equations (\ref{covariant conservation equation in Palatini equivalently hypermomentum}) and (\ref{covariant conservation equation in Palatini equivalently canonical energy-momentum tensor}) are the equations we need to derive the equation of motion. (In fact, as demonstrated in the following text, Eq. (\ref{covariant conservation equation in Palatini equivalently hypermomentum}) is sufficient for the case of torsionless spacetime.)

In Ref. \cite{Damianos Iosifidis}, the authors derived the equation of motion for free particles in a general metric-affine spacetime. They assumed specific forms for $T^{\mu\nu}$, $\mathcal{H}_{\lambda}^{~\mu\nu}$, and $\Sigma^{\mu}_{~\nu}$ in the context of a point particle, as follows:

\textbf{Assumption 0}:
\begin{eqnarray}
	\label{Tmunu in palatini Assumption 0}
	T^{\mu\nu}\left(t,\textbf{x}\right)\!&=&\!
	\frac{1}{\sqrt{-g}}\frac{mu^{\mu}u^{\nu}}{u^{0}}\delta^{3}\left(\textbf{x}-\textbf{X}\left(t\right)\right),\\
	\label{Hlambndamunu in palatini Assumption 0}
	\mathcal{H}_{\lambda}^{~\mu\nu}\left(t,\textbf{x}\right)\!&=&\!
	\frac{1}{\sqrt{-g}}\frac{H_{\lambda}^{~\mu}u^{\nu}}{u^{0}}\delta^{3}\left(\textbf{x}-\textbf{X}\left(t\right)\right),\\
	\label{Sigma in palatini Assumption 0}
	\Sigma^{\mu}_{~\nu}\left(t,\textbf{x}\right)\!&=&
	\!\frac{1}{\sqrt{-g}}\frac{P_{\nu}u^{\mu}}{u^{0}}\delta^{3}\left(\textbf{x}-\textbf{X}\left(t\right)\right).
\end{eqnarray}
Here, Eq. (\ref{Tmunu in palatini Assumption 0}) still assumes the standard form of a structureless point particle \cite{Weinberg2}. Equation (\ref{Sigma in palatini Assumption 0}) assumes that the four-momentum $P^{\mu}$ propagates along the four-velocity $u^{\mu}$ \cite{Damianos Iosifidis}. Regarding the hypermomentum, the authors asserted that the form of Eq. (\ref{Hlambndamunu in palatini Assumption 0}) serves as a valid ansatz in many practical applications \cite{Damianos Iosifidis}. It should be noted that, in a general metric-affine spacetime, the relationship $P^{\mu}=mu^{\mu}$ often requires correction due to Eq. (\ref{relationship canonical energy-momentum tensor}) \cite{Damianos Iosifidis}. For the torsionless spacetime case under consideration, since $\mathcal{H}_{\lambda}^{~\mu\nu}$ is symmetric with respect to $(\mu,\nu)$, Eq. (\ref{Hlambndamunu in palatini Assumption 0}) naturally requires 
\begin{eqnarray}
	\label{Hlambndamunu in torsionless spacetime palatini}
	\mathcal{H}_{\lambda}^{~\mu\nu}\left(t,\textbf{x}\right)=
	\frac{1}{\sqrt{-g}}\frac{H_{\lambda}u^{\mu}u^{\nu}}{u^{0}}\delta^{3}
	\left(\textbf{x}-\textbf{X}\left(t\right)\right).
\end{eqnarray}
﻿In fact, the above assumption can be relaxed without changing anything. A proof of this is provided in Appendix \ref{app: B}. It is worth noting that, in Appendix B, we demonstrate that assumptions (\ref{Tmunu in palatini Assumption 0})–(\ref{Sigma in palatini Assumption 0}) are equivalent to assumptions (\ref{Tmunu in palatini Assumption 2})–(\ref{Sigma in palatini Assumption 2}), which correspond to considering only the monopole moment of the body. Therefore, our assumptions actually correspond to the monopole approximation in the multipole expansion.

In order to derive the equation of motion satisfied by free particles, we rewrite Eq. (\ref{covariant conservation equation in Palatini equivalently hypermomentum}):
\begin{eqnarray}
	\label{rewite covariant conservation equation in Palatini equivalently hypermomentum}
	\frac{1}{\sqrt{-g}}\partial_{\nu}\left(\sqrt{-g}\mathcal{H}_{\lambda}^{~\mu\nu}\right)
	+\widehat{\Gamma}^{\mu}_{~\nu\rho}\mathcal{H}_{\lambda}^{~\rho\nu}
	-\widehat{\Gamma}^{\rho}_{~\nu\lambda}\mathcal{H}_{\rho}^{~\mu\nu}
	=
	2\left(\Sigma^{\mu}_{~\lambda}-T^{\mu}_{~\lambda}\right)
	-N^{\mu}_{~\alpha\beta}\mathcal{H}_{\lambda}^{~\alpha\beta}
	+N^{\alpha}_{~\lambda\beta}\mathcal{H}_{\alpha}^{~\mu\beta}.\nonumber\\
\end{eqnarray}
By substituting Eqs. (\ref{Tmunu in palatini Assumption 0}), (\ref{Sigma in palatini Assumption 0}), and (\ref{Hlambndamunu in torsionless spacetime palatini}) into Eq. (\ref{rewite covariant conservation equation in Palatini equivalently hypermomentum}) and integrating Eq. (\ref{rewite covariant conservation equation in Palatini equivalently hypermomentum}) with $\int d^{3}x \sqrt{-g}$, we can obtain:
\begin{eqnarray}
	\label{integrating rewite covariant conservation equation in Palatini equivalently hypermomentum}
	u^{\mu}\frac{\widehat{D}}{d\tau}H_{\lambda}+H_{\lambda}\frac{\widehat{D}u^{\mu}}{d\tau}
	=
	2\left(u^{\mu}P_{\lambda}-mu^{\mu}u_{\lambda}\right)
	-N^{\mu}_{~\alpha\beta}H_{\lambda}u^{\alpha}u^{\beta}
	+N^{\alpha}_{~\lambda\beta}H_{\alpha}u^{\mu}u^{\beta}.
\end{eqnarray}
This corresponds to the torsionless spacetime case of Eq. (99) in Ref. \cite{Dirk Puetzfeld1}. By multiplying both sides of the equation by $u_{\mu}$ and using $u_{\mu}u^{\mu}=-1$ and $u_{\mu}\widehat{D}u^{\mu}/d\tau=0$, it can be concluded that 
\begin{eqnarray}
	\label{umu integrating rewite covariant conservation equation in Palatini equivalently hypermomentum}
	-\frac{\widehat{D}}{d\tau}H_{\lambda}
	=
	2\left(-P_{\lambda}+mu_{\lambda}\right)
	-N^{\rho}_{~\alpha\beta}H_{\lambda}u^{\alpha}u^{\beta}u_{\rho}
	-N^{\alpha}_{~\lambda\beta}H_{\alpha}u^{\beta}.
\end{eqnarray}
And by substituting Eq. (\ref{umu integrating rewite covariant conservation equation in Palatini equivalently hypermomentum}) into Eq. (\ref{integrating rewite covariant conservation equation in Palatini equivalently hypermomentum}), we have
\begin{eqnarray}
	\label{integrating rewite covariant conservation equation in Palatini equivalently hypermomentum new}
	H_{\lambda}\left(\frac{\widehat{D}u^{\mu}}{d\tau}
	+N^{\mu}_{~\alpha\beta}u^{\alpha}u^{\beta}
	+N^{\rho}_{~\alpha\beta}u^{\alpha}u^{\beta}u_{\rho}u^{\mu}
	\right)=0.
\end{eqnarray}
Since $H_{\lambda} \neq 0$ under our consideration, we can deduce that
\begin{eqnarray}
	\label{equation of motion in Palatini}
	\frac{\widehat{D}u^{\mu}}{d\tau}
	+N^{\mu}_{~\alpha\beta}u^{\alpha}u^{\beta}
	+N^{\rho}_{~\alpha\beta}u^{\alpha}u^{\beta}u_{\rho}u^{\mu}
    =0.
\end{eqnarray}
This is the equation of motion we need. 

From Eq. (\ref{equation of motion in Palatini}), it can be seen that the motion of a free particle is only related to its initial position and velocity, and is independent of the specific values of its charges $m$, $H_{\mu}$, and $P_{\mu}$. This is an interesting property because it suggests that we can still interpret the free motion of particles as geometric effects in this case. One might argue that the independence of motion from the magnitude of the hypermomentum implies that hypermomentum cannot be measured, which could appear physically counterintuitive. However, in our analysis, we consider only the gravitational fields (the metric and the connection) and neglect any external fields. In addition, we work within the monopole approximation. It is therefore plausible that, when external fields are included or higher-order multipole moments are considered, the magnitude of the hypermomentum would influence the motion of the body, rendering the hypermomentum, in principle, measurable. For a general metric-affine spacetime, the motion of particles is often related to the specific values of their charges \cite{Damianos Iosifidis}. Finally, we note that the validity of the monopole approximation requires that, in the equations under consideration, the contributions from hypermomentum and the energy-momentum tensor are of the same order of magnitude. In situations where the hypermomentum contribution is small compared to that of the energy-momentum tensor, it can be neglected at leading order, reducing the equations of motion to Eq. (\ref{geodesic equation}).

In the torsionless spacetime we are considering, $N^{\lambda}_{~\mu\nu}$ depends solely on the non-metricity $Q_{\lambda\mu\nu}\coloneqq \nabla_{\lambda} g_{\mu\nu}$, specifically,  $N^{\lambda}_{~\mu\nu}=\frac{1}{2}Q^{\lambda}_{~\mu\nu}-Q_{(\mu~\nu)}^{~~\lambda}$ \cite{Lavinia Heisenberg2}. Therefore, whether a particle possesses a non-negligible hypermomentum charge $H_{\mu}$ determines whether it can experience the effects of non-metricity. If a particle does not carry a hypermomentum charge $H_{\mu}$ (or if the hypermomentum contribution is negligible) or if spacetime is free from non-metricity ($N^{\lambda}_{~\mu\nu}=0$), the equation of motion of particles will reduce to Eq. (\ref{geodesic equation}), which is identical to the equation in general relativity.

\textcolor{blue}{}.


\section{Shear mode: a new class of gravitational wave polarization modes}
\label{sec: 4}

Gravitational waves are theoretically detected by measuring the relative displacement of two adjacent free test particles. Among them, the polarization modes of gravitational waves are defined by various motion modes of relative displacement \cite{Eardley}. Due to the fact that the equation of motion for particles with hypermomentum in torsionless spacetime is different from Eq. (\ref{geodesic equation}), this may lead to a correction in the relative motion equation between the two particles, which could, in turn, affect the definition of gravitational wave polarization modes. In this section, we will derive the relative motion equations for two particles with hypermomentum and analyze the possible gravitational wave polarization modes in torsionless spacetime.

Let $A$ and $B$ label two adjacent test particles, with trajectories $X^{\mu}\left(\tau\right)$ and $X^{\mu}\left(\tau\right)+\eta^{\mu}\left(\tau\right)$, respectively, where $\eta^{\mu}\left(\tau\right)$ represents the displacement between the two particles. In other words, 
$\eta^{\mu}$ is the displacement vector linking the positions of particles $A$ and $B$ at the same proper time. Both $A$ and $B$, due to their hypermomentum, satisfy Eq. (\ref{equation of motion in Palatini}). Therefore, for particle $A$, the equation of motion can be written as
\begin{eqnarray}
	\label{equation of motion for A}
	\frac{d^2 X^{\mu}}{d\tau^2}
	=
	-\widehat{\Gamma}^{\mu}_{~\nu\rho}\left(X\right)\frac{dX^{\nu}}{d\tau}\frac{dX^{\rho}}{d\tau}
	-N^{\mu}_{~\alpha\beta}\left(X\right)\frac{dX^{\alpha}}{d\tau}\frac{dX^{\beta}}{d\tau}
	-N^{\rho}_{~\alpha\beta}\left(X\right)\frac{dX^{\alpha}}{d\tau}\frac{dX^{\beta}}{d\tau}\frac{dX_{\rho}}{d\tau}\frac{dX^{\mu}}{d\tau}.
\end{eqnarray}
Similarly, for $B$, we have
\begin{eqnarray}
	\label{equation of motion for B}
	\frac{d^2 \left(X^{\mu}+\eta^{\mu}\right)}{d\tau^2}
	\!&=&\!
	-\widehat{\Gamma}^{\mu}_{~\nu\rho}\left(X+\eta\right)\frac{d\left(X^{\nu}+\eta^{\nu}\right)}{d\tau}\frac{d\left(X^{\rho}+\eta^{\rho}\right)}{d\tau}
	\nonumber\\
	\!&-&\!
	N^{\mu}_{~\alpha\beta}\left(X+\eta\right)\frac{d\left(X^{\alpha}+\eta^{\alpha}\right)}{d\tau}\frac{d\left(X^{\beta}+\eta^{\beta}\right)}{d\tau}
	\nonumber\\
	\!&-&\!
	N^{\rho}_{~\alpha\beta}\left(X+\eta\right)\frac{d\left(X^{\alpha}+\eta^{\alpha}\right)}{d\tau}\frac{d\left(X^{\beta}+\eta^{\beta}\right)}{d\tau}\frac{d\left(X_{\rho}+\eta_{\rho}\right)}{d\tau}\frac{d\left(X^{\mu}+\eta^{\mu}\right)}{d\tau}.
\end{eqnarray}

In laboratory settings, it is common to choose the proper detector frame where the position of particle $A$ is $X^{\mu}=(\tau,0,0,0)$, and in the neighborhood of $A$'s trajectory, the metric satisfies \cite{Michele Maggiore,MTW}
\begin{eqnarray}
	\label{proper detector frame}
   ds^{2}\approx 
   \!&-&\!
   dt^{2}\left[1+2a_{i} \delta x^{i}
   +\left(a_{i}\delta x^{i}\right)^2
   +\widehat{R}_{i0j0}\delta x^{i} \delta x^{j}\right]
   \nonumber\\
   \!&+&\!
  2dtdx^{i}\left[-\frac{2}{3}\widehat{R}_{0jik}\delta x^{j} \delta x^{k}\right]
   +
   dx^{i}dx^{j}\left[\delta_{ij}-\frac{1}{3}\widehat{R}_{ikjl}\delta x^{k}\delta x^{l}\right],
\end{eqnarray}
where $\delta x^{i}=x^{i}-X^{i}$, and using Eq. (\ref{equation of motion in Palatini}), $a^{i}=-N^{i}_{~00}\left(X\right)$ (index $i$ is raised and lowered by $\delta_{ij}$). 
Note that $A$ does not satisfy Eq. (\ref{geodesic equation}) \cite{MTW}, and it is typically impossible to achieve $\widehat{\Gamma}^{\lambda}_{~\mu\nu}\left(X\right)=0$ along $A$'s trajectory simultaneously. In the proper detector frame (\ref{proper detector frame}), 
Eq. (\ref{equation of motion for A}) requires
\begin{eqnarray}
	\label{equation of motion for A in tau000}
	\widehat{\Gamma}^{i}_{~00}\left(X\right)
	=
	-N^{i}_{~00}\left(X\right)=a^{i}.
\end{eqnarray}

We assume that, before the arrival of the gravitational waves, the metric and connection in the coordinate system above satisfy Eq. (\ref{flat spacetime background}), and that the two particles are relatively stationary. At this point, we denote the displacement vector, which remains constant over time, as $\bar{\eta}^{\mu}$ and set the starting point of the proper time such that $\bar{\eta}^{0}=0$. Subsequently, a gravitational wave propagates into the region of spacetime where the particles are located, causing a perturbation (\ref{flat spacetime background perturbations}) of order $h$ ($h\ll1$) in the background spacetime. This further implies that $\widehat{\Gamma}^{\lambda}_{~\mu\nu}$ and $N^{\lambda}_{~\mu\nu}$ in Eq. (\ref{equation of motion in Palatini}) are of the same order of magnitude as $h$, and leads to a relative change of order $h$ in the relative displacement $\eta^{\mu}$, i.e., $|\eta^{\mu}-\bar{\eta}^{\mu}| \sim \mathcal{O}\left(h\right)\eta^{\mu}$. 
Based on the above discussion, and assuming that the relative displacement is extremely small (much smaller than the typical scale of variation of the gravitational field), by retaining only the leading-order term, Eq. (\ref{equation of motion for B}) can be approximated as \cite{Michele Maggiore}
\begin{eqnarray}
	\label{equation of motion for B in tau000}
	\frac{d^2 \eta^{\mu}}{dt^2}
	\!&=&\!
	-\partial_{\nu}\widehat{\Gamma}^{\mu}_{~00}\left(X\right)\eta^{\nu}
	-\partial_{\nu}N^{\mu}_{~00}\left(X\right)\eta^{\nu}
	+\partial_{\nu}N^{0}_{~00}\left(X\right)u^{\mu}\eta^{\nu}.
\end{eqnarray}
Here, we use Eq. (\ref{equation of motion for A}) to simplify the above equation. Note that the equation here does not include terms like $\widehat{\Gamma}^{\mu}_{~\nu\lambda}u^{\nu}{d\eta^{\lambda}}/{d\tau}$. This is because both 
$\widehat{\Gamma}^{\mu}_{~\nu\lambda}$ and ${d\eta^{\lambda}}/{d\tau}$ are of the order of $h$, making their combination a small quantity of the order of $h^{2}$. Similarly, since $1+g_{00}$ and 
$\eta^{\mu}-\bar{\eta}^{\mu}$ are of order $h$, and thus ${d^2 \eta^{\mu}}/{dt^2}-{d^2 \eta^{\mu}}/{d\tau^2}$ is a small quantity of order $h^{2}$, we can replace the $d\tau$ on the left-hand side of the equation with $dt$. Assuming the arrival time of gravitational waves is $t_{0}=0$, the initial condition for the equation is given by
\begin{eqnarray}
	\label{initial condition}
	\eta^{\mu}\left(0\right)=\bar{\eta}^{\mu},\quad \frac{d\eta^{\mu}}{dt}\left(0\right)=0.
\end{eqnarray}

Using Eq. (\ref{proper detector frame}), we can further calculate $\partial_{\nu}\widehat{\Gamma}^{\mu}_{~00}\left(X\right)$ (accurate to $h$-order):
\begin{eqnarray}
	\label{PnuGammamu00}
	\partial_{0}\widehat{\Gamma}^{0}_{~00}\left(X\right)=0,\quad
	\partial_{j}\widehat{\Gamma}^{i}_{~00}\left(X\right)=\widehat{R}^{i}_{~0j0}\left(X\right),\nonumber\\
	\partial_{0}\widehat{\Gamma}^{i}_{~00}\left(X\right)=\partial_{i}\widehat{\Gamma}^{0}_{~00}\left(X\right)
	=-\partial_{0}N^{i}_{~00}\left(X\right)=\frac{da^{i}}{dt}.
\end{eqnarray}
By substituting Eq. (\ref{PnuGammamu00}) into Eq. (\ref{equation of motion for B in tau000}), we find that $\eta^{0}$ satisfies
\begin{eqnarray}
	\label{equation of motion for eta0}
	\frac{d^2 \eta^{0}}{dt^2}
	=
	-\frac{da_{i}}{dt}\eta^{i}.
\end{eqnarray}
This implies that $\eta^{0}$ is not always zero in our reference frame. However, this is not surprising. Recall that $\eta^{\mu}$ connects two points, $A$ and $B$, at the same proper time, and that $t$ is strictly equal to the proper time at $A$ in our reference frame. Therefore, consider the two points $X^{\mu}\left(t_{a}\right)$ and $X^{\mu}\left(t_{a}+dt\right)$ at point $A$, and the corresponding points $X^{\mu}\left(t_{a}\right)+\eta^{\mu}\left(t_{a}\right)$ and $X^{\mu}\left(t_{a}+dt\right)+\eta^{\mu}\left(t_{a}+dt\right)$ in $B$, with the same proper time, respectively. It follows that the difference in proper time between the two points at $B$ is $d\tau_{B}=dt$, and the difference in coordinate time between these points is $dt_{B}=dt+d\eta^{0}$ ($d\eta^{0}\ll dt$). Since the trajectory of $B$ deviates only minimally from $\bar{\eta}^{\mu}$, and using Eq. (\ref{proper detector frame}), we obtain 
\begin{eqnarray}
	dt=d\tau_{B} \approx \sqrt{1+2a_{i}\bar{\eta}^{i}}dt_{B} \approx \left(1+a_{i}\bar{\eta}^{i}\right)dt_{B}
	\approx dt+d\eta^{0}+a_{i}\bar{\eta}^{i}dt.
\end{eqnarray}
On the other hand, similarly, due to the small deviation between $\eta^{\mu}$ and $\bar{\eta}^{\mu}$, from Eq. (\ref{equation of motion for eta0}) and the initial condition (\ref{initial condition}), we have 
\begin{eqnarray}
	\left(\frac{d^2 \eta^{0}}{dt^2}
	=
	-\frac{da_{i}}{dt}\eta^{i} \approx
	-\frac{da_{i}}{dt}\bar{\eta}^{i}\right)
	\Rightarrow
	\left(\frac{d\eta^{0}}{dt}=-a_{i}\bar{\eta}^{i}\right)
	\Rightarrow
	\left(d\eta^{0}+a_{i}\bar{\eta}^{i}dt=0\right).
\end{eqnarray}
It is evident that the two equations above describe the same relationship. Therefore, $\eta^{0}$
is not always zero, which implies that the particle $A$ does not satisfy Eq. (\ref{geodesic equation}). In the case of general relativity, where $a^{i}=0$, the equation simplifies to the standard form commonly encountered in textbook examples \cite{Michele Maggiore}.

Let us now focus on the more interesting spatial components of Eq. (\ref{equation of motion for B in tau000}), which define the polarization modes of gravitational waves. Using Eq. (\ref{PnuGammamu00}), we find that $\eta^{i}$ satisfies 
\begin{eqnarray}
	\label{equation of motion for etai}
	\frac{d^2 \eta^{i}}{dt^2}
	=
	-A^{i}_{~j}\eta^{j}
	\coloneqq
	-\widehat{R}^{i}_{~0j0}\eta^{j}-\partial_{j}N^{i}_{~00}\eta^{j}.
\end{eqnarray}
Strictly speaking, since $\eta^{0}\left(t\right)\neq0$, $\eta^{i}\left(t\right)$ obtained from Eq. (\ref{equation of motion for etai}) does not represent the displacement of $B$ relative to $A$ at coordinate time $t$, but rather at coordinate time $t+\eta^{0}\left(t\right)$. However, since the difference between the two is of the order of $h^{2}$ ($\eta^{0}  d\eta^{i}/dt \sim \mathcal{O}\left(h^{2}\right)$), this difference can be safely ignored. In the above sense, $\eta^{i}$ can be interpreted as the spatial displacement of $B$ relative to $A$ at time $t$, within the reference frame under consideration.

Given that $|\eta^{\mu}-\bar{\eta}^{\mu}| \sim \mathcal{O}\left(h\right)\eta^{\mu}$, Eq. (\ref{equation of motion for etai}) can be approximated as
\begin{eqnarray}
	\label{equation of motion for etai approximated}
	\frac{d^2 \eta^{i}}{dt^2}
	=
	-A^{i}_{~j}\bar{\eta}^{j}.
\end{eqnarray}
Furthermore, we can propose an approximate solution as
\begin{eqnarray}
	\label{approximate solution of equation of motion for etai approximated}
	\eta^{i}\left(t\right)
	=
	\bar{\eta}^{i}
	-\bar{\eta}^{j} \int_{0}^{t}dt' \int_{0}^{t'} dt'' A^{i}_{~j} \left(t''\right).
\end{eqnarray}
The initial condition (\ref{initial condition}) has already been applied at this stage. It follows that the relative motion of particles $\eta^{i}-\bar{\eta}^{i}$ depends linearly on the value of $A^{i}_{~j}$. Therefore, the polarization modes of gravitational waves can be defined in terms of $A^{i}_{~j}$.

Let us first review the case where the equation of motion satisfies (\ref{geodesic equation}), which indicates that the test particles do not carry a hypermomentum charge (or the effect of hypermomentum is negligible). At this stage, the equation of relative motion for particles is exactly what we know as \cite{landau,MTW,Michele Maggiore}
\begin{eqnarray}
	\label{geodetic deviation equation}
	\frac{d^2 \eta^{i}}{dt^2}
	=
	-\widehat{R}^{i}_{~0j0}\eta^{j}.
\end{eqnarray}
We consider plane gravitational waves propagating in the $+z$ direction, without loss of generality. This means that
\begin{eqnarray}
	\label{Ri0j0=AEeikx}
	\widehat{R}_{i0j0}=A_{0} E_{ij} e^{ikx}.
\end{eqnarray}
Here, $k^{\mu}$
denotes the four-wave vector, $A_{0}$ represents the wave intensity, and $E_{ij}$ is a $3 \times 3$ matrix encapsulating all polarization information of this plane wave, which satisfies
\begin{eqnarray}
	\label{EE=1}
	E_{ij}E^{ij}=1.
\end{eqnarray}
Owing to the symmetry of $\widehat{R}_{i0j0}$ with respect to the indices $\left(i, j\right)$, $E_{ij}$ possesses only six independent components. Consequently, this allows for the definition of up to six independent gravitational wave polarization modes $P_{1}$ to $P_{6}$ \cite{Eardley}:
\begin{eqnarray}
	\label{P1-P6}
	\widehat{R}_{i0j0}=\begin{pmatrix}
		P_{4}+P_{6} & P_{5} & P_{2}\\
		P_{5}       & -P_{4}+P_{6}  & P_{3}\\
		P_{2}       &  P_{3}   &   P_{1}
	\end{pmatrix}.
\end{eqnarray}

\begin{figure*}[htbp]
	\makebox[\textwidth][c]{\includegraphics[width=1.2\textwidth]{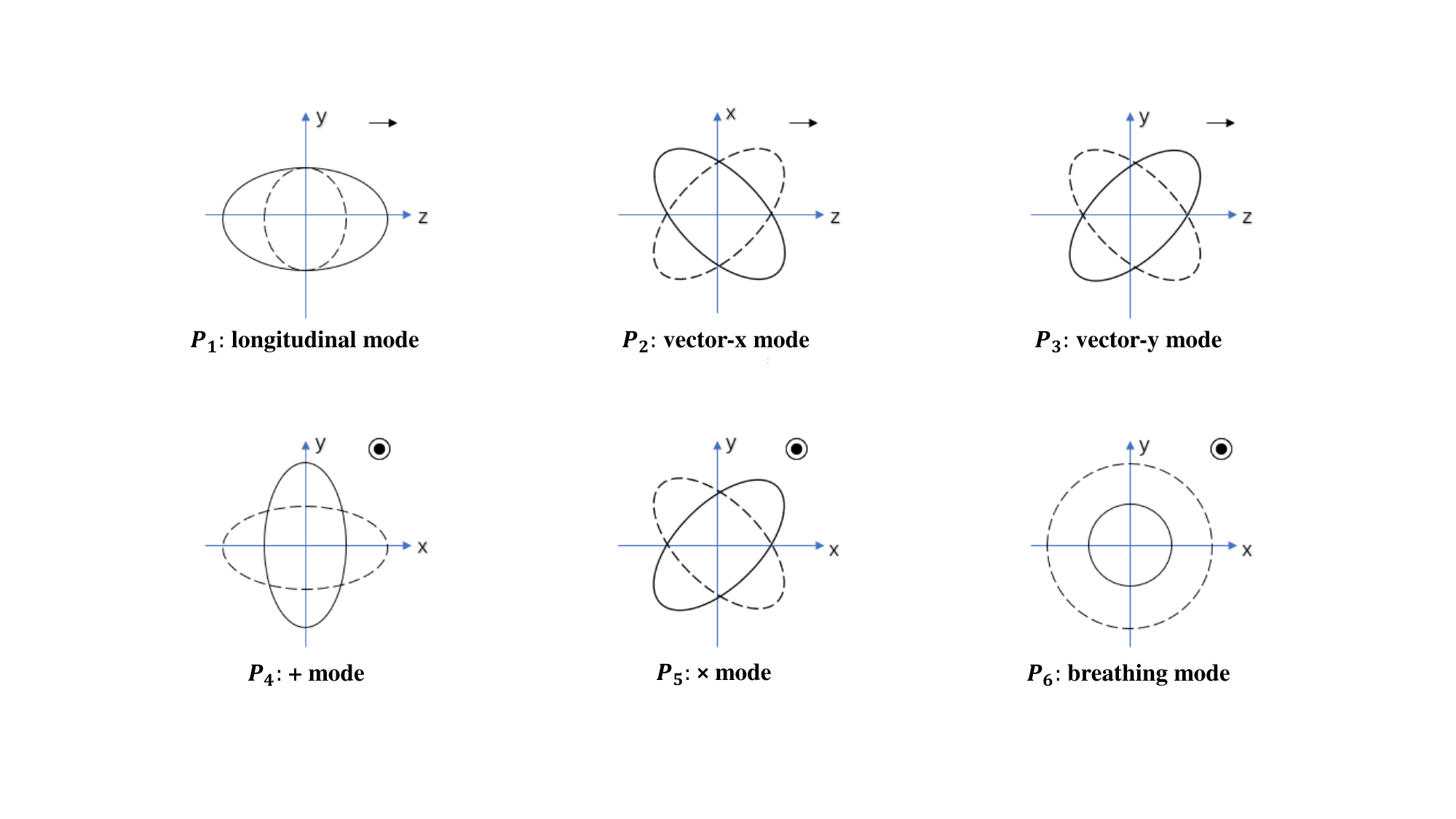}}
	\caption{Six polarization modes of gravitational waves \cite{Eardley}. The gravitational waves propagate in the $+z$ direction. The solid line represents the shape of the test particle array—initially spherical—depicting the motion of particles relative to the central particle of the sphere when the wave phase is $\frac{\pi}{2}$. The dotted line illustrates the shape of the array at a phase of $\frac{3}{2}\pi$. No relative motion occurs between test particles along the third axis, which is not shown in the figure.}
	\label{fig: 1}
\end{figure*}

Any gravitational wave polarization mode can be expressed as a linear superposition of these six independent modes. Similarly, due to linearity, the relative motion of test particles $\eta^{i}-\bar{\eta}^{i}$ can be written as a superposition of six independent motion modes, each corresponding to one of these six polarization modes of gravitational waves. The relative motions of the test particles corresponding to these six polarization modes are shown in Fig. \ref{fig: 1}.

By using Eqs. (\ref{decompose perturbations}), (\ref{gauge invariant second-order spatial tensors}), (\ref{gauge invariant spatial vectors}), and (\ref{gauge invariant spatial scalars}), we can express $\widehat{R}_{i0j0}$ as a gauge invariant:
\begin{eqnarray}
	\label{Ri0j0 gauge invariant}
	\widehat{R}_{i0j0}=-\frac{1}{2}\partial_{0}\partial_{0}h^{TT}_{ij}
	+\frac{1}{2}\partial_{0}\partial_{i}\Xi_{j}
	+\frac{1}{2}\partial_{0}\partial_{j}\Xi_{i}
	+\partial_{i}\partial_{j}\phi
	-\frac{1}{2}\delta_{ij}\partial_{0}\partial_{0}\Theta.
\end{eqnarray}
It follows that the six polarization modes of gravitational waves satisfy
\begin{eqnarray}
	\label{P1-P6 gauge invariant}
	\begin{array}{l}
		P_{1}=\partial_{3}\partial_{3}\phi-\frac{1}{2}\partial_{0}\partial_{0}\Theta, \quad
		P_{2}=\frac{1}{2}\partial_{0}\partial_{3}\Xi_{1},\\
		P_{3}=\frac{1}{2}\partial_{0}\partial_{3}\Xi_{2},  \quad \quad\quad\quad\,\,
		P_{4}=-\frac{1}{2}\partial_{0}\partial_{0}h^{TT}_{11}, \\
		P_{5}=-\frac{1}{2}\partial_{0}\partial_{0}h^{TT}_{12}, \quad\quad~~~
		P_{6}=-\frac{1}{2}\partial_{0}\partial_{0}\Theta.
	\end{array}
\end{eqnarray}
Since $P_{4}$ and $P_{5}$ are induced by spatial tensors, we classify them as tensor modes. Similarly, $P_{2}$ and $P_{3}$ correspond to vector modes, while $P_{1}$ and $P_{6}$ are scalar modes.

The case differs for particles with a hypermomentum charge. At this stage, from Eq. (\ref{equation of motion for etai}), $A_{ij}$ assumes the same role as $\widehat{R}_{i0j0}$ in the previous case. Considering 
\begin{eqnarray}
    \label{Aij=}
     A_{ij}
     \coloneqq
    \widehat{R}_{i0j0}+\partial_{j}N_{i00},
\end{eqnarray}
it becomes apparent that the additional term $\partial_{j}N_{i00}$ breaks the symmetry of $\left(i, j\right)$. Therefore, the number of independent components in $A_{ij}$ exceeds six, implying the existence of more than six independent polarization modes of gravitational waves. Given that the gravitational waves we are considering propagate in the $+z$ direction, only $\partial_{3}N_{i00}$ in 
$\partial_{j}N_{i00}$ can be non-zero. So, we can define the independent polarization modes of gravitational waves as
\begin{eqnarray}
	\label{P1-P8}
	A_{ij}=\begin{pmatrix}
		P_{4}+P_{6} & P_{5} & P_{2}+P_{7}\\
		P_{5}       & -P_{4}+P_{6}  & P_{3}+P_{8}\\
		P_{2}       &  P_{3}   &   P_{1}
	\end{pmatrix}.
\end{eqnarray}
Compared to the previous case (see Eq. (\ref{P1-P6})), two additional modes, $P_{7}$ and $P_{8}$, arise. Thus, there are now eight independent polarization modes of gravitational waves.

$P_{7}$ and $P_{8}$ are solely related to $\partial_{j}N_{i00}$, reflecting the deviation of spacetime geometry from Riemannian geometry. In the torsionless spacetime case we consider, $P_{7}$ and $P_{8}$ arise from non-metricity, offering a method for verifying the presence of non-metricity effects through the detection of gravitational wave polarization modes.

\begin{figure*}[htbp]
	\makebox[\textwidth][c]{\includegraphics[width=0.77\textwidth]{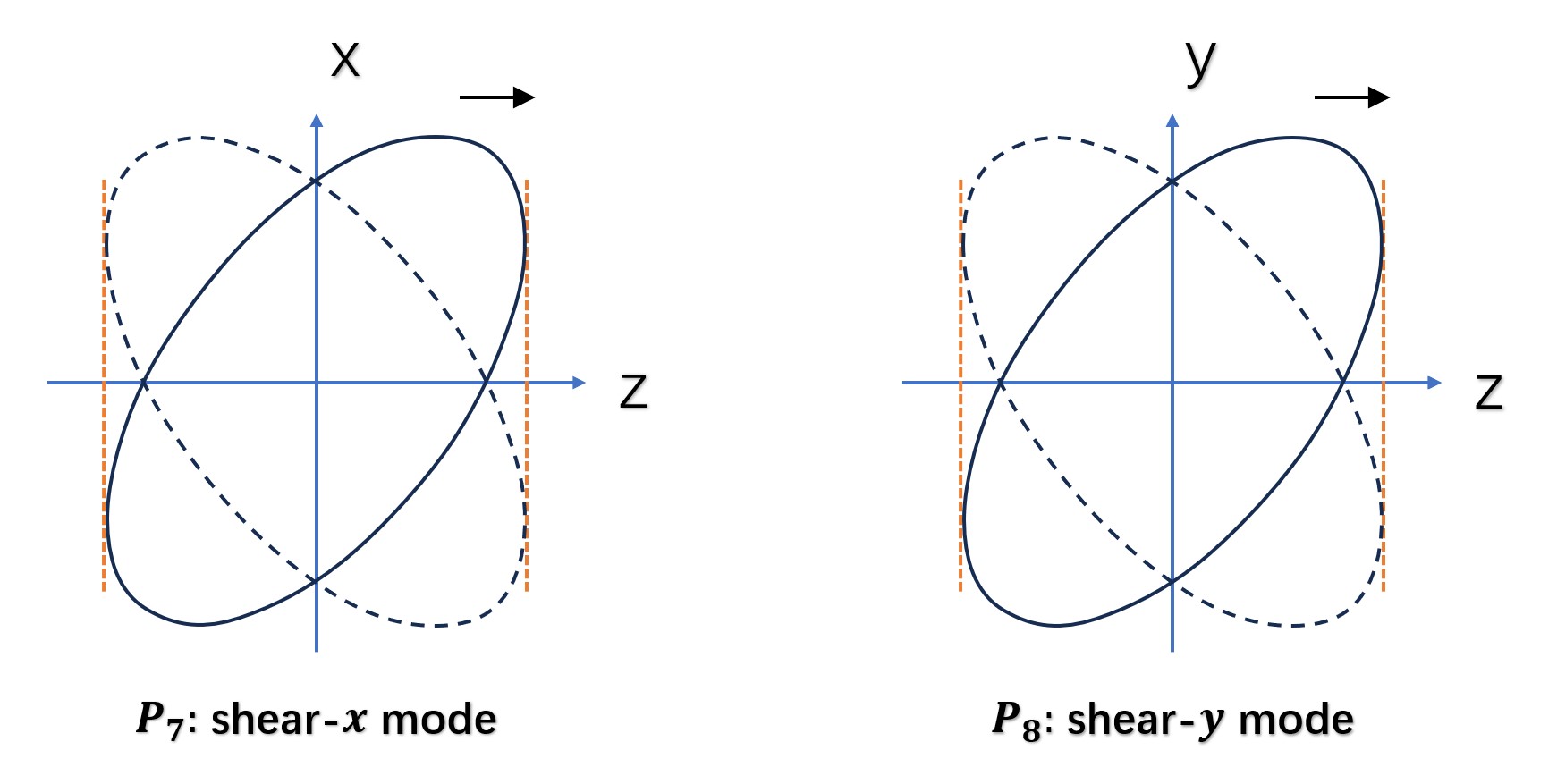}}
	\caption{The two new shear modes. The gravitational waves in the figure propagate in the $+z$ direction. In the shear-$x$ and shear-$y$ modes, the periodic tangential motion of a test particle array occurs along the $x$ and $y$ directions, respectively. In addition, the amplitude of this motion increases with the relative distance in the $z$-direction between the test particles. The solid and dashed lines represent the cases where the phases are labeled as 
	$\frac{\pi}{2}$ and $\frac{3}{2}\pi$, respectively.}
	\label{fig: 2}
\end{figure*}

\begin{figure*}[htbp]
	\makebox[\textwidth][c]{\includegraphics[width=0.9\textwidth]{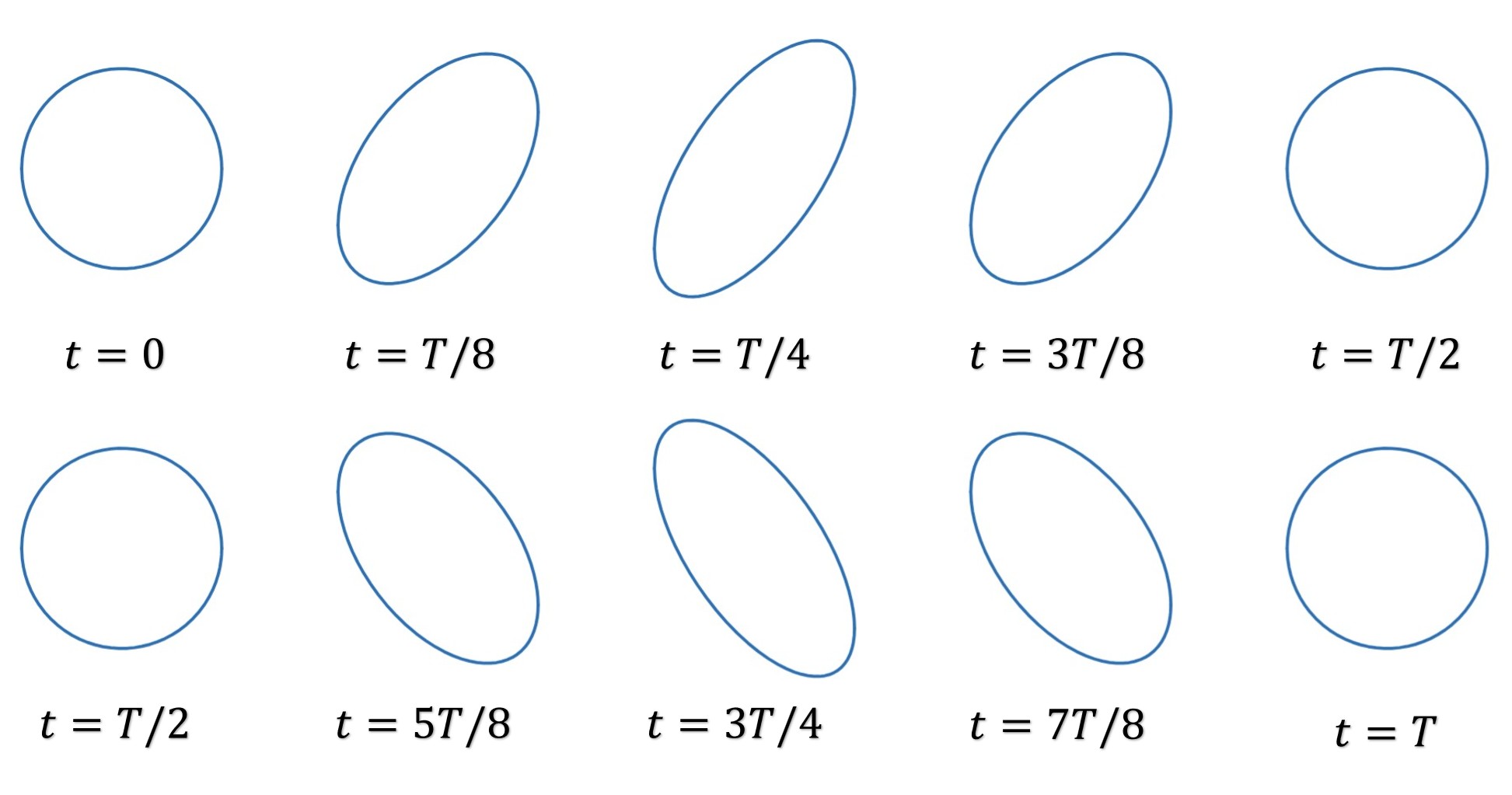}}
	\caption{Schematic diagram illustrating the relative motion of test particles over one period under shear-mode gravitational waves.}
	\label{fig: 3}
\end{figure*}

We use $P_{7}$ as an example to illustrate how the test particles move in this new mode. Note that by rotating the $x-y$ plane $90$ degrees, $P_{7}$ is transformed into $P_{8}$, so the motion modes of $P_{8}$ and $P_{7}$ are identical. Given that only $P_{7} \neq 0$ among all polarization modes, substituting Eq. (\ref{P1-P8}) into Eq. (\ref{equation of motion for etai}) yields 
\begin{eqnarray}
	\label{equation of motion for etai in P7}
	\frac{d^2 \eta^{1}}{dt^2}=-P_{7}\left(t\right)\eta^{3},\quad
	\frac{d^2 \eta^{2}}{dt^2}=\frac{d^3 \eta^{3}}{dt^2}=0.
\end{eqnarray}
It is evident that a $P_{7}$-mode gravitational wave propagating along the $+z$ direction does not induce relative changes in the $y$ and $z$ directions of the test particles. However, it will cause periodic relative motion in the $x$ direction for test particles with an initial relative separation in the $z$ direction. This motion mode is tangential, so we refer to $P_{7}$ and $P_{8}$ as shear modes, specifically the shear-$x$ mode and shear-$y$ mode, respectively. In Fig. \ref{fig: 2} and Fig. \ref{fig: 3}, we show the motion of the test particles in these two new modes.

Similar to Eq. (\ref{Ri0j0 gauge invariant}), $A_{ij}$ can be expressed in a gauge-invariant form using Eqs. (\ref{decompose perturbations}), (\ref{gauge invariant second-order spatial tensors}), (\ref{gauge invariant spatial vectors}), and (\ref{gauge invariant spatial scalars}):
\begin{eqnarray}
	\label{Aij gauge invariant}
	A_{ij}\!&=&\!-\frac{1}{2}\partial_{0}\partial_{0}h^{TT}_{ij}
	+\frac{1}{2}\partial_{0}\partial_{i}\Xi_{j}
	+\frac{1}{2}\partial_{0}\partial_{j}\Xi_{i}
	+\partial_{i}\partial_{j}\phi
	-\frac{1}{2}\delta_{ij}\partial_{0}\partial_{0}\Theta
	\nonumber \\
	\!&-&\!\partial_{j}\Omega_{i}
	-\partial_{0}\partial_{j}K_{i}
	-\partial_{i}\partial_{j}\Pi
	-\partial_{0}\partial_{i}\partial_{j}\bar{\Pi}
	-\frac{\partial_{0}^{2}}{2\Delta}\partial_{i}\partial_{j}L
	+\frac{3\partial_{0}^{2}}{2\Delta}\partial_{i}\partial_{j}\Theta
	-\partial_{i}\partial_{j}\phi.
\end{eqnarray}
Therefore, the eight polarization modes of gravitational waves satisfy
\begin{eqnarray}
	\label{P1-P8 gauge invariant}
		&P_{1}=\partial_{0}\partial_{0}\Theta
		-\partial_{3}\partial_{3}\Pi-\partial_{0}\partial_{3}\partial_{3}\bar{\Pi}
		-\frac{1}{2}\partial_{0}\partial_{0}L,~
		P_{6}=-\frac{1}{2}\partial_{0}\partial_{0}\Theta, \nonumber \\
		&P_{2}=\frac{1}{2}\partial_{0}\partial_{3}\Xi_{1},\quad P_{3}=\frac{1}{2}\partial_{0}\partial_{3}\Xi_{2},  \nonumber \\
		&P_{4}=-\frac{1}{2}\partial_{0}\partial_{0}h^{TT}_{11}, \quad
		P_{5}=-\frac{1}{2}\partial_{0}\partial_{0}h^{TT}_{12}, \nonumber \\
		&P_{7}=\partial_{3}\Omega_{1}
		-\partial_{0}\partial_{3}K_{1},\quad
		P_{8}=\partial_{3}\Omega_{2}
		-\partial_{0}\partial_{3}K_{2}.
\end{eqnarray}
It can be seen that non-metricity not only generates the new $P_{7}$ and $P_{8}$ modes but also modifies the longitudinal mode $P_{1}$. The two shear modes are excited by the gauge-invariant spatial vectors $\Omega_{i}$ and $K_{i}$, making them vector modes. 

Finally, it is important to note that not all gauge invariants (\ref{gauge invariant third-order spatial tensors})-(\ref{gauge invariant spatial scalars}) derived from perturbation combinations contribute to the polarization modes of gravitational waves. From Eq. (\ref{P1-P8 gauge invariant}), it is clear that no transverse-traceless third-order spatial tensor contributes. Among the 7 transverse-traceless second-order spatial tensors, only $h^{TT}_{ij}$ contributes. Of the 10 transverse spatial vectors, $\Xi_{i}$, $\Omega_{i}$, and $K_{i}$ contribute. Finally, of the 12 spatial scalars, $\Theta, \Pi, \bar{\Pi}$, and $L$ contribute. However, $\phi$, which contributes in Eq. (\ref{P1-P6 gauge invariant}), no longer contributes.

{It should be noted that polarization modes of gravitational waves and the propagating degrees of freedom of perturbations are distinct concepts. Polarization modes are defined via the relative motions of test particles and therefore represent modes of motion. From this perspective, the number of independent polarization modes in a theory does not directly determine the number of propagating degrees of freedom of the perturbations. The presence of a given polarization mode only indicates that the corresponding perturbation component can propagate; to determine the actual number of propagating degrees of freedom in a theory, one must solve the field equations.}

\section{shear modes and field equations: a specific example}
\label{sec: 5}

In the previous analysis, we demonstrate that novel shear modes can theoretically exist in torsionless non-Riemannian spacetimes. In this section, we provide a concrete example to further show that shear modes are indeed permitted by the field equations in certain theories. To meet the needs of some readers, Appendices \ref{app: C} and \ref{app: D} offer a comprehensive analysis of the polarization modes of gravitational waves in the most general Palatini theory.

The specific theory that we present as an example is as follows:
\begin{eqnarray}
	\label{antion of example theory}
	S=\int d^4 x  \sqrt{-g} \left[ 
	\frac{c^{3}}{16\pi G}\widehat{R}
	+\alpha R
	-\frac{\alpha}{2} Q^{\lambda\mu}_{~~~\mu}Q_{\rho\lambda}^{~~\rho}
	+\frac{\alpha}{4} Q^{\lambda\mu}_{~~~\mu}Q_{\lambda~\nu}^{~\nu}
	-\frac{\beta}{2}  \bar{R}^{\mu\nu}\bar{R}_{\mu\nu}
	\right].
\end{eqnarray}
Here, $\bar{R}_{\mu\nu} \coloneqq R^{\lambda}_{~\lambda\mu\nu}$. The term $\widehat{R}$ retains the structure of the Einstein-Hilbert action, while $\alpha$ and $\beta$ represent deviations from general relativity.

The analysis of gravitational wave polarization modes in this theory requires solving the linearized field equations. As shown in Appendix \ref{app: D} and Refs. \cite{Y.Dong5,Y.Dong4,Y.Dong3}, these equations can be simplified by decoupling them using gauge invariants and transforming them into algebraic equations in momentum space, thereby rendering the system more tractable. By applying these methods, we find that for tensor modes, the $+$ and $\times$ polarizations always propagate at the speed of light, regardless of whether the particle carries hypermomentum. 

However, for the vector modes, the determinant of the coefficient matrix of the corresponding algebraic equation system is zero, leading to a mass solution
\begin{eqnarray}
	\label{mass of vector part another example}
	m^{2}=\frac{\alpha}{\beta}.
\end{eqnarray}
{To avoid the emergence of an imaginary mass, we require that $\alpha$ and $\beta$ share the same sign.} For this solution, the corresponding gauge invariant vector perturbations satisfy
\begin{eqnarray}
	\label{vector mode another example}
	N_{i}=C_{i}=\bar{C}_{i}=f_{i}=q_{i}=\Xi_{i}=\bar{\Omega}_{i}=K_{i}=0,
	~ \Omega_{i}=h_{i}.
\end{eqnarray}
Since $\Xi_{i}=0$, no vector polarization mode can be detected for test particles without hypermomentum charge (or the effect of hypermomentum is negligible), as indicated by Eq. (\ref{P1-P6 gauge invariant}). However, for test particles with hypermomentum charges, Eq. (\ref{P1-P8 gauge invariant}) reveals the existence of shear-$x$ and shear-$y$ modes with mass $m$ that satisfy Eq. (\ref{mass of vector part another example}) at this time.

For the scalar modes, the determinant of the coefficient matrix of the corresponding algebraic equation system is zero, which also leads to a mass 
\begin{eqnarray}
	\label{mass of scalar part another example}
	m^{2}=\frac{\alpha}{\beta}.
\end{eqnarray}
This mass is the same as that of vector mode gravitational waves, and the corresponding gauge invariant scalar perturbations satisfy
\begin{eqnarray}
	\label{scalar mode another example}
	&\Theta=\phi=L=K=\bar{A}=n=\bar{\Psi}=\bar{\Pi}=0,\nonumber\\
	&A=-\Psi=-\frac{\Delta}{\partial_{0}}m,~
	\Pi=-m.
\end{eqnarray}
From Eq. (\ref{P1-P6 gauge invariant}), it follows that, due to $\Theta=\phi=0$, test particles without hypermomentum charge (or the effect of hypermomentum is negligible) cannot detect any scalar modes. In contrast, according to Eq. (\ref{P1-P8 gauge invariant}), test particles with a hypermomentum charge will detect a longitudinal mode with mass $m$ satisfying Eq. (\ref{mass of scalar part another example}).

In our simple example, test particles without hypermomentum charge (or the effect of hypermomentum is negligible) can only detect tensor modes propagating at the speed of light, similar to general relativity. However, for particles with a hypermomentum charge, additional shear and longitudinal modes with mass $m^{2}=\alpha/\beta$ may also be detected.

\section{Conclusion}
\label{sec: 6}

In this paper, we investigated the polarization properties of gravitational waves in a torsionless spacetime. In such a spacetime, both the gravitational field action $S_{g}$ and the matter field action $S_{m}$ generally depend on the metric and the connection. 

We first examined the action of the matter field $S_{m}$ and derived the equation of motion for free particles in a torsionless spacetime using generalized covariance. In a torsionless spacetime, free particles without a hypermomentum charge (or the effect of hypermomentum is negligible) follow the same equation of motion  (\ref{geodesic equation}) as in general relativity. However, for particles carrying hypermomentum charges, the equation of motion must be modified (see Eq. (\ref{equation of motion in Palatini})). In this modified equation, the particle's charges $m$, $P^{\mu}$, and $H^{\mu}$ do not appear. Therefore, the trajectory of the particles depends only on their initial position and velocity, and is independent of the specific values of the charges they carry. From this perspective, the effect of a hypermomentum charge is solely to introduce non-metricity interactions in spacetime, while the free motion of particles remains a manifestation of geometric effects. This is particularly interesting because, in a general metric-affine spacetime, the trajectory of particles would depend on the specific values of the charges they carry \cite{Damianos Iosifidis}. It is also important to note that the equation of motion for a free particle does not correspond to the geodesic equation in a metric-affine geometry.

After deriving the equation of motion for a free particle, we redefined the polarization modes of gravitational waves that can exist in a torsionless spacetime by analyzing the relative motion of test particles. In contrast to the commonly recognized six independent polarization modes, two entirely new gravitational wave polarization modes have emerged in the torsionless spacetime. Based on the relative motion of test particles in these two modes, we named them the shear-$x$ and shear-$y$ modes. These modes are excited by gauge invariant vector perturbations related to non-metricity, and therefore belong to the vector modes. Detecting these shear modes could provide a novel approach to using gravitational wave polarization to probe the underlying spacetime geometry.

Having identified, from the perspective of the equations of motion, that non-metricity induces two novel gravitational wave polarization modes, we proceeded to examine the field equations of the theory. Through a concrete example, we demonstrated that these new polarization modes can indeed exist in certain theories. In the example theory we considered, test particles without a hypermomentum charge can only detect the $+$ and $\times$ modes, which propagate at the speed of light, as in general relativity. However, for particles with a hypermomentum charge, shear and longitudinal modes with mass $m^{2}=\alpha/\beta$ can also be detected. This example highlights two important aspects of gravitational wave polarization in a torsionless spacetime. First, the polarization effect differs between particles that carry hypermomentum and those that do not. Second, it suggests that new shear modes may indeed exist.

Additionally, in Appendices \ref{app: C} and \ref{app: D}, we provided a systematic analysis of the polarization modes of gravitational waves in the most general Palatini theory by constructing a general analytical framework. When analyzing a specific Palatini theory, one can directly obtain the correspondence between the parameters of the theory and our framework from Appendix \ref{app: F}, and then derive the decoupled equations from Appendix \ref{app: E}. Together with the results from Appendix \ref{app: D}, this enables the determination of the gravitational wave polarization properties of the theory.

Finally, it is necessary to discuss the relationship between nonmetricity and Lorentz invariance. Although nonmetricity is often presumed to be subject to strict constraints from Lorentz symmetry, the two concepts are not intrinsically or logically connected. This belief may stem from two main viewpoints. One possible viewpoint is that, similar to vector-tensor theories, assigning a nonzero background value to nonmetricity can introduce a preferred direction in spacetime. As a result, fields propagating on such a background may exhibit Lorentz symmetry breaking, which could further lead to CPT violation. However, in our work, the background field is defined as in Eq. (\ref{flat spacetime background}); that is, the background value of nonmetricity is taken to be zero. Therefore, Lorentz invariance is preserved, and CPT symmetry is consequently maintained. This is also evident from the most general second-order perturbation action given in Eq. (\ref{the most general second-order perturbation action in general Palatini theory}). Another viewpoint holds that nonmetricity modifies the inner product of vectors under parallel transport, allowing a null vector to become time-like and thereby distorting the light cone structure. This argument, however, rests on the assumption that light propagation is governed by the parallel transport of its wave vector—an assumption valid in general relativity but not necessarily applicable to modified gravity theories. In fact, parallel transport is a mathematical notion, and the physical trajectory of light must be derived from the electromagnetic action in a given background through the field equations and geometric optics approximation. Therefore, the existence of nonmetricity does not inherently imply a breakdown of the light cone structure, and consequently, does not necessarily lead to Lorentz symmetry violation. Notably, as demonstrated in Ref. \cite{Lavinia Heisenberg1}, it is possible to construct a theory involving nonmetricity that remains physically equivalent to general relativity. Based on the above discussion, the result of this paper are consistent with the constraints imposed by current experiments \cite{S. Herrmann,S. Navas}.

Our work indicates that if spacetime is torsionless but exhibits non-metricity, gravitational waves may possess novel shear polarization modes that are distinct from the six known modes. These modes do not appear in spacetimes based on Riemannian geometry. Consequently, they could serve as a new diagnostic tool to test spacetime geometry and gravitational theories, providing deeper insights into the fundamental nature of gravity. However, how these shear modes will specifically affect the signals detected by various gravitational wave detectors remains an open question that warrants further investigation.

\section*{Acknowledgments}
This work is supported in part by the National Key Research and Development Program of China (Grant No. 2020YFC2201503), the National Natural Science Foundation of China (Grants No. 123B2074, No. 12475056,  and No. 12247101), Gansu Province's Top Leading Talent Support Plan, the Fundamental Research Funds for the Central Universities (Grant No. lzujbky-2024-jdzx06), the Natural Science Foundation of Gansu Province (No. 22JR5RA389), and the `111 Center' under Grant No. B20063.

\appendix
\section{Derivation of the geodesic equation in general relativity}
\label{app: A}
In this appendix, we illustrate how to use Eqs. (\ref{Tmunu in point particle}) and (\ref{covariant conservation equation of T in GR}) to derive the geodesic equation. Using the relationship \cite{landau}
\begin{eqnarray}
	\label{gamma mumulambda=partialg}
	\widehat{\Gamma}^{\mu}_{~\lambda\mu}
	=
	\frac{1}{2g}\partial_{\lambda}g
	=\partial_{\lambda} \ln\sqrt{-g},
\end{eqnarray}
Eq. (\ref{covariant conservation equation of T in GR}) can be further written as 
\begin{eqnarray}
	\label{covariant conservation equation of T in GR new}
	\widehat{\nabla}_{\mu}T^{\mu\nu}=
	\frac{1}{\sqrt{-g}}\partial_{\mu}\left(\sqrt{-g}T^{\mu\nu}\right)
	+\widehat{\Gamma}^{\nu}_{~\mu\lambda}T^{\mu\lambda}=0.
\end{eqnarray}

In fact, although we do not know the specific form of $S_{m}$, we can use Eq. (\ref{covariant conservation equation of T in GR new}) to constrain the possible forms of $M^{\mu\nu}$ in Eq. (\ref{Tmunu in point particle}). For this, we calculate $\partial_{\mu}\left(x^{\lambda}\sqrt{-g}T^{\mu\nu}\right)/\sqrt{-g}$. From Eq. (\ref{covariant conservation equation of T in GR new}), we have 
\begin{eqnarray}
	\label{PxT1}
	\frac{1}{\sqrt{-g}}\partial_{\mu}\left(x^{\lambda} \sqrt{-g}T^{\mu\nu}\right)
	=T^{\lambda\nu}-x^{\lambda}\widehat{\Gamma}^{\nu}_{~\mu\rho}T^{\mu\rho}.
\end{eqnarray}
By substituting Eq. (\ref{Tmunu in point particle}) into the above equation and integrating $\int d^{3}x \sqrt{-g}$ on both sides simultaneously, we obtain
\begin{eqnarray}
	\label{PxT2}
	\frac{d}{dt}\left(X^{\lambda} \frac{M^{0\nu}}{u^{0}}\right)
	=\frac{M^{\lambda\nu}}{u^{0}}
	-X^{\lambda}\widehat{\Gamma}^{\nu}_{~\mu\rho}\frac{M^{\mu\rho}}{u^{0}}.
\end{eqnarray}
Similarly, by substituting Eq. (\ref{Tmunu in point particle}) into the covariant conservation equation (\ref{covariant conservation equation of T in GR new}) and integrating $\int d^{3}x \sqrt{-g}$ on both sides simultaneously, we have
\begin{eqnarray}
	\label{PT1}
	\frac{d}{dt}\left(\frac{M^{0\nu}}{u^{0}}\right)
	+\widehat{\Gamma}^{\nu}_{~\mu\lambda}\frac{M^{\mu\lambda}}{u^{0}}=0.
\end{eqnarray}
Using the above equation, Eq. (\ref{PxT2}) can be further simplified to 
\begin{eqnarray}
	\label{PxT3}
	u^{\mu} \frac{M^{0\nu}}{u^{0}}
	=M^{\mu\nu}.
\end{eqnarray}
It can be seen that $M^{\mu\nu}$ should have the form $M^{\mu\nu}=u^{\mu}M^{\nu}$, where $M^{\nu}=M^{0\nu}/u^{0}$ is a vector. Furthermore, since $T^{\mu\nu}$ is a symmetric tensor, $M^{\mu\nu}$ should also be symmetric, which requires 
\begin{eqnarray}
	\label{Mmunu in GR}
	M^{\mu\nu}=m u^{\mu} u^{\nu}.
\end{eqnarray}
Here, $m$ is a scalar. By substituting Eq. (\ref{Mmunu in GR}) into Eq. (\ref{PT1}), it can be concluded that
\begin{eqnarray}
	\label{PT2}
	m \frac{\widehat{D}u^{\mu}}{d\tau}+ u^{\mu}\frac{dm}{d\tau}=0.
\end{eqnarray}

By multiplying both sides by $u_{\mu}$ and using $u_{\mu}\widehat{D}u^{\mu}/d\tau=0$ (This is because $u^{\mu}u_{\mu}=-1$), we find that $m$ is conserved:
\begin{eqnarray}
	\label{m  conserved}
	\frac{dm}{d\tau}=0.
\end{eqnarray}
Therefore, Eq. (\ref{PT2}) becomes the well-known geodesic equation
\begin{eqnarray}
	\frac{\widehat{D}u^{\mu}}{d\tau}=0.
\end{eqnarray}

\section{Proof of the equivalence of assumptions}
\label{app: B}
In fact, we point out that Assumption $0$ in Sec. \ref{The equation of motion in a torsionless spacetime} can be relaxed without changing anything. To demonstrate this, we will show that Assumption $0$ is equivalent to one of the following two assumptions:

\textbf{Assumption a}:
\begin{eqnarray}
	\label{Tmunu in palatini Assumption 1}
	T^{\mu\nu}\left(t,\textbf{x}\right)\!&=&\!
	\frac{1}{\sqrt{-g}}\frac{M^{\mu\nu}}{u^{0}}\delta^{3}\left(\textbf{x}-\textbf{X}\left(t\right)\right),\\
	\label{Hlambndamunu in palatini Assumption 1}
	\mathcal{H}_{\lambda}^{~\mu\nu}\left(t,\textbf{x}\right)\!&=&\!
	\frac{1}{\sqrt{-g}}\frac{H_{\lambda}u^{\mu}u^{\nu}}{u^{0}}\delta^{3}\left(\textbf{x}-\textbf{X}\left(t\right)\right).
\end{eqnarray}

\textbf{Assumption b}:
\begin{eqnarray}
	\label{Tmunu in palatini Assumption 2}
	T^{\mu\nu}\left(t,\textbf{x}\right)\!&=&\!
	\frac{1}{\sqrt{-g}}\frac{M^{\mu\nu}}{u^{0}}\delta^{3}\left(\textbf{x}-\textbf{X}\left(t\right)\right),\\
	\label{Hlambndamunu in palatini Assumption 2}
	\mathcal{H}_{\lambda}^{~\mu\nu}\left(t,\textbf{x}\right)\!&=&\!
	\frac{1}{\sqrt{-g}}\frac{H_{\lambda}^{~\mu\nu}}{u^{0}}\delta^{3}\left(\textbf{x}-\textbf{X}\left(t\right)\right),\\
	\label{Sigma in palatini Assumption 2}
	\Sigma^{\mu}_{~\nu}\left(t,\textbf{x}\right)\!&=&
	\!\frac{1}{\sqrt{-g}}\frac{P^{\mu}_{~\nu}}{u^{0}}\delta^{3}\left(\textbf{x}-\textbf{X}\left(t\right)\right).
\end{eqnarray}
Here, Assumption a assumes the structure of $\mathcal{H}_{\lambda}^{~\mu\nu}$ and requires $T^{\mu\nu}$ to follow a delta function distribution, without any requirements for $\Sigma^{\mu}_{~\nu}$. Assumption b only requires $T^{\mu\nu}$, $\mathcal{H}_{\lambda}^{~\mu\nu}$, and $\Sigma^{\mu}_{~\nu}$ to be delta distributions. That is, it does not include any derivative terms of the delta function, i.e., terms that vary rapidly over a small range, such as those describing point dipoles.

Note that Assumption $0$ can lead to Assumptions a and b. To prove their equivalence, it is only necessary to show that Assumptions a and b can lead to Assumption $0$. To complete the proof, we need to first introduce a lemma. 

We start by introducing the following symbol. For the delta function $\delta\left(x\right)$, we denote its $n$-th derivative as 
\begin{eqnarray}
	\label{derivative of delta}
	\delta^{(n)}\left(x\right) \coloneqq
	\frac{d^{n}}{dx^{n}}\delta\left(x\right).
\end{eqnarray}
It is defined by the following integral:
\begin{eqnarray}
	\label{derivative of delta defined by}
	\int dx f(x) \delta^{(n)} (x)
	=(-1)^{n} f^{(n)}(0),
\end{eqnarray}
where $f(x)$ is a smooth function, and $f^{(n)}(x)$ is its $n$-th derivative. With this symbol, our lemma can be expressed as:

\textbf{Lemma 1}: For a series of $a_{lmn}$ ($l,m,n \in \mathbb{N}_{0}$) that do not depend on $(x, y, z)$, if 
\begin{eqnarray}
	\label{ax+by=0}
	p(x,y,z)\coloneqq\!\!\sum_{l,m,n\in \mathbb{N}_{0}} a_{lmn}~\delta^{(l)} (x)\delta^{(m)} (y)\delta^{(n)} (z)=0,
\end{eqnarray}
then 
\begin{eqnarray}
	\label{almn=0}
	a_{lmn}=0.
\end{eqnarray}

\textbf{Proof}: Using Eq. (\ref{derivative of delta defined by}), we have
\begin{eqnarray}
	\label{exp}
	q\left(\lambda_{1},\lambda_{2},\lambda_{3}\right)\!&\coloneqq&\!
	\int d^{3}x~ e^{\lambda_{1}x+\lambda_{2}y+\lambda_{3}z}~p(x,y,z)\nonumber\\
	\!&=&\!\! \sum_{l,m,n\in \mathbb{N}_{0}} a_{lmn} \int d^{3}x~ e^{\lambda_{1}x+\lambda_{2}y+\lambda_{3}z} \delta^{(l)} (x)\delta^{(m)} (y)\delta^{(n)} (z)\nonumber\\
	\!&=&\!\! \sum_{l,m,n\in \mathbb{N}_{0}} a_{lmn} 
	\left(\int dx e^{\lambda_{1}x}\delta^{(l)}(x)\right)
	\left(\int dy e^{\lambda_{2}y}\delta^{(m)}(y)\right)
	\left(\int dz e^{\lambda_{3}z}\delta^{(n)}(z)\right)\nonumber \\
	\!&=&\!\! \sum_{l,m,n\in \mathbb{N}_{0}} (-1)^{l+m+n} a_{lmn} ~\lambda_{1}^{l}\lambda_{2}^{m}\lambda_{3}^{n}\nonumber \\
	\!&=&\! 0.
\end{eqnarray}
Note that $q\left(\lambda_{1},\lambda_{2},\lambda_{3}\right)$ is a power series, and due to the arbitrariness of $\left(\lambda_{1},\lambda_{2},\lambda_{3}\right)$, it can be seen that $(-1)^{l+m+n} a_{lmn}=0$. Thus, $a_{lmn}=0$. $\hfill\square$

Now we prove the equivalence of the three assumptions. 

\textbf{Proof}: Since $T^{\mu\nu}$, $\mathcal{H}_{\lambda}^{~\mu\nu}$, and $\Sigma^{\mu}_{~\nu}$ are covariant tensors, we can always choose a special coordinate system to simplify our analysis without loss of generality. For this, we choose a coordinate system such that the coordinates of the free particle are $X^{\mu}=\left(\tau,0,0,0\right)$. This further leads to the four-velocity of the particle being $u^{\mu}=\left(1,0,0,0\right)$. In this coordinate system, 
\begin{eqnarray}
	\label{delta in coordinate system}
	\frac{1}{\sqrt{-g}}\frac{1}{u^{0}}\delta^{3}\left(\textbf{x}-\textbf{X}\left(t\right)\right)
	\rightarrow
	\frac{1}{\sqrt{-g}}\delta^{3}\left(\textbf{x}\right)
	=
	\frac{1}{\sqrt{-g}}\delta(x)\delta(y)\delta(z).
\end{eqnarray}

We consider Assumption a. In the aforementioned special coordinate system, by substituting Eqs. (\ref{Tmunu in palatini Assumption 1}) and (\ref{Hlambndamunu in palatini Assumption 1}) into Eq. (\ref{covariant conservation equation in Palatini equivalently hypermomentum}), we find that $\partial_{\nu}\mathcal{H}_{\lambda}^{~\mu\nu}$ introduces a term containing the first derivative of the delta function in the expression. It is 
\begin{eqnarray}
	\label{Hlambndamunu containing d delta }
	\frac{1}{\sqrt{-g}}H_{\lambda}u^{\mu}u^{\nu}\partial_{\nu}\delta^{3}\left(\textbf{x}\right).
\end{eqnarray}
Note that $\partial_{0}\delta^{3}\left(\textbf{x}\right)=0$ and $u^{i}=0$, so the above expression is zero. This indicates that $\Sigma^{\mu}_{~\nu}$ can be expressed using an expression that only contains the delta function without its derivative, i.e., it can be expressed as 
\begin{eqnarray}
	\Sigma^{\mu}_{~\nu}=
	\frac{1}{\sqrt{-g}}P^{\mu}_{~\nu}\delta^{3}\left(\textbf{x}\right)
\end{eqnarray}
in the special coordinate system. Since $\Sigma^{\mu}_{~\nu}$ is a tensor, the above equation indicates that $\Sigma^{\mu}_{~\nu}$ can be expressed in the form of Eq. (\ref{Sigma in palatini Assumption 2}) for any coordinate system. In this way, we have proven that Assumption a can lead to Assumption b.

To complete the final proof, we only need to prove that Assumption b can lead to Assumption $0$. We also use the coordinate system where $X^{\mu}=\left(\tau,0,0,0\right)$. By substituting Eqs. (\ref{Tmunu in palatini Assumption 2}), (\ref{Hlambndamunu in palatini Assumption 2}), and (\ref{Sigma in palatini Assumption 2}) into Eq. (\ref{covariant conservation equation in Palatini equivalently hypermomentum}), we find that the term containing the first derivative of the delta function in the equation is only
\begin{eqnarray}
	\label{Hlambndamunu containing d delta Assumption 2}
	\frac{1}{\sqrt{-g}}H_{\lambda}^{~\mu\nu}\partial_{\nu}\delta^{3}\left(\textbf{x}\right).
\end{eqnarray}
Due to Lemma $1$, this requires
\begin{eqnarray}
	\label{Hlambndamunu containing d delta Assumption 2=0}
	H_{\lambda}^{~\mu i}=0.
\end{eqnarray}
In the coordinate system we have chosen $u^{\mu}=(1,0,0,0)$. Therefore, from Eq. (\ref{Hlambndamunu containing d delta Assumption 2=0}), $H_{\lambda}^{~\mu\nu}$ can be expressed as 
$H_{\lambda}^{~\mu}u^{\nu}$, where $H_{\lambda}^{~\mu}=H_{\lambda}^{~\mu 0}$. Furthermore, based on the symmetry of the indices, $H_{\lambda}^{~\mu\nu}$ can be further expressed as
\begin{eqnarray}
	\label{Hlambndamunu  Assumption 2}
	H_{\lambda}^{~\mu\nu}=H_{\lambda}u^{\mu}u^{\nu}.
\end{eqnarray}
Here $H_{\lambda}=H_{\lambda}^{~00}$. Considering that $H_{\lambda}^{~\mu\nu}$ is a tensor, 
$\mathcal{H}_{\lambda}^{~\mu\nu}$ retains the form of Eq. (\ref{Hlambndamunu in palatini Assumption 0}) in any coordinate system.

The process of proving that $T^{\mu\nu}$ satisfies the form of Eq. (\ref{Tmunu in palatini Assumption 0}) is very similar. By substituting Eqs. (\ref{Tmunu in palatini Assumption 2}) and (\ref{Hlambndamunu in torsionless spacetime palatini}) into Eq. (\ref{covariant conservation equation in Palatini}), we find that the term containing the derivative of the delta function is only 
\begin{eqnarray}
	\label{2MPdelta}
	2M^{\mu}_{~\nu}\partial_{\mu}\delta^{3}\left(\textbf{x}\right).
\end{eqnarray}
Similar reasoning requires $M^{\mu\nu}=mu^{\mu}u^{\nu}$.

Finally, we need to prove that $\Sigma^{\mu}_{~\nu}$ satisfies Eq. (\ref{Sigma in palatini Assumption 0}). By substituting Eqs. (\ref{Sigma in palatini Assumption 2}) and (\ref{Hlambndamunu in torsionless spacetime palatini}) into Eq. (\ref{covariant conservation equation in Palatini equivalently canonical energy-momentum tensor}), the term containing the derivative of the delta function is nonzero only for
\begin{eqnarray}
	\label{Sigma containing d delta }
	\frac{1}{\sqrt{-g}}P^{\mu}_{~\alpha}\partial_{\mu}\delta^{3}\left(\textbf{x}\right).
\end{eqnarray}
As before, this indicates that $P^{\mu}_{~\nu}$ has the form $P^{\mu}_{~\nu}=P_{\nu}u^{\mu}$, which is the form of Eq. (\ref{Sigma in palatini Assumption 0}), thus completing the proof. $\hfill\square$

It is worth noting that if this proof method is used in the context of general relativity, Eq. (\ref{Mmunu in GR}) can be easily derived using Eqs. (\ref{Tmunu in point particle}) and (\ref{covariant conservation equation of T in GR}).

\section{Construction of the second-order perturbation action for the most general Palatini theory}
\label{app: C}

In this section, we focus on the gravitational part $S_{g}$ of the action (\ref{Palatini action}), specifically examining the polarization modes of gravitational waves predicted by the most general Palatini theory and their corresponding dispersion relations. To achieve this, we need to solve the linear perturbation equations for the most general Palatini theory. These equations are derived by constructing the most general second-order perturbation action on a flat spacetime background (\ref{flat spacetime background}) \cite{Y.Dong4,Y.Dong5}. Therefore, we first construct the most general second-order perturbation action, where the perturbations satisfy Eq. (\ref{flat spacetime background perturbations}). In this section and the following parts, since we are considering linearized gravity theory, indices are raised and lowered by $\eta_{\mu\nu}$, as in the first section.

We consider that the second-order perturbation action of general Palatini theory satisfies the following assumptions:
\begin{itemize}
	\item [(1)] The spacetime is four-dimensional.
	\item [(2)] The second-order perturbation action is gauge invariant.
	\item [(3)] The field equations are second-order.
	\item [(4)] Each term in the second-order perturbation action can be represented as a combination of $\eta_{\mu\nu}$, $h_{\mu\nu}$, $\Sigma^{\lambda}_{~\mu\nu}$, $\partial_{\mu}$, and the theoretical parameters.
\end{itemize}
In Assumption (1), we exclude the possibility of extra dimensions in spacetime. Assuming general covariance of the Palatini theory, we deduce that the second-order perturbation action must satisfy Assumption (2). For Assumption (3), we assume the field equation to be second-order to avoid the Ostrogradsky instability typically present in higher-order derivative field equations \cite{M.Ostrogradsky,R.P.Woodard,H.Motohashi,A.Ganz}. In Assumption (4), for simplicity, we neglect terms involving the four-dimensional antisymmetric tensor $\epsilon_{\mu\nu\lambda\rho}$, which could lead to parity symmetry breaking.

Considering the symmetry of indices $\left(\mu, \nu\right)$ in $\eta_{\mu\nu}$, $h_{\mu\nu}$, and $\Sigma^{\lambda}_{~\mu\nu}$, and accounting for all possible combinations, we can express the most general second-order perturbation action satisfying the above assumptions as 
(By ``most general," we mean that any second-order perturbation action satisfying these assumptions can be transformed into the following form via integration by parts.)
\begin{eqnarray}
	\label{the most general second-order perturbation action in general Palatini theory}
	S_{g}^{(2)}=S^{(2)}_{0}+S^{(2)}_{1}+S^{(2)}_{2}=\int d^4 x  \left(\mathcal{L}_{0}+\mathcal{L}_{1}+\mathcal{L}_{2}\right),
\end{eqnarray}
where
\begin{eqnarray}
	\label{L0 new}
	\mathcal{L}_{0}
	\!&=&\!
	\left(2B_{(1)}+E_{(1)}\right)\eta^{\alpha\beta}\Sigma^{\lambda}_{~\lambda\mu}\Sigma^{\mu}_{~\alpha\beta}
	+\left(B_{(1)}+D_{(1)}+\frac{1}{2}E_{(1)}\right)\eta^{\mu\nu}\Sigma^{\lambda}_{~\lambda\mu}\Sigma^{\rho}_{~\rho\nu}
	\nonumber \\
	\!&+&\!
	\frac{1}{2}C_{(1)}\eta_{\lambda\rho}\eta^{\mu\alpha}\eta^{\nu\beta}\Sigma^{\lambda}_{~\mu\nu}\Sigma^{\rho}_{~\alpha\beta}
	+\frac{1}{2}E_{(1)}\eta_{\lambda\rho}\eta^{\mu\nu}\eta^{\alpha\beta}\Sigma^{\lambda}_{~\mu\nu}\Sigma^{\rho}_{~\alpha\beta}
	\nonumber \\
	\!&+&\!
	\left(A_{(1)}+\frac{1}{2}C_{(1)}\right)\eta^{\nu\rho}\Sigma^{\lambda}_{~\mu\nu}\Sigma^{\mu}_{~\lambda\rho},
\end{eqnarray}
\begin{eqnarray}
	\label{L1 new}
	\mathcal{L}_{1}
	\!&=&\!
	A_{(1)}\eta^{\mu\alpha}\eta^{\nu\beta}\left(\partial_{\lambda}h_{\mu\nu}\right)\Sigma^{\lambda}_{~\alpha\beta}
	+B_{(1)}\eta^{\alpha\beta}\left(\partial_{\lambda}h\right)\Sigma^{\lambda}_{~\alpha\beta}
	+C_{(1)}\eta^{\nu\alpha}\left(\partial^{\beta}h_{\mu\nu}\right)\Sigma^{\mu}_{~\alpha\beta}
	\nonumber\\
	\!&+&\!
	D_{(1)}\eta^{\lambda\rho}\left(\partial_{\lambda}h\right)\Sigma^{\mu}_{~\mu\rho}
	+E_{(1)}\eta^{\lambda\nu}\eta^{\alpha\beta}\left(\partial_{\lambda}h_{\mu\nu}\right)\Sigma^{\mu}_{~\alpha\beta}
	\nonumber \\
	\!&+&\!
	\left(2B_{(1)}+E_{(1)}\right)\eta^{\mu\rho}\eta^{\lambda\nu}\left(\partial_{\lambda}h_{\mu\nu}\right)\Sigma^{\sigma}_{~\rho\sigma},
\end{eqnarray}
\begin{eqnarray}
	\label{L2 new}
	\mathcal{L}_{2}
	\!&=&\!
	A_{(2)}\left(\Box h_{\mu\nu}\right)h^{\mu\nu}
	+
	\left(-A_{(2)}-\frac{1}{4}C_{(1)}+\frac{1}{4}A_{(1)}+\frac{1}{4}B_{(1)}-\frac{1}{4}D_{(1)}\right)\left(\Box h\right)h
	\nonumber \\
	\!&+&\!
	\left(2A_{(2)}-\frac{1}{2}A_{(1)}+\frac{1}{2}C_{(1)}-B_{(1)}\right)\left(\partial_{\mu}\partial_{\nu}h\right)h^{\mu\nu}
	\nonumber \\
	\!&+&\!
	\left(-2A_{(2)}-\frac{1}{2}C_{(1)}-\frac{1}{2}E_{(1)}\right)\left(\partial_{\mu}\partial_{\nu}h^{\mu\rho}\right)h^{\nu}_{~\rho}
	\nonumber \\
	\!&+&\!
	B_{(2)}\eta^{\alpha\beta}\left(\Box \Sigma^{\lambda}_{~\lambda\mu}\right)\Sigma^{\mu}_{~\alpha\beta}
	+C_{(2)}\eta^{\mu\nu}\left(\Box\Sigma^{\lambda}_{~\lambda\mu}\right)\Sigma^{\rho}_{~\rho\nu}
	+D_{(2)}\eta_{\lambda\rho}\eta^{\mu\alpha}\eta^{\nu\beta}\left(\Box\Sigma^{\lambda}_{~\mu\nu}\right)\Sigma^{\rho}_{~\alpha\beta}
	\nonumber \\
	\!&+&\!
	E_{(2)}\eta_{\lambda\rho}\eta^{\mu\nu}\eta^{\alpha\beta}\left(\Box\Sigma^{\lambda}_{~\mu\nu}\right)\Sigma^{\rho}_{~\alpha\beta}
	+F_{(2)}\eta^{\nu\rho}\left(\Box\Sigma^{\lambda}_{~\mu\nu}\right)\Sigma^{\mu}_{~\lambda\rho}
	\nonumber \\
	\!&-&\!
	2E_{(2)}\eta^{\alpha\beta}\eta_{\lambda\rho}\left(\partial^{\mu}\partial^{\nu}\Sigma^{\lambda}_{~\mu\nu}\right)\Sigma^{\rho}_{~\alpha\beta}
	-B_{(2)}\left(\partial^{\mu}\partial^{\nu}\Sigma^{\lambda}_{~\mu\nu}\right)\Sigma^{\rho}_{~\lambda\rho}
	\nonumber \\
	\!&+&\!
	H_{(2)}\eta^{\lambda\rho}\left(\partial_{\mu}\partial^{\nu}\Sigma^{\mu}_{~\nu\lambda}\right)\Sigma^{\alpha}_{~\rho\alpha}
	+I_{(2)}\eta^{\alpha\beta}\left(\partial_{\mu}\partial^{\nu}\Sigma^{\mu}_{~\nu\lambda}\right)\Sigma^{\lambda}_{~\alpha\beta}
	+J_{(2)}\eta^{\lambda\rho}\eta^{\alpha\beta}\left(\partial_{\mu}\partial_{\nu}\Sigma^{\mu}_{~\lambda\rho}\right)\Sigma^{\nu}_{~\alpha\beta}
	\nonumber \\
	\!&+&\!
	\left(\frac{1}{2}B_{(2)}-\frac{1}{2}H_{(2)}+F_{(2)}+\frac{1}{2}I_{(2)}+K_{(2)}\right)\eta^{\lambda\alpha}\eta^{\rho\beta}\left(\partial_{\mu}\partial_{\nu}\Sigma^{\mu}_{~\lambda\rho}\right)\Sigma^{\nu}_{~\alpha\beta}
	\nonumber \\
	\!&+&\!
	\left(-D_{(2)}+E_{(2)}\right)\eta_{\lambda\rho}\eta^{\alpha\beta}\left(\partial^{\mu}\partial^{\nu}\Sigma^{\lambda}_{~\mu\alpha}\right)\Sigma^{\rho}_{~\nu\beta}
	+K_{(2)}\left(\partial^{\mu}\partial^{\nu}\Sigma^{\lambda}_{~\mu\rho}\right)\Sigma^{\rho}_{~\nu\lambda}
	\nonumber \\
	\!&+&\!
	\left(\frac{1}{2}B_{(2)}+\frac{1}{2}I_{(2)}+J_{(2)}-\frac{1}{2}H_{(2)}-C_{(2)}\right)\left(\partial^{\mu}\partial^{\nu}\Sigma^{\lambda}_{~\mu\lambda}\right)\Sigma^{\rho}_{~\nu\rho}
	\nonumber \\
	\!&+&\!
	\left(-B_{(2)}-I_{(2)}-2J_{(2)}\right)\eta^{\lambda\rho}\left(\partial_{\mu}\partial^{\nu}\Sigma^{\mu}_{~\lambda\rho}\right)\Sigma^{\alpha}_{~\nu\alpha}
	\nonumber \\
	\!&+&\!
	\left(-2F_{(2)}-I_{(2)}-2K_{(2)}\right)\eta^{\alpha\beta}\left(\partial_{\mu}\partial^{\nu}\Sigma^{\mu}_{~\lambda\alpha}\right)\Sigma^{\lambda}_{~\nu\beta}.
\end{eqnarray}
Here, $h \coloneqq \eta^{\mu\nu}h_{\mu\nu}$ and a series of $A, B, C$ with subscripts, such as $A_{(1)}$, are constants. 

The linear perturbation equations of the general Palatini theory are derived by varying the action (\ref{the most general second-order perturbation action in general Palatini theory}) with respect to the perturbations. Specifically, varying the action with respect to $h^{\mu\nu}$ yields
\begin{eqnarray}
	\label{linear perturbation equation h}
	\!&-&\! A_{(1)}\delta_{\mu}^{~\alpha}\delta_{\nu}^{~\beta}\partial_{\lambda}\Sigma^{\lambda}_{~\alpha\beta}
	-B_{(1)}\eta_{\mu\nu}\eta^{\alpha\beta}\partial_{\lambda}\Sigma^{\lambda}_{~\alpha\beta}
	-\frac{1}{2}C_{(1)}\delta_{\nu}^{~\alpha}\eta_{\mu\sigma}\partial^{\beta}\Sigma^{\sigma}_{~\alpha\beta}
	-\frac{1}{2}C_{(1)}\delta_{\mu}^{~\alpha}\eta_{\nu\sigma}\partial^{\beta}\Sigma^{\sigma}_{~\alpha\beta}
	\nonumber \\
	\!&-&\!
	D_{(1)}\eta_{\mu\nu}\eta^{\lambda\rho}\partial_{\lambda}\Sigma^{\sigma}_{~\sigma\rho}
	-\frac{1}{2}E_{(1)}\delta^{\lambda}_{~\nu}\eta_{\mu\sigma}\eta^{\alpha\beta}\partial_{\lambda}\Sigma^{\sigma}_{~\alpha\beta}
	-\frac{1}{2}E_{(1)}\delta^{\lambda}_{~\mu}\eta_{\nu\sigma}\eta^{\alpha\beta}\partial_{\lambda}\Sigma^{\sigma}_{~\alpha\beta}
	\nonumber \\
	\!&-&\!\frac{1}{2}\left(2B_{(1)}+E_{(1)}\right)\delta_{\mu}^{~\rho}\delta_{\nu}^{~\lambda}\partial_{\lambda}\Sigma^{\sigma}_{~\rho\sigma}
	-\frac{1}{2}\left(2B_{(1)}+E_{(1)}\right)\delta_{\nu}^{~\rho}\delta_{\mu}^{~\lambda}\partial_{\lambda}\Sigma^{\sigma}_{~\rho\sigma}
	\nonumber \\
	\!&+&\!
	2A_{(2)}\Box h_{\mu\nu}
	+2\left(-A_{(2)}-\frac{1}{4}C_{(1)}+\frac{1}{4}A_{(1)}+\frac{1}{4}B_{(1)}-\frac{1}{4}D_{(1)}\right)\eta_{\mu\nu}\Box h
	\nonumber \\
	\!&+&\!
	\left(2A_{(2)}-\frac{1}{2}A_{(1)}+\frac{1}{2}C_{(1)}-B_{(1)}\right)\left(\partial_{\mu}\partial_{\nu}h\right)
	\nonumber \\
	\!&+&\!
	\left(2A_{(2)}-\frac{1}{2}A_{(1)}+\frac{1}{2}C_{(1)}-B_{(1)}\right)\eta_{\mu\nu}\left(\partial_{\lambda}\partial_{\rho}h^{\lambda\rho}\right)
	\nonumber \\
	\!&+&\!
	\left(-2A_{(2)}-\frac{1}{2}C_{(1)}-\frac{1}{2}E_{(1)}\right)\left(\partial_{\nu}\partial^{\lambda}h_{\mu\lambda}\right)
	\nonumber \\
	\!&+&\!
	\left(-2A_{(2)}-\frac{1}{2}C_{(1)}-\frac{1}{2}E_{(1)}\right)\left(\partial_{\mu}\partial^{\lambda}h_{\nu\lambda}\right)=0.
\end{eqnarray}
And varying the action with respect to $\Sigma^{\lambda}_{~\mu\nu}$ yields
\begin{eqnarray}
	\label{linear perturbation equation Sigma}
	\!&&\!
	\left(2B_{(1)}+E_{(1)}\right)\eta^{\mu\nu}\Sigma^{\rho}_{~\rho\lambda}
	+\frac{1}{2}\left(2B_{(1)}+E_{(1)}\right)\eta^{\alpha\beta}\delta^{\nu}_{~\lambda}\Sigma^{\mu}_{~\alpha\beta}
	+\frac{1}{2}\left(2B_{(1)}+E_{(1)}\right)\eta^{\alpha\beta}\delta^{\mu}_{~\lambda}\Sigma^{\nu}_{~\alpha\beta}
	\nonumber \\
	\!&+&\!
	\left(B_{(1)}-D_{(1)}+\frac{1}{2}E_{(1)}\right)\eta^{\mu\sigma}\delta^{\nu}_{~\lambda}\Sigma^{\rho}_{~\rho\sigma}
	+\left(B_{(1)}-D_{(1)}+\frac{1}{2}E_{(1)}\right)\eta^{\nu\sigma}\delta^{\mu}_{~\lambda}\Sigma^{\rho}_{~\rho\sigma}
	\nonumber \\
	\!&+&\!
	C_{(1)}\eta_{\lambda\rho}\eta^{\mu\alpha}\eta^{\nu\beta}\Sigma^{\rho}_{~\alpha\beta}
	+E_{(1)}\eta_{\lambda\rho}\eta^{\mu\nu}\eta^{\alpha\beta}\Sigma^{\rho}_{~\alpha\beta}
	\nonumber \\
	\!&+&\!
	\left(A_{(1)}+\frac{1}{2}C_{(1)}\right)\eta^{\nu\rho}\Sigma^{\mu}_{~\lambda\rho}
	+\left(A_{(1)}+\frac{1}{2}C_{(1)}\right)\eta^{\mu\rho}\Sigma^{\nu}_{~\lambda\rho}
	\nonumber \\
	\!&+&\!
	A_{(1)}\eta^{\mu\alpha}\eta^{\nu\beta}\left(\partial_{\lambda}h_{\alpha\beta}\right)
	+B_{(1)}\eta^{\mu\nu}\left(\partial_{\lambda}h\right)
	+\frac{1}{2}C_{(1)}\eta^{\rho\mu}\left(\partial^{\nu}h_{\lambda\rho}\right)
	+\frac{1}{2}C_{(1)}\eta^{\rho\nu}\left(\partial^{\mu}h_{\lambda\rho}\right)
	\nonumber \\
	\!&+&\!
	\frac{1}{2}D_{(1)}\eta^{\rho\nu}\delta^{\mu}_{~\lambda}\left(\partial_{\rho}h\right)
	+\frac{1}{2}D_{(1)}\eta^{\rho\mu}\delta^{\nu}_{~\lambda}\left(\partial_{\rho}h\right)
	+E_{(1)}\eta^{\rho\sigma}\eta^{\mu\nu}\left(\partial_{\rho}h_{\lambda\sigma}\right)
	\nonumber \\
	\!&+&\!
	\frac{1}{2}\left(2B_{(1)}+E_{(1)}\right)\eta^{\mu\rho}\eta^{\alpha\beta}\delta^{\nu}_{~\lambda}\left(\partial_{\alpha}h_{\rho\beta}\right)
	+\frac{1}{2}\left(2B_{(1)}+E_{(1)}\right)\eta^{\nu\rho}\eta^{\alpha\beta}\delta^{\mu}_{~\lambda}\left(\partial_{\alpha}h_{\rho\beta}\right)
	\nonumber \\
	\!&+&\!
	B_{(2)}\eta^{\mu\nu}\left(\Box \Sigma^{\rho}_{~\lambda\rho}\right)
	+\frac{1}{2}B_{(2)}\eta^{\alpha\beta}\delta^{\nu}_{~\lambda}\left(\Box \Sigma^{\mu}_{~\alpha\beta}\right)
	+\frac{1}{2}B_{(2)}\eta^{\alpha\beta}\delta^{\mu}_{~\lambda}\left(\Box \Sigma^{\nu}_{~\alpha\beta}\right)
	\nonumber \\
	\!&+&\!
	C_{(2)}\eta^{\rho\nu}\delta^{\mu}_{~\lambda}\left(\Box \Sigma^{\sigma}_{~\sigma\rho}\right)
	+C_{(2)}\eta^{\rho\mu}\delta^{\nu}_{~\lambda}\left(\Box \Sigma^{\sigma}_{~\sigma\rho}\right)
	+2D_{(2)}\eta_{\lambda\rho}\eta^{\mu\alpha}\eta^{\nu\beta}\left(\Box \Sigma^{\rho}_{~\alpha\beta}\right)
	\nonumber \\
	\!&+&\!
	2E_{(2)}\eta_{\lambda\rho}\eta^{\mu\nu}\eta^{\alpha\beta}\left(\Box \Sigma^{\rho}_{~\alpha\beta}\right)
	+F_{(2)}\eta^{\nu\rho}\left(\Box \Sigma^{\mu}_{~\lambda\rho}\right)
	+F_{(2)}\eta^{\mu\rho}\left(\Box \Sigma^{\nu}_{~\lambda\rho}\right)    
	\nonumber \\
	\!&-&\!    
	2E_{(2)}\eta^{\alpha\beta}\eta_{\lambda\rho}\left(\partial^{\mu}\partial^{\nu}\Sigma^{\rho}_{~\alpha\beta}\right)
	-2E_{(2)}\eta^{\mu\nu}\eta_{\lambda\rho}\left(\partial^{\alpha}\partial^{\beta}\Sigma^{\rho}_{\alpha\beta}\right)
	\nonumber \\
	\!&-&\! 
	B_{(2)}\left(\partial^{\mu}\partial^{\nu}\Sigma^{\rho}_{~\lambda\rho}\right)
	-\frac{1}{2}B_{(2)}\delta^{\nu}_{~\lambda}\left(\partial^{\alpha}\partial^{\beta}\Sigma^{\mu}_{~\alpha\beta}\right)
	-\frac{1}{2}B_{(2)}\delta^{\mu}_{~\lambda}\left(\partial^{\alpha}\partial^{\beta}\Sigma^{\nu}_{~\alpha\beta}\right)  
	\nonumber \\
	\!&+&\!
	\frac{1}{2}H_{(2)}\eta^{\sigma\mu}\delta^{\nu}_{~\lambda}\left(\partial_{\alpha}\partial^{\beta}\Sigma^{\alpha}_{~\beta\sigma}\right)
	+\frac{1}{2}H_{(2)}\eta^{\sigma\nu}\delta^{\mu}_{~\lambda}\left(\partial_{\alpha}\partial^{\beta}\Sigma^{\alpha}_{~\beta\sigma}\right)
	\nonumber \\
	\!&+&\!
	\frac{1}{2}H_{(2)}\eta^{\nu\rho}\left(\partial_{\lambda}\partial^{\mu}\Sigma^{\alpha}_{~\rho\alpha}\right)
	+\frac{1}{2}H_{(2)}\eta^{\mu\rho}\left(\partial_{\lambda}\partial^{\nu}\Sigma^{\alpha}_{~\rho\alpha}\right)
	\nonumber \\
	\!&+&\!
	I_{(2)}\eta^{\mu\nu}\left(\partial_{\alpha}\partial^{\beta}\Sigma^{\alpha}_{~\beta\lambda}\right)
	+\frac{1}{2}I_{(2)}\eta^{\alpha\beta}\left(\partial_{\lambda}\partial^{\nu}\Sigma^{\mu}_{~\alpha\beta}\right)
	+\frac{1}{2}I_{(2)}\eta^{\alpha\beta}\left(\partial_{\lambda}\partial^{\mu}\Sigma^{\nu}_{~\alpha\beta}\right)
	\nonumber \\
	\!&+&\!
	2J_{(2)}\eta^{\mu\nu}\eta^{\alpha\beta}\left(\partial_{\lambda}\partial_{\rho}\Sigma^{\rho}_{~\alpha\beta}\right)
	\nonumber \\
	\!&+&\!
	2\left(\frac{1}{2}B_{(2)}-\frac{1}{2}H_{(2)}+F_{(2)}+\frac{1}{2}I_{(2)}+K_{(2)}\right)\eta^{\sigma\mu}\eta^{\rho\nu}\left(\partial_{\gamma}\partial_{\lambda}\Sigma^{\gamma}_{~\sigma\rho}\right)
	\nonumber \\
	\!&+&\!
	\left(-D_{(2)}+E_{(2)}\right)\eta_{\lambda\rho}\eta^{\nu\beta}\left(\partial^{\mu}\partial^{\sigma}\Sigma^{\rho}_{~\sigma\beta}\right)
	+\left(-D_{(2)}+E_{(2)}\right)\eta_{\lambda\rho}\eta^{\mu\beta}\left(\partial^{\nu}\partial^{\sigma}\Sigma^{\rho}_{~\sigma\beta}\right)
	\nonumber \\
	\!&+&\!
	K_{(2)}\left(\partial^{\rho}\partial^{\nu}\Sigma^{\mu}_{~\rho\lambda}\right)
	+K_{(2)}\left(\partial^{\rho}\partial^{\mu}\Sigma^{\nu}_{~\rho\lambda}\right)
	\nonumber \\
	\!&+&\!
	\left(\frac{1}{2}B_{(2)}+\frac{1}{2}I_{(2)}+J_{(2)}-\frac{1}{2}H_{(2)}-C_{(2)}\right)\delta^{\mu}_{~\lambda}\left(\partial^{\rho}\partial^{\nu}\Sigma^{\sigma}_{~\rho\sigma}\right)
	\nonumber \\
	\!&+&\!
	\left(\frac{1}{2}B_{(2)}+\frac{1}{2}I_{(2)}+J_{(2)}-\frac{1}{2}H_{(2)}-C_{(2)}\right)\delta^{\nu}_{~\lambda}\left(\partial^{\rho}\partial^{\mu}\Sigma^{\sigma}_{~\rho\sigma}\right)
	\nonumber \\
	\!&+&\!
	\frac{1}{2}\left(-B_{(2)}-I_{(2)}-2J_{(2)}\right)\eta^{\alpha\beta}\delta^{\mu}_{~\lambda}\left(\partial_{\rho}\partial^{\nu}\Sigma^{\rho}_{~\alpha\beta}\right)
	\nonumber\\
	\!&+&\!
	\frac{1}{2}\left(-B_{(2)}-I_{(2)}-2J_{(2)}\right)\eta^{\alpha\beta}\delta^{\nu}_{~\lambda}\left(\partial_{\rho}\partial^{\mu}\Sigma^{\rho}_{~\alpha\beta}\right)
	\nonumber \\
	\!&+&\!
	\left(-B_{(2)}-I_{(2)}-2J_{(2)}\right)\eta^{\mu\nu}\left(\partial_{\lambda}\partial^{\rho}\Sigma^{\alpha}_{~\rho\alpha}\right)
	\nonumber \\
	\!&+&\!
	\frac{1}{2}\left(-2F_{(2)}-I_{(2)}-2K_{(2)}\right)\eta^{\alpha\mu}\left(\partial_{\rho}\partial^{\nu}\Sigma^{\rho}_{~\lambda\alpha}\right)
	+\frac{1}{2}\left(-2F_{(2)}-I_{(2)}-2K_{(2)}\right)\eta^{\alpha\nu}\left(\partial_{\rho}\partial^{\mu}\Sigma^{\rho}_{~\lambda\alpha}\right)
	\nonumber \\
	\!&+&\!
	\frac{1}{2}\left(-2F_{(2)}-I_{(2)}-2K_{(2)}\right)\eta^{\mu\beta}\left(\partial_{\lambda}\partial^{\alpha}\Sigma^{\nu}_{~\alpha\beta}\right)
	+\frac{1}{2}\left(-F_{(2)}-I_{(2)}-2K_{(2)}\right)\eta^{\nu\beta}\left(\partial_{\lambda}\partial^{\alpha}\Sigma^{\mu}_{~\alpha\beta}\right)
	\nonumber \\
	\!&=&\!
	0.
\end{eqnarray}

Note that linear perturbation equations (\ref{linear perturbation equation h}) and (\ref{linear perturbation equation Sigma}) have the form of ``tensor$=0$", so we can uniquely decompose them in the same way as Eq. (\ref{decompose perturbations}). Owing to the uniqueness of the decomposition, the spatial tensor, vector, and scalar components in the linear perturbation equations must each independently equal $0$. By substituting Eq. (\ref{decompose perturbations}) into these equations and expressing them in terms of gauge invariants via Eqs. (\ref{gauge invariant third-order spatial tensors})-(\ref{gauge invariant spatial scalars}), we find that the spatial tensor, vector, and scalar equations depend exclusively on the gauge-invariant spatial tensors, vectors, and scalars. Thus, the equations have been decoupled. This enables independent consideration of tensor, vector, and scalar perturbations, corresponding to the case of tensor, vector, and scalar mode gravitational waves. Given the large number and complexity of these equations, their explicit forms are provided in Appendix \ref{app: E}. These decoupled equations are not fully independent; the number of independent equations is the same as the number of independent gauge invariant perturbations (\ref{gauge invariant third-order spatial tensors})-(\ref{gauge invariant spatial scalars}). Solving these decoupled equations forms the basis for analyzing the polarization modes of gravitational waves in.

\section{Polarization modes of gravitational waves in the most general Palatini theory}
\label{app: D}

In this section, we examine four different parameter selection cases to investigate the polarization characteristics of gravitational waves in general Palatini theory. 
For the convenience of readers who wish to directly apply the framework developed in this paper to their specific gravity theories of interest, we provide in Appendix \ref{app: F} a detailed list of the relationships between all possible terms in the Lagrangian density, constructed algebraically from $g_{\mu\nu}$, $R^{\mu}_{~\nu\lambda\rho}$, $Q_{\lambda\mu\nu}$, and the parameters in Eq. (\ref{the most general second-order perturbation action in general Palatini theory}).

\subsection{Case 1: homogeneous field equations}

Some readers may be particularly interested in the special case where the field equations are homogeneous (At this point, all possible gravitational waves can only propagate at the speed of light.), which corresponds to the parameter setting
\begin{eqnarray}
	\label{case0}
	A_{(1)}=B_{(1)}=C_{(1)}=D_{(1)}=E_{(1)}=0.
\end{eqnarray}
From Eq. (\ref{L2 new}), it can be seen that for the above conditions, $h_{\mu\nu}$ and $\Sigma^{\lambda}_{~\mu\nu}$ are decoupled, meaning that their motions do not influence each other. From Appendix \ref{app: F}, it can be observed that if the Lagrangian density is constructed using only the quadratic term of the curvature tensor and $\widehat{R}$, the resulting theory falls into this case.

In this case, if the test particles do not possess hypermomentum (or the effect of hypermomentum is negligible), then the presence of $\Sigma^{\lambda}_{~\mu\nu}$ will have no impact on the problem under consideration, and the situation will revert to general relativity. When the particles possess hypermomentum, although the $\Sigma^{\lambda}_{~\mu\nu}$ does not influence the motion of $h_{\mu\nu}$, it still impacts the relative motion between the test particles, thereby affects the polarization modes of gravitational waves. However, assuming the existence of a hypermomentum charge in this case would result in physical inconsistency. By noting that the gauge invariant $K_{i}$ in Eq. (\ref{gauge invariant spatial vectors}) is composed of $h_{\mu\nu}$ and $\Sigma^{\lambda}_{~\mu\nu}$, we can conclude that the gauge-invariant field equations in this case impose no constraints on $K_{i}$. However, as indicated by Eq. (\ref{P1-P8 gauge invariant}), $K_{i}$ can lead to observable polarization effects, making this situation unreasonable.

\subsection{Case 2: the connection as a constraint variable}

In this case, the action (\ref{the most general second-order perturbation action in general Palatini theory}) depends only on the metric, its derivative, and the connection, without involving the derivative of the connection. Specifically, we consider the case where the parameters satisfy 
\begin{eqnarray}
	\label{case1}
	B_{(2)}=C_{(2)}=D_{(2)}=E_{(2)}=F_{(2)}=H_{(2)}=I_{(2)}=J_{(2)}=K_{(2)}=0.
\end{eqnarray}
At this case, there is no derivative of the connection in Eq. (\ref{linear perturbation equation Sigma}). Moreover, the connection can generally be solved algebraically from the decoupled equations in Eq. (\ref{linear perturbation equation Sigma}), allowing us to compute
\begin{eqnarray}
	\label{connection in case 1}
	&B^{i}_{~jk}=0, \nonumber \\
	&V_{ij}=E_{ij}=0,~
	U_{ij}=S_{ij}=-\frac{1}{2}\partial_{0}h^{TT}_{ij},~
	C_{ij}=-D_{ij}=\frac{1}{2}h^{TT}_{ij},
	\nonumber \\
	&f_{i}=N_{i}=h_{i}=q_{i}=0,~
	C_{i}=\bar{C}_{i}=K_{i}=-\bar{\Omega}_{i}=\frac{1}{2}\Xi_{i},~
	\Omega_{i}=-\frac{1}{2}\partial_{0}\Xi_{i},
	\nonumber \\
	&K=\Psi=\bar{\Pi}=0,~
	\Pi=\bar{\Psi}=-\phi,~
	L=3\Theta,~
	m=-n=\frac{1}{2}\Theta,~
	A=\bar{A}=-\frac{1}{2}\partial_{0}\Theta.
\end{eqnarray}
When the Einstein-Hilbert action is expressed in the Palatini formalism, the derivative term of the connection takes the form $h\partial\Sigma$ in the second-order perturbation action. After integration by parts $h\partial\Sigma \rightarrow -\Sigma \partial h$, it becomes evident that the action falls into this case. Additionally, as discussed in Appendix \ref{app: F}, if we introduce a correction term to $R$ solely using the non-metricity $Q_{\lambda\mu\nu}$, without incorporating the quadratic term of the curvature tensor, the resulting theory still falls within this case.

Furthermore, the relationship between the connection and the metric, as given by Eq. (\ref{connection in case 1}), is independent of the coefficient parameters in the action (\ref{the most general second-order perturbation action in general Palatini theory}). Therefore, it can be seen that Eq. (\ref{connection in case 1}) is equivalent to the linear order part of the equation
\begin{eqnarray}
	\label{LC connection}
	\Gamma^{\lambda}_{~\mu\nu}=\widehat{\Gamma}^{\lambda}_{~\mu\nu}.
\end{eqnarray}
This is equivalent to stating that, in the linear approximation, the connection reduces to the Levi-Civita connection.

Substituting Eq. (\ref{connection in case 1}) into the decoupled equations from Eq. (\ref{linear perturbation equation h}) leads to the conclusion that the tensor mode equation satisfies
\begin{eqnarray}
	\label{tensor mode case 1}
	-\frac{1}{2} \left(A_{(1)}-4A_{(2)}-C_{(1)}\right)\Box h^{TT}_{ij}=0.
\end{eqnarray}
From this equation and Eq. (\ref{P1-P8 gauge invariant}), it follows that tensor modes—specifically, the $+$ and $\times$ modes—propagate at the speed of light.

For vector mode, the equation is
\begin{eqnarray}
	\label{vector mode case 1}
	-\frac{1}{2} \left(A_{(1)}-4A_{(2)}-C_{(1)}\right)\Delta \Xi_{i}=0.
\end{eqnarray}
It can be seen that there is no vector mode gravitational wave propagation in this case. 

For scalar mode, the equations are
\begin{eqnarray}
	\label{scalar mode case 1}
	-\left(A_{(1)}-4A_{(2)}-C_{(1)}\right)\Delta \Theta\!&=&\!0,\nonumber \\
	-\frac{1}{2} \left(A_{(1)}-4A_{(2)}-C_{(1)}\right)\left(2\phi+\Theta\right)\!&=&\!0.
\end{eqnarray}
Similarly, there are no scalar mode gravitational waves.

By comparing the above equations with linearized general relativity \cite{Michele Maggiore}, for the theory in Case 1 to recover the Newtonian limit, it is required that
\begin{eqnarray}
	\label{Newtonian limit case 1}
	A_{(1)}-4A_{(2)}-C_{(1)}=-\frac{c^{3}}{16\pi G}.
\end{eqnarray}
This also renders the theory in Case 1 equivalent to general relativity in the linear approximation. However, at nonlinear orders, the two are generally not equivalent. We demonstrate this by calculating the effective energy-momentum tensor of gravitational waves in the Isaacson picture. References \cite{Y.Dong5,Lavinia Heisenberg3} demonstrated how to derive the effective energy-momentum tensor of gravitational waves $t_{\mu\nu}$ by varying the action with respect to the background metric $\eta_{\mu\nu}$. (In the case where the action of the matter field $S_{m}$ is independent of the connection, $t_{\mu\nu}+T_{\mu\nu}$ is conserved under the linear approximation.) Using this approach, we obtain
\begin{eqnarray}
	\label{effective energy-momentum tensor of gravitational waves case 1}
	t_{\mu\nu}
	= 
	-2\left\langle\frac{\delta S_{g}^{(2)}}{\delta \eta^{\mu\nu}}\right\rangle
	=
	\left(2A_{(2)}-\frac{1}{2}A_{(1)}-\frac{3}{2}C_{(1)}\right)
	\left\langle\partial_{\mu}h_{TT}^{ij}\partial_{\nu}h^{TT}_{ij}\right\rangle.
\end{eqnarray}
From Eq. (\ref{Newtonian limit case 1}), it follows that as long as $C_{(1)} \neq 0$, the effective energy-momentum tensor of gravitational waves, as given by Eq. (\ref{effective energy-momentum tensor of gravitational waves case 1}), differs from that in general relativity. This difference is expected to influence the waveform of gravitational waves by altering the energy loss rate of binary star mergers, allowing relevant parameters to be constrained through gravitational wave detection.

\subsection{Case 3: the metric as a constraint variable}

In this case, the action (\ref{the most general second-order perturbation action in general Palatini theory}) depends only on the connection, its derivative, and the metric, without involving the derivative of the metric. In other words,
\begin{eqnarray}
	\label{case2}
	&A_{(2)}=0,~
	-A_{(2)}-\frac{1}{4}C_{(1)}+\frac{1}{4}A_{(1)}+\frac{1}{4}B_{(1)}-\frac{1}{4}D_{(1)}=0,\nonumber \\
	&2A_{(2)}-\frac{1}{2}A_{(1)}+\frac{1}{2}C_{(1)}-B_{(1)}=0,~
	-2A_{(2)}-\frac{1}{2}C_{(1)}-\frac{1}{2}E_{(1)}=0.
\end{eqnarray}
The Einstein-Hilbert action in the Palatini formalism falls into this category.

Now, the decoupled equations to be solved (given in Appendix \ref{app: E}) are quite complex. However, since the equations are linear, they can be solved in momentum space. This approach actually assumes that gauge invariant perturbations take the form of plane gravitational waves with a four-dimensional wave vector $k^{\mu}$, such as 
\begin{eqnarray}
	\label{plane wave}
	h^{TT}_{ij}=\mathring{h}^{TT}_{ij} e^{ikx},~ S_{ij}=\mathring{S}_{ij} e^{ikx}, ....
\end{eqnarray}
Here, $\mathring{h}^{TT}_{ij}$ and $\mathring{S}_{ij}$ are constants. By substituting these plane wave forms into the decoupled equations, we can transform the system of differential equations into a linear algebraic system (system of linear equations), due to the linearity of the equations and their constant coefficients. At this stage, the possible wave vector $k^{\mu}$ can be determined by the condition that the determinant of the coefficient matrix is zero, yielding the dispersion relation of gravitational waves. For the corresponding $k^{\mu}$, solving the linear system provides the relationship between gauge invariant perturbations and determines the polarization modes of gravitational waves. Additionally, due to spatial rotational symmetry, we can assume that gravitational waves propagate in the $+z$ direction without loss of generality, simplifying the equations to be solved. Henceforth, we assume gravitational waves propagate along the $+z$ direction.

$h_{\mu\nu}$ and $\Sigma^{\lambda}_{~\mu\nu}$ can be formally treated as tensors under Lorentz transformations, making the second-order perturbation action (\ref{the most general second-order perturbation action in general Palatini theory}) Lorentz invariant. Consequently, $k^{2}\coloneqq \eta^{\mu\nu}k_{\mu}k_{\nu}$ must be constant, implying that the dispersion relation takes the form of a massive graviton or a constant speed of light.

Solving the decoupled equations, we find that the third-order tensor $B^{i}_{~jk}$ can propagate with a mass $m$ which satisfies
\begin{eqnarray}
	\label{mass third order tensor in case 2}
	m^{2}=-\frac{A_{(1)}+C_{(1)}}{D_{(2)}+F_{(2)}}.
\end{eqnarray}
However, due to Eq. (\ref{P1-P8 gauge invariant}), it does not contribute to the polarization modes of gravitational waves.

For the second-order antisymmetric tensor parts ($V_{ij}, E_{ij}$), there are two possible masses: 
\begin{eqnarray}
	\label{mass second-order antisymmetric tensor in case 2}
	m_{1}^{2}=\frac{2A_{(1)}-C_{(1)}}{4D_{(2)}-2F_{(2)}},~
	m_{2}^{2}=\frac{2A_{(1)}-C_{(1)}}{2\left(D_{(2)}+E_{(2)}-F_{(2)}-K_{(2)}\right)}.
\end{eqnarray}
However, they also do not contribute to the polarization modes of gravitational waves. For the second-order symmetric tensor parts, there are four possible masses:
\begin{eqnarray}
	\label{mass second-order symmetric tensor in case 2}
	m_{1}^{2}=0,~
	m_{2}^{2}=-\frac{A_{(1)}+C_{(1)}}{D_{(2)}+F_{(2)}},~
	m_{3}^{2}=\frac{2A_{(1)}-C_{(1)}}{4D_{(2)}-2F_{(2)}},~
	m_{4}^{2}=\frac{\mathcal{P}}{\mathcal{Q}},
\end{eqnarray}
where,
\begin{eqnarray}
	\mathcal{P}\!&=&\!-2A_{(1)}^{3}+A_{(1)}^{2}C_{(1)}+2A_{(1)}C_{(1)}^{2}-C_{(1)}^{3},
	\nonumber \\
	\mathcal{Q}\!&=&\!
	B_{(2)}C_{(1)}^{2}+2A_{(1)}^{2}\left(D_{(2)}+E_{(2)}+F_{(2)}+K_{(2)}\right)
	+2A_{(1)}C_{(1)}\left(I_{(2)}2K_{(2)}\right)
	\nonumber \\
	\!&+&\!C_{(1)}^{2}\left(2D_{(2)}+2F_{(2)} -H_{(2)}+I_{(2)}+2K_{(2)}\right).
\end{eqnarray}
According to Eq. (\ref{P1-P8 gauge invariant}), only $h^{TT}_{ij}$ contributes to the polarization modes. However, in the solutions corresponding to $m_{2}$ and $m_{3}$ in Eq. (\ref{mass second-order symmetric tensor in case 2}), one can obtian $h^{TT}_{ij}=0$, which means these solutions do not excite tensor mode gravitational waves. Therefore, tensor modes, specifically the $+$ and $\times$ modes, can only propagate as massless modes at the speed of light or propagate with mass $m_{4}$. It is important to note that tensor mode gravitational waves, propagating at the speed of light, always exist.

For the vector and scalar components, solving possible dispersion relations (using the determinant of the coefficient matrix is zero) is transformed into solving an $n$-th order equation with $n \geq 5$. Such equations typically lack closed-form solutions, making generalized analysis difficult. However, it is expected that vector and scalar mode gravitational waves will exhibit diverse properties depending on the parameter choices. For specific theories, the relevant parameters can be substituted into the decoupling equations in Appendix \ref {app: C}, and solved as a system of linear equations using the program outlined in this section.

\subsection{Case 4: no constraint variables}

This case corresponds to the most general action (\ref{the most general second-order perturbation action in general Palatini theory}) without any parameter constraint. At this point, the properties of the third-order tensor and the second-order antisymmetric tensor parts remain unchanged from Case 3. For the second-order symmetric tensor parts, the number of possible masses is one more than in Case 3:
\begin{eqnarray}
	\label{mass second-order symmetric tensor in case 3}
	m_{1}^{2}=0,~
	m_{2}^{2}=-\frac{A_{(1)}+C_{(1)}}{D_{(2)}+F_{(2)}},~
	m_{3}^{2}=\frac{2A_{(1)}-C_{(1)}}{4D_{(2)}-2F_{(2)}},~
	m_{4}^{2}=\frac{\mathcal{A}}{\mathcal{B}},~
	m_{5}^{2}=\frac{\mathcal{C}}{\mathcal{D}}.
\end{eqnarray}
Here, the expressions for $\mathcal{A}$, $\mathcal{B}$, $\mathcal{C}$, and $\mathcal{D}$ are exceptionally complex, so they will not be listed here. Similarly, in the solutions corresponding to $m_{2}$ and $m_{3}$, $h^{TT}_{ij}=0$, so it does not actually contribute to tensor mode gravitational waves. The $+$ and $\times$ modes that propagate at the speed of light always exist. The difference between the vector and scalar modes and Case 3 is that they possess more degrees of freedom. For the same reasons, a general analysis remains challenging.

Finally, regarding the general framework we have developed for analyzing gravitational wave polarization modes, further research is needed, particularly in the following two areas: (1) Can the second-order perturbation action (\ref{the most general second-order perturbation action in general Palatini theory}) be generalized to the strong gravitational regime, similar to how the Einstein-Hilbert action is derived from the second-order action in general relativity? (2) It is crucial to identify and exclude parameters in the second-order perturbation action that could lead to physical instability of the theory. This process can be facilitated by the approach outlined in Ref. \cite{Will Barker}.

\section{Decoupled equations}
\label{app: E}
\begin{itemize}
	\item Third-order tensor part:
\end{itemize}

The equation for the transverse traceless third-order spatial tensor part is obtained as	
\begin{eqnarray}
	\label{transverse traceless third-order spatial tensor part equation gauge invariant}
	&C_{(1)}B^{k}_{~ij}+\left(A_{(1)}+\frac{1}{2}C_{(1)}\right)B^{i}_{~kj}+\left(A_{(1)}+\frac{1}{2}C_{(1)}\right)B^{j}_{~ki}
	\nonumber \\
	&+2D_{(2)}\Box B^{k}_{~ij}+F_{(2)}\Box B^{i}_{~kj}+F_{(2)}\Box B^{j}_{~ki}=0.
\end{eqnarray}

\begin{itemize}
	\item Second-order tensor part:
\end{itemize}

The equations for the transverse traceless second-order spatial tensor part are
\begin{eqnarray}
	\label{transverse traceless second-order spatial tensor part equation gauge invariant 1}
	-A_{(1)}\partial_{0}S_{ij}-A_{(1)}\Delta C_{ij}+C_{(1)}\partial_{0} U_{ij}-C_{(1)}\Delta D_{ij}+2A_{(2)}\Box h^{TT}_{ij}=0,
\end{eqnarray}	
\begin{eqnarray}
	\label{transverse traceless second-order spatial tensor part equation gauge invariant 2}
	&-C_{(1)}S_{ij}+\left(2A_{(1)}+C_{(1)}\right)U_{ij}+A_{(1)}\partial_{0}h^{TT}_{ij}
	-2D_{(2)}\Box S_{ij}+2F_{(2)}\Box U_{ij}
	\nonumber \\
	&+\left(B_{(2)}-H_{(2)}+2F_{(2)}+I_{(2)}+2K_{(2)}\right)\partial_{0}^{2}S_{ij}
	\nonumber \\
	&+\left(B_{(2)}-H_{(2)}+2F_{(2)}+I_{(2)}+2K_{(2)}\right)\partial_{0}\Delta C_{ij}
	\nonumber \\
	&-\left(-2F_{(2)}-I_{(2)}-2K_{(2)}\right)\partial_{0}^{2}U_{ij}
	+\left(-2F_{(2)}-I_{(2)}-2K_{(2)}\right)\partial_{0}\Delta D_{ij}=0,
\end{eqnarray}	
\begin{eqnarray}
	\label{transverse traceless second-order spatial tensor part equation gauge invariant 3}
	&-C_{(1)}U_{ij}+\left(A_{(1)}+\frac{1}{2}C_{(1)}\right)S_{ij}-\left(A_{(1)}+\frac{1}{2}C_{(1)}\right)U_{ij}
	-\frac{1}{2}C_{(1)}\partial_{0}h^{TT}_{ij}
	\nonumber \\
	&-2D_{(2)}\Box U_{ij}-F_{(2)}\Box U_{ij}+F_{(2)}\Box S_{ij}+\left(-D_{(2)}+E_{(2)}\right)\partial_{0}^{2}U_{ij}
	-\left(-D_{(2)}+E_{(2)}\right)\partial_{0}\Delta D_{ij}
	\nonumber \\
	&+K_{(2)}\partial_{0}^{2}U_{ij}
	-K_{(2)}\partial_{0}\Delta D_{ij}
	\nonumber \\
	&-\frac{1}{2}\left(-2F_{(2)}-I_{(2)}-2K_{(2)}\right)\partial_{0}^{2}S_{ij}
	-\frac{1}{2}\left(-2F_{(2)}-I_{(2)}-2K_{(2)}\right)\partial_{0}\Delta C_{ij}=0,
\end{eqnarray}		
\begin{eqnarray}
	\label{transverse traceless second-order spatial tensor part equation gauge invariant 4}
	&-C_{(1)}V_{ij}+\left(A_{(1)}+\frac{1}{2}C_{(1)}\right)V_{ij}
	-2D_{(2)}\Box V_{ij}+F_{(2)}\Box V_{ij}
	+\left(-D_{(2)}+E_{(2)}\right)\partial_{0}^{2}V_{ij}
	\nonumber \\
	&-\left(-D_{(2)}+E_{(2)}\right)\partial_{0}\Delta E_{ij}
	-K_{(2)}\partial_{0}^{2}V_{ij}+K_{(2)}\partial_{0}\Delta E_{ij}=0,
\end{eqnarray}		
\begin{eqnarray}
	\label{transverse traceless second-order spatial tensor part equation gauge invariant 5}
	&C_{(1)}C_{ij}+\left(2A_{(1)}+C_{(1)}\right)D_{ij}+A_{(1)}h^{TT}_{ij}
	+2D_{(2)}\Box C_{ij}
	+2F_{(2)}\Box D_{ij}
	\nonumber \\
	&+\left(B_{(2)}-H_{(2)}+2F_{(2)}+I_{(2)}+2K_{(2)}\right)\partial_{0}S_{ij}
	+\left(B_{(2)}-H_{(2)}+2F_{(2)}+I_{(2)}+2K_{(2)}\right)\Delta C_{ij}
	\nonumber \\
	&-\left(-2F_{(2)}-I_{(2)}-2K_{(2)}\right)\partial_{0}U_{ij}
	+\left(-2F_{(2)}-I_{(2)}-2K_{(2)}\right)\Delta D_{ij}=0,
\end{eqnarray}			
\begin{eqnarray}
	\label{transverse traceless second-order spatial tensor part equation gauge invariant 6}
	&C_{(1)}D_{ij}+\left(A_{(1)}+\frac{1}{2}C_{(1)}\right)\left(D_{ij}+C_{ij}\right)+\frac{1}{2}C_{(1)}h^{TT}_{ij}
	+2D_{(2)}\Box D_{ij}+F_{(2)}\Box C_{ij}+F_{(2)}\Box D_{ij}	
	\nonumber \\
	&-\left(-D_{(2)}+E_{(2)}\right)\partial_{0}U_{ij}
	+\left(-D_{(2)}+E_{(2)}\right)\Delta D_{ij}
	-K_{(2)}\partial_{0}U_{ij}
	+K_{(2)}\Delta D_{ij}
	\nonumber \\
	&+\frac{1}{2}\left(-2F_{(2)}-I_{(2)}-2K_{(2)}\right)\partial_{0}S_{ij}
	+\frac{1}{2}\left(-2F_{(2)}-I_{(2)}-2K_{(2)}\right)\Delta C_{ij}=0,
\end{eqnarray}		
\begin{eqnarray}
	\label{transverse traceless second-order spatial tensor part equation gauge invariant 7}
	&\left(A_{(1)}-\frac{1}{2}C_{(1)}\right)E_{ij}-2D_{(2)}\Box E_{ij}+F_{(2)}\Box E_{ij}
	\nonumber \\
	&+\left(-D_{(2)}+E_{(2)}\right)\partial_{0}V_{ij}
	-\left(-D_{(2)}+E_{(2)}\right)\Delta E_{ij}
	-K_{(2)}\partial_{0}V_{ij}
	+K_{(2)}\Delta E_{ij}=0.
\end{eqnarray}		

\begin{itemize}
	\item Vector part:
\end{itemize}

The equations for the transverse spatial vector part are
\begin{eqnarray}
	\label{transverse spatial vector part equation gauge invariant 1}
	&\left(-A_{(1)}-\frac{1}{2}C_{(1)}-B_{(1)}-\frac{1}{2}E_{(1)}\right)\partial_{0}N_{i}
	-A_{(1)}\Delta \bar{C}_{i}
	+\frac{1}{2}C_{(1)}\Delta C_{i}
	+\left(-2E_{(1)}-B_{(1)}\right)\partial_{0}h_{i} 
	\nonumber \\
	&+\left(-3E_{(1)}-4B_{(1)}\right)\partial_{0}q_{i}
	+\left(-B_{(1)}-\frac{1}{2}E_{(1)}\right)\partial_{0}\Delta f_{i}
	+\left(\frac{1}{2}C_{(1)}+\frac{1}{2}E_{(1)}\right)\left(\partial_{0}\Omega_{i}+\partial_{0}^{2}K_{i}\right)
	\nonumber \\
	&+2A_{(2)}\Delta \Xi_{i}+\frac{1}{2}C_{(1)} \Delta \left(\Xi_{i}-K_{i}\right)
	-\frac{1}{2}E_{(1)}\Delta \left(K_{i}-\bar{\Omega}_{i}-\Xi_{i}\right)
	=0,
\end{eqnarray}	
\begin{eqnarray}
	\label{transverse spatial vector part equation gauge invariant 2}
	&-A_{(1)}\partial_{0}C_{i}
	+\left(-A_{(1)}-\frac{1}{2}C_{(1)}-B_{(1)}-\frac{1}{2}E_{(1)}\right)\Delta f_{i}
	+\left(-A_{(1)}-\frac{1}{2}C_{(1)}-3E_{(1)}-4B_{(1)}\right)q_{i}
	\nonumber \\
	&+\frac{1}{2}C_{(1)}\partial_{0}\bar{C}_{i}
	+\left(-\frac{1}{2}C_{(1)}-2E_{(1)}-B_{(1)}\right)h_{i}
	+\left(-B_{(1)}-\frac{1}{2}E_{(1)}\right)N_{i}
	+\frac{1}{2}C_{(1)}\partial_{0}K_{i}
	\nonumber \\
	&+\left(-\frac{1}{2}C_{(1)}-\frac{1}{2}E_{(1)}\right)\frac{\Delta}{\partial_{0}}\left(K_{i}-\bar{\Omega}_{i}-\Xi_{i}\right)
	+\frac{1}{2}E_{(1)}\left(\Omega_{i}+\partial_{0}K_{i}\right)
	+2A_{(2)}\partial_{0}\Xi_{i}
	=0,
\end{eqnarray}		
\begin{eqnarray}
	\label{transverse spatial vector part equation gauge invariant 3}
	&-\left(2A_{(1)}+2B_{(1)}+\frac{1}{2}C_{(1)}+E_{(1)}\right)N_{i}
	-\left(2B_{(1)}+4E_{(1)}\right)h_{i}
	-\left(8B_{(1)}+6E_{(1)}\right)q_{i}
	\nonumber \\
	&+\left(C_{(1)}+E_{(1)}\right)\left(\Omega_{i}+\partial_{0}K_{i}\right)
	-E_{(1)}\frac{\Delta}{\partial_{0}}\left(K_{i}-\bar{\Omega}_{i}-\Xi_{i}\right)
	\nonumber \\
	&+\Delta\left(
	-2F_{(2)}\partial_{0}\bar{C}_{i}
	-I_{(2)}\partial_{0}C_{i}
	-2K_{(2)}\partial_{0}\bar{C}_{i}
	-2K_{(2)}\partial_{0}C_{i}
	-I_{(2)}\Delta f_{i}
	-2F_{(2)}N_{i}
	+2D_{(2)}\Omega_{i}
	\right.
	\nonumber \\
	&\left.
	-4E_{(2)}h_{i}
	+2E_{(2)}\Omega_{i}
	-4E_{(2)}q_{i}
	-I_{(2)}q_{i}
	-B_{(2)}h_{i}
	-B_{(2)}\Delta f_{i}
	-B_{(2)}N_{i}
	-4B_{(2)}q_{i}
	\right)
	\nonumber\\
	&=0,
\end{eqnarray}	
\begin{eqnarray}
	\label{transverse spatial vector part equation gauge invariant 4}
	&\left(4B_{(1)}+2E_{(1)}+D_{(1)}\right)h_{i}
	+\left(6B_{(1)}+3E_{(1)}+4D_{(1)}\right)q_{i}
	+\left(B_{(1)}+D_{(1)}+\frac{1}{2}E_{(1)}\right)\Delta f_{i}
	\nonumber \\
	&+\left(A_{(1)}+B_{(1)}+\frac{3}{2}C_{(1)}+D_{(1)}+\frac{1}{2}E_{(1)}\right)N_{i}
	-\left(2B_{(1)}-\frac{1}{2}E_{(1)}-A_{(1)}-\frac{1}{2}C_{(1)}\right)\left(\Omega_{i}+\partial_{0}K_{i}\right)
	\nonumber \\
	&+\frac{1}{2}\left(2B_{(1)}+E_{(1)}\right)\frac{\Delta}{\partial_{0}}\left(K_{i}-\bar{\Omega}_{i}-\Xi_{i}\right)
	\nonumber \\
	&+\frac{1}{2}\left[
	-H_{(2)}\partial_{0}^{2}h_{i}
	-3I_{(2)}\partial_{0}^{2}h_{i}
	-2D_{(2)}\partial_{0}\Delta C_{i}
	+2E_{(2)}\partial_{0}\Delta C_{i}
	-2F_{(2)}\partial_{0}\Delta \bar{C}_{i}
	-2F_{(2)}\partial_{0}\Delta C_{i}
	\right.
	\nonumber \\
	&+H_{(2)}\partial_{0}\Delta \bar{C}_{i}
	+H_{(2)}\partial_{0}\Delta C_{i}
	-I_{(2)}\partial_{0}\Delta \bar{C}_{i}
	-I_{(2)}\partial_{0}\Delta C_{i}
	-H_{(2)}\partial_{0}^{2}\Delta f_{i}
	-2K_{(2)}\partial_{0}\Delta \bar{C}_{i}
	\nonumber \\
	&-2K_{(2)}\partial_{0}\Delta C_{i}
	+H_{(2)}\Delta^{2} f_{i}
	-2D_{(2)}\partial_{0}^{2}N_{i}
	-2E_{(2)}\partial_{0}^{2}N_{i}
	-2F_{(2)}\partial_{0}^{2}N_{i}
	+4D_{(2)}\Delta N_{i}
	\nonumber \\
	&+2F_{(2)}\Delta N_{i}
	+I_{(2)}\partial_{0}\Delta \bar{\Omega}_{i}
	-2F_{(2)}\Delta \Omega_{i}
	-4H_{(2)}\partial_{0}^{2}q_{i}
	-2I_{(2)}\partial_{0}^{2}q_{i}
	+H_{(2)}\Delta q_{i}
	\nonumber \\
	&+2C_{(2)}\Box\left(h_{i}+\Delta f_{i}+N_{i}+4q_{i}\right)
	\nonumber \\
	&\left.
	+B_{(2)}\left(
	-3\partial_{0}^{2}h_{i}
	+2\Delta h_{i}
	-2\partial_{0}\Delta\bar{C}_{i}
	-2\partial_{0}^{2}N_{i}
	+\partial_{0}\Delta \bar{\Omega}_{i}
	-\Delta \Omega_{i}
	-2\partial_{0}^{2}q_{i}
	+2\Delta q_{i}
	\right)
	\right]
	\nonumber \\
	&=0,
\end{eqnarray}		
\begin{eqnarray}
	\label{transverse spatial vector part equation gauge invariant 5}
	&-C_{(1)}C_{i}
	+\left(A_{(1)}+\frac{1}{2}C_{(1)}\right)\bar{C}_{i}
	+A_{(1)}\left(K_{i}-\Xi_{i}\right)
	+\frac{1}{2}C_{(1)}K_{i}
	\nonumber \\
	&-\frac{1}{2}\left[
	-2I_{(2)}\partial_{0}h_{i}
	-2B_{(2)}\partial_{0}^{2}C_{i}
	-4D_{(2)}\partial_{0}^{2}C_{i}
	-I_{(2)}\partial_{0}^{2}\bar{C}_{i}
	-2I_{(2)}\partial_{0}^{2}C_{i}
	+2K_{(2)}\partial_{0}h_{i}
	-2K_{(2)}\partial_{0}^{2}\bar{C}_{i}
	\right.
	\nonumber \\
	&-4K_{(2)}\partial_{0}^{2}C_{i}
	+2D_{(2)}\Delta C_{i}
	+2E_{(2)}\Delta C_{i}
	+I_{(2)} \Delta \bar{C}_{i}
	-2B_{(2)}\partial_{0}\Delta f_{i}
	-I_{(2)}\partial_{0}\Delta f_{i}
	+2K_{(2)}\Delta \bar{C}_{i}
	\nonumber \\
	&-2K_{(2)}\partial_{0}\Delta f_{i}
	+2D_{(2)}\partial_{0}N_{i}
	-2E_{(2)}\partial_{0}N_{i}
	+I_{(2)}\partial_{0}N_{i}
	+2K_{(2)}\partial_{0}N_{i}
	-2K_{(2)}\Delta \bar{\Omega}_{i}
	+I_{(2)}\partial_{0}\Omega_{i}
	\nonumber \\
	&+2K_{(2)}\partial_{0}\Omega_{i}
	-2B_{(2)}\partial_{0}q_{i}
	-3I_{(2)}\partial_{0}q_{i}
	-2K_{(2)}\partial_{0}q_{i}
	-H_{(2)}\partial_{0}
	\left(
	h_{i}-2\partial_{0}C_{i}-\Delta f_{i}+N_{i}+2q_{i}
	\right)
	\nonumber \\
	&\left.
	+2F_{(2)}
	\left(
	\partial_{0}h_{i}
	-2\partial_{0}^{2}C_{i}
	-\partial_{0}\Delta f_{i}
	+\partial_{0}N_{i}
	-\Delta \bar{\Omega}_{i}
	-\partial_{0}q_{i}
	\right)
	\right]=0,
\end{eqnarray}		
\begin{eqnarray}
	\label{transverse spatial vector part equation gauge invariant 6}
	&-C_{(1)}\bar{C}_{i}+\left(A_{(1)}+\frac{1}{2}C_{(1)}\right)C_{i}-A_{(1)}K_{i}-\frac{1}{2}C_{(1)}\left(K_{i}-\Xi_{i}\right)
	\nonumber \\
	&-\frac{1}{2}\left[
	-2D_{(2)}\partial_{0}^{2}\bar{C}_{i}
	-2E_{(2)}\partial_{0}^{2}\bar{C}_{i}
	-I_{(2)}\partial_{0}^{2}C_{i}
	-2K_{(2)}\partial_{0}^{2}C_{i}
	+\left(3I_{(2)}+2K_{(2)}\right)\partial_{0}h_{i}
	+2B_{(2)}\Delta \bar{C}_{i}
	\right.
	\nonumber \\
	&+4D_{(2)}\Delta \bar{C}_{i}
	+4F_{(2)}\Delta \bar{C}_{i}
	+2I_{(2)}\Delta \bar{C}_{i}
	+I_{(2)}\Delta C_{i}
	-2D_{(2)}\partial_{0}\Delta f_{i}
	+2E_{(2)}\partial_{0}\Delta f_{i}
	-2F_{(2)}\partial_{0}\Delta f_{i}
	\nonumber \\
	&-I_{(2)}\partial_{0}\Delta f_{i}
	+4K_{(2)}\Delta \bar{C}_{i}
	+2K_{(2)}\Delta C_{i}
	-2K_{(2)}\partial_{0}\Delta f_{i}
	+2B_{(2)}\partial_{0}N_{i}
	+2F_{(2)}\partial_{0}N_{i}
	+I_{(2)}\partial_{0}N_{i}
	\nonumber \\
	&+2K_{(2)}\partial_{0}N_{i}
	-I_{(2)}\Delta \bar{\Omega}_{i}
	-2K_{(2)}\Delta \bar{\Omega}_{i}
	+2F_{(2)}\partial_{0}\Omega_{i}
	+2K_{(2)}\partial_{0}\Omega_{i}
	\nonumber \\
	&+H_{(2)}
	\left(
	\partial_{0}h_{i}-2\Delta \bar{C}_{i}+\partial_{0}\Delta f_{i}-\partial_{0}N_{i}+4\partial_{0}q_{i}
	\right)
	\nonumber \\
	&\left.
	-2D_{(2)}\partial_{0}q_{i}
	+2E_{(2)}\partial_{0}q_{i}
	-2F_{(2)}\partial_{0}q_{i}
	+I_{(2)}\partial_{0}q_{i}
	-2K_{(2)}\partial_{0}q_{i}
	\right]=0,
\end{eqnarray}	
\begin{eqnarray}
	\label{transverse spatial vector part equation gauge invariant 7}
	&\left(A_{(1)}+\frac{1}{2}C_{(1)}\right)C_{i}
	-\left(A_{(1)}+\frac{1}{2}C_{(1)}\right)\bar{C}_{i}
	-\frac{1}{2}C_{(1)}K_{i}-\frac{1}{2}C_{(1)}\left(K_{i}-\Xi_{i}\right)
	\nonumber \\
	&+\frac{1}{2}\left(
	10E_{(2)}\partial_{0}h_{i}
	+2F_{(2)}\partial_{0}^{2}\bar{C}_{i}
	+I_{(2)}\partial_{0}^{2}C_{i}
	+2K_{(2)}\partial_{0}^{2}\bar{C}_{i}
	+2K_{(2)}\partial_{0}^{2}C_{i}
	+2F_{(2)}\Delta C_{i}
	+I_{(2)}\Delta \bar{C}_{i}
	\right.
	\nonumber \\
	&+2F_{(2)}\partial_{0}\Delta f_{i}
	+I_{(2)}\partial_{0}\Delta f_{i}
	+2K_{(2)}\Delta \bar{C}_{i}
	+2K_{(2)}\Delta C_{i}
	+2F_{(2)}\partial_{0}N_{i}
	+I_{(2)}\partial_{0}N_{i}
	-2E_{(2)}\Delta \bar{\Omega}_{i}
	\nonumber \\
	&-2E_{(2)}\partial_{0}\Omega_{i}
	+2D_{(2)}\partial_{0}h_{i}
	-2D_{(2)}\Delta \bar{\Omega}_{i}
	-2D_{(2)}\partial_{0}\Omega_{i}
	+8E_{(2)}\partial_{0}q_{i}
	+2F_{(2)}\partial_{0}q_{i}
	+I_{(2)}\partial_{0}q_{i}
	\nonumber \\
	&\left.
	+2B_{(2)}\partial_{0}h_{i}
	+2B_{(2)}\partial_{0}N_{i}
	+8B_{(2)}\partial_{0}q_{i}
	+B_{(2)}\partial_{0}\Delta f_{i}
	\right)=0,
\end{eqnarray}		
\begin{eqnarray}
	\label{transverse spatial vector part equation gauge invariant 8}
	&\left(A_{(1)}+\frac{3}{2}C_{(1)}\right)f_{i}
	+\left(A_{(1)}+\frac{1}{2}C_{(1)}\right)\frac{1}{\partial_{0}}\left(K_{i}-\bar{\Omega}_{i}-\Xi_{i}\right)
	\nonumber \\
	&+\frac{1}{2}\left(
	2I_{(2)}h_{i}
	+2B_{(2)}\partial_{0}C_{i}
	+2D_{(2)\partial_{0}}\bar{C}_{i}
	-2E_{(2)}\partial_{0}\bar{C}_{i}
	+I_{(2)}\partial_{0}\bar{C}_{i}
	+I_{(2)}\partial_{0}C_{i}
	-4D_{(2)}\partial_{0}^{2}f_{i}
	\right.
	\nonumber \\
	&+2K_{(2)}\partial_{0}\bar{C}_{i}
	+2K_{(2)}\partial_{0}C_{i}
	+2B_{(2)}\Delta f_{i}
	+2D_{(2)}\Delta f_{i}
	+2E_{(2)}\Delta f_{i}
	-2F_{(2)}h_{i}
	+2F_{(2)}\partial_{0}\bar{C}_{i}
	\nonumber\\
	&+2F_{(2)}\partial_{0}C_{i}
	-2F_{(2)}\partial_{0}^{2}f_{i}
	+2F_{(2)}\Delta f_{i}
	+2F_{(2)}\partial_{0}\bar{\Omega}_{i}
	-I_{(2)}\Omega_{i}
	+2B_{(2)}q_{i}
	-2D_{(2)}q_{i}
	+2E_{(2)}q_{i}
	\nonumber \\
	&\left.
	+2I_{(2)}q_{i}
	+H_{(2)}h_{i}
	-2H_{(2)}\partial_{0}C_{i}
	-H_{(2)}\Delta f_{i}
	+H_{(2)}N_{i}
	+2H_{(2)}q_{i}
	\right)=0,
\end{eqnarray}
\begin{eqnarray}
	\label{transverse spatial vector part equation gauge invariant 9}
	&\left(2A_{(1)}+C_{(1)}\right)f_{i}+\frac{C_{(1)}}{\partial_{0}}\left(K_{i}-\bar{\Omega}_{i}-\Xi_{i}\right)
	\nonumber \\
	&-2D_{(2)}h_{i}
	-4E_{(2)}h_{i}
	-2F_{(2)}\partial_{0}C_{i}
	-I_{(2)}\partial_{0}C_{i}
	-2F_{(2)}\partial_{0}^{2}f_{i}
	-2K_{(2)}\partial_{0}\bar{C}_{i}
	\nonumber \\
	&-2K_{(2)}\partial_{0}C_{i}
	-I_{(2)}\partial_{0}^{2}f_{i}
	+2D_{(2)}\partial_{0}\bar{\Omega}_{i}
	+2E_{(2)}\Omega_{i}
	-4E_{(2)}q_{i}
	-2F_{(2)}q_{i}
	\nonumber \\
	&-I_{(2)}q_{i}
	-B_{(2)}h_{i}
	-B_{(2)}\Delta f_{i}
	-B_{(2)}N_{i}
	-4B_{(2)}q_{i}
	=0,
\end{eqnarray}	
\begin{eqnarray}
	\label{transverse spatial vector part equation gauge invariant 10}
	&\left(2B_{(1)}+E_{(1)}\right)N_{i}
	+\left(2B_{(1)}+E_{(1)}\right)\Delta f_{i}
	+\left(2B_{(1)}+4E_{(1)}+C_{(1)}\right)h_{i}
	\nonumber \\
	&+\left(8B_{(1)}+6E_{(1)}+2A_{(1)}+C_{(1)}\right)q_{i}
	-E_{(1)}\left(\Omega_{i}+\partial_{0}K_{i}\right)
	+E_{(1)}\frac{\Delta}{\partial_{0}}\left(K_{i}-\bar{\Omega}_{i}-\Xi_{i}\right)
	\nonumber \\
	&-6E_{(2)}\partial_{0}^{2}h_{i}
	+4E_{(2)}\Delta h_{i}
	-I_{(2)}\partial_{0}\Delta\bar{C}_{i}
	+I_{(2)}\partial_{0}\Delta C_{i}
	+I_{(2)}\Delta^{2}f_{i}
	+2D_{(2)}\Box h_{i}
	\nonumber \\
	&-I_{(2)}\partial_{0}^{2}N_{i}
	+2E_{(2)}\partial_{0}\Delta\bar{\Omega}_{i}
	-2E_{(2)}\Delta \Omega_{i}
	-4E_{(2)}\partial_{0}^{2}q_{i}
	-2F_{(2)}\partial_{0}^{2}q_{i}
	+4E_{(2)}\Delta q_{i}
	\nonumber \\
	&+2F_{(2)}\Delta q_{i}
	+I_{(2)}\Delta q_{i}
	+B_{(2)}\Box
	\left(
	h_{i}+\Delta f_{i}+N_{i}+4q_{i}
	\right)
	=0,
\end{eqnarray}		
\begin{eqnarray}
	\label{transverse spatial vector part equation gauge invariant 11}
	&\left(2B_{(1)}+E_{(1)}+D_{(1)}+A_{(1)}+\frac{1}{2}C_{(1)}\right)h_{i}
	+\left(6B_{(1)}+3E_{(1)}+4D_{(1)}+A_{(1)}+\frac{3}{2}C_{(1)}\right)q_{i}
	\nonumber \\
	&+\left(B_{(1)}+D_{(1)}+\frac{1}{2}E_{(1)}\right)\Delta f_{i}
	+\left(B_{(1)}+D_{(1)}+\frac{1}{2}E_{(1)}\right)N_{i}
	\nonumber \\
	&+\left(-B_{(1)}+\frac{1}{2}E_{(1)}\right)\left(\Omega_{i}+\partial_{0}K_{i}\right)
	+\left(B_{(1)}+\frac{1}{2}E_{(1)}\right)\frac{\Delta}{\partial_{0}}
	\left(K_{i}-\bar{\Omega}_{i}-\Xi_{i}\right)
	\nonumber \\
	&+\frac{1}{2}\left[
	-2F_{(2)}\partial_{0}^{2}h_{i}
	+2F_{(2)}\Delta h_{i}
	-H_{(2)}\partial_{0}\Delta \bar{C}_{i}
	+H_{(2)}\partial_{0}\Delta C_{i}
	+H_{(2)}\Delta^{2}f_{i}
	-H_{(2)}\partial_{0}^{2}N_{i}
	\right.
	\nonumber \\
	&-4D_{(2)}\partial_{0}^{2}q_{i}
	-2F_{(2)}\partial_{0}^{2}q_{i}
	+4D_{(2)}\Delta q_{i}
	+2F_{(2)}\Delta q_{i}
	+H_{(2)}\Delta q_{i}
	\nonumber \\
	&
	+2C_{(2)}\Box\left(h_{i}+\Delta f_{i}+N_{i}+4q_{i}\right)
	+B_{(2)}\left(-3\partial_{0}^{2}+2\Delta\right)h_{i}
	\nonumber \\
	&\left.
	+B_{(2)}\partial_{0}\Delta \bar{\Omega}_{i}
	-B_{(2)}\Delta \left(\Omega_{i}-2q_{i}\right)
	-2B_{(2)}\partial_{0}^{2}q_{i}
	\right]=0.
\end{eqnarray}			

\begin{itemize}
	\item Scalar part:
\end{itemize}

The equations for the spatial scalar part are	
\begin{eqnarray}
	\label{spatial scalar part equation gauge invariant 1}
	&\left(3B_{(1)}+3E_{(1)}\right)\partial_{0}A
	+\left(3B_{(1)}+D_{(1)}\right)\Delta m
	+\left(2B_{(1)}+4D_{(1)}\right)\Delta n
	\nonumber \\
	&+\left(-3D_{(1)}-6B_{(1)}-3E_{(1)}\right)\partial_{0}\bar{A}
	-\left(B_{(1)}+E_{(1)}\right)\Delta \bar{\Psi}
	\nonumber \\
	&+\left(-A_{(1)}-3B_{(1)}-C_{(1)}-D_{(1)}-2E_{(1)}\right)\partial_{0}\Psi
	\nonumber \\
	&+\left(-\frac{1}{2}A_{(1)}-\frac{3}{2}B_{(1)}-\frac{1}{2}C_{(1)}-\frac{1}{2}D_{(1)}-E_{(1)}\right)\partial_{0}^{2}K
	\nonumber \\
	&+\left(\frac{1}{2}B_{(1)}+\frac{1}{2}C_{(1)}+\frac{1}{2}D_{(1)}+\frac{1}{2}E_{(1)}\right)\Delta K
	+\left(-A_{(1)}-2B_{(1)}-E_{(1)}\right)\Delta \phi
	\nonumber\\
	&+\left(-A_{(1)}-B_{(1)}\right)\Delta \Pi
	-\left(D_{(1)}-2B_{(1)}-E_{(1)}\right)\partial_{0}\Delta\bar{\Pi}
	+\left(-A_{(1)}-B_{(1)}\right)\Delta\partial_{0}\bar{\Pi}
	\nonumber\\
	&+\left(-\frac{1}{2}A_{(1)}-\frac{3}{2}B_{(1)}-\frac{1}{2}D_{(1)}-\frac{1}{2}E_{(1)}\right)\partial_{0}^{2}L
	+\left(\frac{1}{2}B_{(1)}+\frac{1}{2}D_{(1)}\right)\Delta L
	\nonumber \\
	&+\left(\frac{3}{2}A_{(1)}+3B_{(1)}+\frac{3}{2}E_{(1)}\right)\partial_{0}^{2}\Theta
	+\left(4A_{(2)}+C_{(1)}-A_{(1)}-2B_{(1)}\right)\Delta \Theta=0,
\end{eqnarray}				
\begin{eqnarray}
	\label{spatial scalar part equation gauge invariant 2}
	&\left(-A_{(1)}-\frac{1}{2}C_{(1)}-3B_{(1)}-\frac{3}{2}E_{(1)}\right)\bar{A}
	+\left(\frac{1}{2}C_{(1)}+\frac{3}{2}E_{(1)}\right)A
	\nonumber \\
	&+\left(-2E_{(1)}-B_{(1)}\right)\partial_{0}m
	+\left(-3E_{(1)}-4B_{(1)}\right)\partial_{0}n
	\nonumber \\
	&+\left(-E_{(1)}-B_{(1)}\right)\Psi
	-\left(\frac{1}{2}C_{(1)}+\frac{1}{2}E_{(1)}\right)\frac{\Delta}{\partial_{0}}\bar{\Psi}
	+\left(-\frac{1}{2}A_{(1)}-\frac{1}{4}C_{(1)}-\frac{3}{4}E_{(1)}-B_{(1)}\right)\partial_{0}K
	\nonumber \\
	&+\left(\frac{1}{4}C_{(1)}+\frac{1}{4}E_{(1)}\right)\frac{\Delta}{\partial_{0}}K
	+\left(\frac{1}{2}C_{(1)}+\frac{1}{2}E_{(1)}\right)\partial_{0}\phi
	+\left(-\frac{1}{2}C_{(1)}-\frac{1}{2}E_{(1)}\right)\frac{\Delta}{\partial_{0}}\phi
	\nonumber \\
	&+\left(\frac{1}{2}C_{(1)}+\frac{1}{2}E_{(1)}\right)\partial_{0}\Pi
	+\left(\frac{1}{2}C_{(1)}+\frac{1}{2}E_{(1)}\right)\partial_{0}^{2}\bar{\Pi}
	+\left(-A_{(1)}-\frac{1}{2}C_{(1)}-B_{(1)}-\frac{1}{2}E_{(1)}\right)\Delta \bar{\Pi}
	\nonumber \\
	&+\left(\frac{1}{4}C_{(1)}+\frac{1}{4}E_{(1)}\right)\frac{\partial_{0}^{3}}{\Delta}L
	+\left(-\frac{1}{2}A_{(1)}-\frac{1}{4}C_{(1)}-B_{(1)}-\frac{3}{4}E_{(1)}\right)\partial_{0}L
	\nonumber \\
	&-\left(\frac{3}{4}C_{(1)}+\frac{3}{4}E_{(1)}\right)\frac{\partial_{0}^{3}}{\Delta}\Theta
	+\left(4A_{(2)}+\frac{7}{4}C_{(1)}+\frac{7}{4}E_{(1)}\right)\partial_{0}\Theta=0,
\end{eqnarray}			
\begin{eqnarray}
	\label{spatial scalar part equation gauge invariant 3}
	&\left(-2A_{(1)}-C_{(1)}-6E_{(1)}-8B_{(1)}\right)n
	+\left(-C_{(1)}-4E_{(1)}-2B_{(1)}\right)m
	\nonumber \\
	&+A_{(1)}\bar{\Psi}+\left(-\frac{1}{2}A_{(1)}-B_{(1)}-\frac{1}{2}E_{(1)}\right)K
	+\left(4A_{(2)}+C_{(1)}+E_{(1)}\right)\phi+E_{(1)}\Pi
	\nonumber \\
	&+\left(C_{(1)}+E_{(1)}\right)\partial_{0}\bar{\Pi}
	+\left(\frac{1}{2}C_{(1)}+\frac{1}{2}E_{(1)}\right)\frac{\partial_{0}^{2}}{\Delta}L
	+\left(-\frac{1}{2}A_{(1)}-\frac{1}{2}C_{(1)}-E_{(1)}-B_{(1)}\right)L
	\nonumber \\
	&+\left(-\frac{3}{2}C_{(1)}-\frac{3}{2}E_{(1)}\right)\frac{\partial_{0}^{2}}{\Delta}\Theta
	+\left(2A_{(2)}+2C_{(1)}+2E_{(1)}\right)\Theta=0,
\end{eqnarray}				
\begin{eqnarray}
	\label{spatial scalar part equation gauge invariant 4}
	&\left(-A_{(1)}-3B_{(1)}\right)\partial_{0}A
	+\left(C_{(1)}+3D_{(1)}\right)\partial_{0}\bar{A}
	\nonumber \\
	&+\left(-A_{(1)}-3B_{(1)}-D_{(1)}\right)\Delta m
	+\left(-2B_{(1)}-C_{(1)}-4D_{(1)}\right)\Delta n
	\nonumber \\
	&+\left(B_{(1)}+D_{(1)}\right)\partial_{0}\Psi
	+B_{(1)}\Delta \bar{\Psi}
	+\left(\frac{1}{2}B_{(1)}+\frac{1}{2}D_{(1)}\right)\partial_{0}^{2}K
	+\left(-\frac{1}{2}B_{(1)}-\frac{1}{2}D_{(1)}\right)\Delta K
	\nonumber \\
	&+\left(-4A_{(2)}-C_{(1)}+A_{(1)}+2B_{(1)}\right)\Delta\phi
	+B_{(1)}\Delta\Pi
	+\left(B_{(1)}+D_{(1)}\right)\Delta\partial_{0}\bar{\Pi}
	\nonumber \\
	&+\left(\frac{1}{2}B_{(1)}+\frac{1}{2}D_{(1)}\right)\partial_{0}^{2}L
	+\left(-\frac{1}{2}B_{(1)}-\frac{1}{2}D_{(1)}\right)\Delta L
	\nonumber \\
	&+\left(4A_{(2)}+\frac{3}{2}C_{(1)}-\frac{3}{2}A_{(1)}-3B_{(1)}\right)\partial_{0}^{2}\Theta
	+\left(-2A_{(2)}-C_{(1)}+A_{(1)}+2B_{(1)}\right)\Delta\Theta
	\nonumber\\
	&=0,
\end{eqnarray}		
\begin{eqnarray}
	\label{spatial scalar part equation gauge invariant 5}
	&\left(-12B_{(1)}-6E_{(1)}-6D_{(1)}\right)\bar{A}
	+\left(6B_{(1)}+6E_{(1)}\right)A
	\nonumber \\
	&+\left(-6B_{(1)}-4E_{(1)}-2D_{(1)}-2C_{(1)}-2A_{(1)}\right)\Psi
	\nonumber \\
	&-\left(2B_{(1)}+2E_{(1)}\right)\frac{\Delta}{\partial_{0}}\bar{\Psi}
	+\left(-3B_{(1)}-2E_{(1)}-D_{(1)}-C_{(1)}-A_{(1)}\right)\partial_{0}K
	\nonumber \\
	&+\left(B_{(1)}+E_{(1)}\right)\frac{\Delta}{\partial_{0}}K
	-\left(2B_{(1)}+2E_{(1)}\right)\frac{\Delta}{\partial_{0}}\phi
	+3\left(B_{(1)}+E_{(1)}\right)\partial_{0}\Theta
	\nonumber \\
	&+\left(-4B_{(1)}-2E_{(1)}-2D_{(1)}\right)\Delta \bar{\Pi}
	+\left(-2B_{(1)}-E_{(1)}-D_{(1)}\right)\partial_{0}L
	\nonumber \\
	&+\Delta\left[
	2\left(B_{(2)}+2E_{(2)}\right)A
	-\left(3B_{(2)}+6C_{(2)}+H_{(2)}+I_{(2)}\right)\bar{A}
	\right.
	\nonumber \\
	&+3B_{(2)}\partial_{0}m
	+2C_{(2)}\partial_{0}m
	+H_{(2)}\partial_{0}m
	+3I_{(2)}\partial_{0}m
	+2B_{(2)}\partial_{0}n
	+8C_{(2)}\partial_{0}n
	\nonumber \\
	&+4H_{(2)}\partial_{0}n
	+2I_{(2)}\partial_{0}n
	-B_{(2)}\Delta\bar{\Pi}
	-2C_{(2)}\Delta\bar{\Pi}
	-H_{(2)}\Delta\bar{\Pi}
	-I_{(2)}\Delta\bar{\Pi}
	\nonumber \\
	&\left.
	-2B_{(2)}\Psi
	-2C_{(2)}\Psi
	-2D_{(2)}\Psi
	-2E_{(2)}\Psi
	-2F_{(2)}\Psi
	\right]=0,
\end{eqnarray}		
\begin{eqnarray}
	\label{spatial scalar part equation gauge invariant 6}
	&\left(-2B_{(1)}-4E_{(1)}\right)m
	+\left(-8B_{(1)}-6E_{(1)}\right)n
	+\left(-A_{(1)}-B_{(1)}-\frac{1}{2}C_{(1)}-\frac{1}{2}E_{(1)}\right)K
	\nonumber \\
	&+\left(C_{(1)}+E_{(1)}\right)\phi
	+\left(C_{(1)}+E_{(1)}\right)\Pi
	+\left(C_{(1)}+E_{(1)}\right)\partial_{0}\bar{\Pi}
	\nonumber \\
	&+\left(\frac{1}{2}C_{(1)}+\frac{1}{2}E_{(1)}\right)\frac{\partial_{0}^{2}}{\Delta}L
	+\left(-B_{(1)}-E_{(1)}\right)L
	+2E_{(1)}\Theta
	-\frac{3}{2}\left(C_{(1)}+E_{(1)}\right)\frac{\partial_{0}^{2}}{\Delta}\Theta
	\nonumber \\
	&-2\left(2I_{(2)}+3J_{(2)}+K_{(2)}\right)\partial_{0}A
	\nonumber \\
	&-\left(3B_{(2)}+2D_{(2)}+2E_{(2)}+2F_{(2)}-3H_{(2)}+3I_{(2)}+6J_{(2)}+2K_{(2)}\right)\partial_{0}\bar{A}\nonumber \\
	&-\Delta\left[
	4J_{(2)}m-2I_{(2)}n
	+E_{(2)}\left(4m-2\Pi\right)
	-B_{(2)}\Pi
	-2D_{(2)}\Pi
	-2F_{(2)}\Pi
	+H_{(2)}\Pi
	\right.
	\nonumber \\
	&\left.
	-I_{(2)}\Pi
	-2K_{(2)}\Pi
	-2I_{(2)}\bar{\Psi}
	-2K_{(2)}\bar{\Psi}
	-2J_{(2)}\left(2n+\Pi+\bar{\Psi}\right)
	\right]=0,
\end{eqnarray}			
\begin{eqnarray}
	\label{spatial scalar part equation gauge invariant 7}
	&\left(4B_{(1)}+2E_{(1)}+D_{(1)}\right)m
	+\left(6B_{(1)}+3E_{(1)}+4D_{(1)}\right)n
	\nonumber \\
	&+\left(\frac{1}{2}B_{(1)}+\frac{1}{2}D_{(1)}+\frac{1}{4}E_{(1)}+\frac{3}{4}C_{(1)}+\frac{1}{2}A_{(1)}\right)K
	\nonumber \\
	&+\left(-B_{(1)}-\frac{1}{2}E_{(1)}-\frac{1}{2}C_{(1)}-A_{(1)}\right)\phi
	+\left(-B_{(1)}-\frac{1}{2}E_{(1)}-A_{(1)}-\frac{1}{2}C_{(1)}\right)\Pi
	\nonumber \\
	&+\left(-B_{(1)}-\frac{1}{2}E_{(1)}-A_{(1)}-\frac{1}{2}C_{(1)}\right)\partial_{0}\bar{\Pi}
	+\left(-\frac{1}{2}B_{(1)}-\frac{1}{4}E_{(1)}-\frac{1}{2}A_{(1)}-\frac{1}{4}C_{(1)}\right)\frac{\partial_{0}^{2}}{\Delta}L
	\nonumber \\
	&+\left(B_{(1)}+\frac{1}{2}D_{(1)}+\frac{1}{2}E_{(1)}\right)L
	+\left(\frac{3}{2}B_{(1)}+\frac{3}{4}E_{(1)}+\frac{3}{2}A_{(1)}+\frac{3}{4}C_{(1)}\right)\frac{\partial_{0}^{2}}{\Delta}\Theta
	\nonumber \\
	&+\left(-2B_{(1)}-E_{(1)}\right)\Theta
	+\frac{1}{2}\left[
	\left(3B_{(2)}+6C_{(2)}+H_{(2)}-3I_{(2)}-6J_{(2)}-2K_{(2)}\right)\partial_{0}\bar{A}
	\right.
	\nonumber \\
	&-\left(3B_{(2)}+2D_{(2)}+10E_{(2)}+2F_{(2)}-H_{(2)}+I_{(2)}+6J_{(2)}+2K_{(2)}\right)\partial_{0}A
	\nonumber \\
	&-3B_{(2)}\partial_{0}^{2}m
	-2C_{(2)}\partial_{0}^{2}m
	-H_{(2)}\partial_{0}^{2}m
	-3I_{(2)}\partial_{0}^{2}m
	-2I_{(2)}\Delta m
	-4J_{(2)}\Delta m
	-2B_{(2)}\partial_{0}^{2}n
	\nonumber \\
	&-8C_{(2)}\partial_{0}^{2}n
	-4H_{(2)}\partial_{0}^{2}n
	-2I_{(2)}\partial_{0}^{2}n
	+2B_{(2)}\Delta n
	-2H_{(2)}\Delta n
	+2I_{(2)}\Delta n
	+4J_{(2)}\Delta n
	\nonumber \\
	&+2I_{(2)}\Delta \Pi
	+2J_{(2)}\Delta \Pi
	+2K_{(2)}\Delta \Pi
	+B_{(2)}\partial_{0}\Delta\bar{\Pi}
	+2C_{(2)}\partial_{0}\Delta\bar{\Pi}
	+H_{(2)}\partial_{0}\Delta\bar{\Pi}
	+I_{(2)}\partial_{0}\Delta\bar{\Pi}
	\nonumber \\
	&+2B_{(2)}\partial_{0}\Psi
	+2C_{(2)}\partial_{0}\Psi
	+2D_{(2)}\partial_{0}\Psi
	+2E_{(2)}\partial_{0}\Psi
	+2F_{(2)}\partial_{0}\Psi
	\nonumber \\
	&\left.
	+\left(
	B_{(2)}+2D_{(2)}+2E_{(2)}+2F_{(2)}-H_{(2)}+I_{(2)}+2J_{(2)}+2K_{(2)}
	\right)\Delta\bar{\Psi}
	\right]=0,
\end{eqnarray}			
\begin{eqnarray}
	\label{spatial scalar part equation gauge invariant 8}
	&C_{(1)}\frac{1}{\partial_{0}}\bar{\Psi}
	-\frac{1}{2}C_{(1)}\frac{1}{\partial_{0}}K
	+C_{(1)}\frac{1}{\partial_{0}}\phi
	+\left(2A_{(1)}+C_{(1)}\right)\bar{\Pi}
	+\left(A_{(1)}+\frac{1}{2}C_{(1)}\right)\frac{\partial_{0}}{\Delta}L
	\nonumber \\
	&+\left(-3A_{(1)}-\frac{3}{2}C_{(1)}\right)\frac{\partial_{0}}{\Delta}\Theta
	+2\left(D_{(2)}+2E_{(2)}\right)A
	-\left(3B_{(2)}+2F_{(2)}+I_{(2)}\right)\bar{A}
	\nonumber \\
	&-\left[
	-2I_{(2)}\partial_{0}m
	+2K_{(2)}\partial_{0}m
	-2B_{(2)}\partial_{0}n
	-3I_{(2)}\partial_{0}n
	-2K_{(2)}\partial_{0}n
	\right.
	\nonumber \\
	&+I_{(2)}\partial_{0}\Pi
	+2K_{(2)}\partial_{0}\Pi
	+B_{(2)}\Delta\bar{\Pi}
	+I_{(2)}\Delta\bar{\Pi}
	+B_{(2)}\Psi
	\nonumber \\
	&+2E_{(2)}\Psi
	+B_{(2)}\partial_{0}\bar{\Psi}
	+2D_{(2)}\partial_{0}\bar{\Psi}
	+I_{(2)}\partial_{0}\bar{\Psi}
	+2K_{(2)}\partial_{0}\bar{\Psi}
	\nonumber \\
	&\left.
	+2F_{(2)}\partial_{0}\left(m-n+\bar{\Psi}\right)
	-H_{(2)}\partial_{0}\left(m+2n+\bar{\Psi}\right)
	\right]=0,
\end{eqnarray}	
\begin{eqnarray}
	\label{spatial scalar part equation gauge invariant 9}
	&\left(6B_{(1)}+3E_{(1)}+2A_{(1)}+C_{(1)}\right)\bar{A}
	+\left(-C_{(1)}-3E_{(1)}\right)A
	\nonumber \\
	&+\left(2B_{(1)}+2E_{(1)}\right)\Psi
	+E_{(1)}\frac{\Delta}{\partial_{0}}\bar{\Psi}
	+\left(B_{(1)}+E_{(1)}\right)\partial_{0}K
	-\frac{1}{2}E_{(1)}\frac{\Delta}{\partial_{0}}K
	\nonumber \\
	&+E_{(1)}\frac{\Delta}{\partial_{0}}\phi
	+\left(2B_{(1)}+E_{(1)}\right)\Delta\bar{\Pi}
	+\left(B_{(1)}+\frac{1}{2}E_{(1)}\right)\partial_{0}L
	+\left(A_{(1)}-\frac{3}{2}E_{(1)}\right)\partial_{0}\Theta
	\nonumber \\
	&+\left(
	B_{(2)}\partial_{0}^{2}
	+2D_{(2)}\partial_{0}^{2}
	+6E_{(2)}\partial_{0}^{2}
	+2F_{(2)}\partial_{0}^{2}
	-H_{(2)}\partial_{0}^{2}
	+I_{(2)}\partial_{0}^{2}
	+6J_{(2)}\partial_{0}^{2}
	\right.
	\nonumber \\
	&\left.
	+2K_{(2)}\partial_{0}^{2}
	-2D_{(2)}\Delta
	-4E_{(2)}\Delta
	\right)A
	+\left(
	6J_{(2)}\partial_{0}^{2}
	+2K_{(2)}\partial_{0}^{2}
	+3B_{(2)}\Delta
	+2F_{(2)}\Delta
	\right.
	\nonumber \\
	&\left.
	+4I_{(2)}\partial_{0}^{2}
	+I_{(2)}\Delta
	\right)\bar{A}
	-\Delta\left[
	H_{(2)}\partial_{0}m
	-4J_{(2)}\partial_{0}m
	-2K_{(2)}\partial_{0}m
	+4B_{(2)}\partial_{0}n
	\right.
	\nonumber \\
	&+5I_{(2)}\partial_{0}n
	+4J_{(2)}\partial_{0}n
	+2K_{(2)}\partial_{0}n
	+2F_{(2)}\partial_{0}\left(-m+n\right)
	+I_{(2)}\partial_{0}\Pi
	+2J_{(2)}\partial_{0}\Pi
	\nonumber \\
	&\left.
	-B_{(2)}\Delta\bar{\Pi}
	-I_{(2)}\Delta\bar{\Pi}
	-B_{(2)}\Psi
	-2E_{(2)}\Psi
	+2E_{(2)}\partial_{0}\bar{\Psi}
	+2J_{(2)}\partial_{0}\bar{\Psi}
	\right]=0,
\end{eqnarray}		
\begin{eqnarray}
	\label{spatial scalar part equation gauge invariant 10}
	&\left(3B_{(1)}+\frac{3}{2}E_{(1)}+A_{(1)}+\frac{1}{2}C_{(1)}\right)A
	+\left(-3B_{(1)}-3D_{(1)}-\frac{3}{2}E_{(1)}-\frac{3}{2}C_{(1)}-A_{(1)}\right)\bar{A}
	\nonumber \\
	&+\left(-2B_{(1)}-E_{(1)}-D_{(1)}\right)\Psi
	-\left(B_{(1)}+\frac{1}{2}E_{(1)}\right)\frac{\Delta}{\partial_{0}}\bar{\Psi}
	\nonumber \\
	&+\left(-B_{(1)}-\frac{1}{2}E_{(1)}-\frac{1}{2}D_{(1)}\right)\partial_{0}K
	+\left(\frac{1}{2}B_{(1)}
	+\frac{1}{4}E_{(1)}\right)\frac{\Delta}{\partial_{0}}K
	-\left(B_{(1)}+\frac{1}{2}E_{(1)}\right)\frac{\Delta}{\partial_{0}}\phi
	\nonumber \\
	&+\left(-B_{(1)}-D_{(1)}-\frac{1}{2}E_{(1)}\right)\Delta\bar{\Pi}
	+\left(-\frac{1}{2}B_{(1)}-\frac{1}{2}D_{(1)}-\frac{1}{4}E_{(1)}\right)\partial_{0}L
	\nonumber \\
	&+\left(\frac{3}{2}B_{(1)}+\frac{3}{4}E_{(1)}-\frac{1}{2}C_{(1)}\right)\partial_{0}\Theta
	\nonumber \\
	&+\frac{1}{2}\left[
	\left(
	3B_{(2)}+2D_{(2)}+2E_{(2)}+2F_{(2)}-3H_{(2)}+3I_{(2)}+6J_{(2)}+2K_{(2)}
	\right)\partial_{0}^{2}\bar{A}
	\right.
	\nonumber \\
	&+2\left(2I_{(2)}+3J_{(2)}+K_{(2)}\right)\partial_{0}^{2}A
	-\left(6C_{(2)}+4D_{(2)}+2F_{(2)}+H_{(2)}\right)\Delta \bar{A}
	\nonumber \\
	&+2\left(B_{(2)}+F_{(2)}\right)\Delta A
	+\Delta\left(
	2F_{(2)}\partial_{0}m
	+H_{(2)}\partial_{0}m
	+3I_{(2)}\partial_{0}m
	+4J_{(2)}\partial_{0}m
	\right.
	\nonumber \\
	&+2K_{(2)}\partial_{0}m
	+2D_{(2)}\partial_{0}n
	-2E_{(2)}\partial_{0}n
	+4H_{(2)}\partial_{0}n
	-2I_{(2)}\partial_{0}n
	-4J_{(2)}\partial_{0}n
	\nonumber \\
	&-2K_{(2)}\partial_{0}n
	+2B_{(2)}\partial_{0}m
	-2B_{(2)}\partial_{0}n
	-B_{(2)}\partial_{0}\Pi
	+H_{(2)}\partial_{0}\Pi
	-I_{(2)}\partial_{0}\Pi
	\nonumber \\
	&-2J_{(2)}\partial_{0}\Pi
	-H_{(2)}\Delta\bar{\Pi}
	+2C_{(2)}\partial_{0}m
	+8C_{(2)}\partial_{0}n
	-2C_{(2)}\Delta\bar{\Pi}
	-2C_{(2)}\Psi
	\nonumber \\
	&\left.\left.
	-B_{(2)}\Psi
	-I_{(2)}\partial_{0}\bar{\Psi}
	-2J_{(2)}\partial_{0}\bar{\Psi}
	\right)\right]
	=0,
\end{eqnarray}			
\begin{eqnarray}
	\label{spatial scalar part equation gauge invariant 11}
	&-\left(A_{(1)}+\frac{1}{2}C_{(1)}\right)\frac{1}{\partial_{0}}\bar{\Psi}
	+\left(\frac{1}{2}A_{(1)}+\frac{1}{4}C_{(1)}\right)\frac{1}{\partial_{0}}K
	-\left(A_{(1)}+\frac{1}{2}C_{(1)}\right)\frac{1}{\partial_{0}}\phi
	\nonumber \\
	&+\left(-A_{(1)}-\frac{3}{2}C_{(1)}\right)\bar{\Pi}
	+\left(-\frac{1}{2}A_{(1)}-\frac{3}{4}C_{(1)}\right)\frac{\partial_{0}}{\Delta}L
	+\left(\frac{3}{2}A_{(1)}+\frac{9}{4}C_{(1)}\right)\frac{\partial_{0}}{\Delta}\Theta
	\nonumber \\
	&+\frac{1}{2}\left[
	-\left(2B_{(2)}-2D_{(2)}+2E_{(2)}+H_{(2)}\right)\bar{A}
	+2\left(-F_{(2)}+I_{(2)}\right)A
	\right.
	\nonumber \\
	&+\left(2B_{(2)}+2D_{(2)}+10E_{(2)}-H_{(2)}\right)\partial_{0}m
	-H_{(2)}\Psi
	-I_{(2)}\Psi
	\nonumber \\
	&-\Delta\left(2B_{(2)}+2D_{(2)}+2E_{(2)}+2F_{(2)}-H_{(2)}\right)\bar{\Pi}
	-I_{(2)}\partial_{0}\left(3m+\bar{\Psi}\right)
	\nonumber \\
	&-2\partial_{0}\left(
	-\left(4B_{(2)}+D_{(2)}+3E_{(2)}+2F_{(2)}-2H_{(2)}\right)n
	\right.
	\nonumber \\
	&\left.\left.
	+\left(D_{(2)}+E_{(2)}+F_{(2)}\right)\Pi
	+K_{(2)}\left(m-n+\Pi+\bar{\Psi}\right)
	\right)\right]=0,
\end{eqnarray}	
\begin{eqnarray}
	\label{spatial scalar part equation gauge invariant 12}
	&\left(A_{(1)}+C_{(1)}\right)\frac{1}{\Delta}L
	-\left(3A_{(1)}+3C_{(1)}\right)\frac{1}{\Delta}\Theta
	\nonumber \\
	&-4E_{(2)}m
	-2F_{(2)}m
	+H_{(2)}m
	+2I_{(2)}m
	-2E_{(2)}n
	-2F_{(2)}n
	+2H_{(2)}n
	\nonumber \\
	&+I_{(2)}n
	+2E_{(2)}\Pi
	-I_{(2)}\Pi
	+2F_{(2)}\partial_{0}\bar{\Pi}
	\nonumber \\
	&-2D_{(2)}\left(m+n-\partial_{0}\bar{\Pi}\right)
	+H_{(2)}\bar{\Psi}
	-B_{(2)}\left(m+2n+\bar{\Psi}\right)
	=0,
\end{eqnarray}	
\begin{eqnarray}
	\label{spatial scalar part equation gauge invariant 13}
	&\left(2B_{(1)}+4E_{(1)}+C_{(1)}\right)m
	+\left(8B_{(1)}+6E_{(1)}+2A_{(1)}+C_{(1)}\right)n
	\nonumber \\
	&+\left(B_{(1)}+\frac{1}{2}E_{(1)}\right)K
	-E_{(1)}\phi
	-E_{(1)}\Pi
	-E_{(1)}\partial_{0}\bar{\Pi}
	-\frac{1}{2}E_{(1)}\frac{\partial_{0}^{2}}{\Delta}L
	\nonumber \\
	&+\left(B_{(1)}+E_{(1)}\right)L
	+\frac{3}{2}E_{(1)}\frac{\partial_{0}^{2}}{\Delta}\Theta
	+\left(A_{(1)}-2E_{(1)}\right)\Theta
	\nonumber \\
	&+\left(3B_{(2)}+4E_{(2)}+2F_{(2)}+3I_{(2)}+6J_{(2)}+2K_{(2)}\right)\partial_{0}\bar{A}
	\nonumber \\
	&+\left(B_{(2)}+2F_{(2)}-H_{(2)}+2I_{(2)}+6J_{(2)}+2K_{(2)}\right)\partial_{0}A
	\nonumber \\
	&-B_{(2)}\partial_{0}^{2}m
	-2D_{(2)}\partial_{0}^{2}m
	-6E_{(2)}\partial_{0}^{2}m
	+B_{(2)}\Delta m
	+2D_{(2)}\Delta m
	\nonumber \\
	&+4E_{(2)}\Delta m
	+2F_{(2)}\Delta m
	-H_{(2)}\Delta m
	+I_{(2)} \Delta m
	+4J_{(2)} \Delta m
	+2K_{(2)} \Delta m
	\nonumber \\
	&-4B_{(2)}\partial_{0}^{2}n
	-2F_{(2)}\partial_{0}^{2}n
	-3I_{(2)}\Delta n
	-4J_{(2)}\Delta n
	-2K_{(2)}\Delta n
	\nonumber \\
	&-2E_{(2)}\Delta \Pi
	-2J_{(2)}\Delta \Pi
	+B_{(2)}\Delta \partial_{0}\bar{\Pi}
	+2E_{(2)}\Delta \partial_{0}\bar{\Pi}
	\nonumber \\
	&+B_{(2)}\partial_{0}\Psi
	+I_{(2)}\partial_{0}\Psi
	-\left(I_{(2)}
	+2J_{(2)}\right)\Delta\bar{\Psi}
	=0, 
\end{eqnarray}		
\begin{eqnarray}
	\label{spatial scalar part equation gauge invariant 14}
	&\left(4B_{(1)}+2E_{(1)}+D_{(1)}+A_{(1)}+\frac{1}{2}C_{(1)}\right)m
	+\left(6B_{(1)}+3E_{(1)}+4D_{(1)}+A_{(1)}+\frac{3}{2}C_{(1)}\right)n
	\nonumber \\
	&+\left(\frac{1}{2}B_{(1)}+\frac{1}{2}D_{(1)}+\frac{1}{4}E_{(1)}\right)K
	-\left(B_{(1)}+\frac{1}{2}E_{(1)}\right)\phi
	+\left(-B_{(1)}-\frac{1}{2}E_{(1)}\right)\Pi
	\nonumber \\
	&+\left(-B_{(1)}-\frac{1}{2}E_{(1)}\right)\partial_{0}\bar{\Pi}
	+\left(-\frac{1}{2}B_{(1)}-\frac{1}{4}E_{(1)}\right)\frac{\partial_{0}^{2}}{\Delta}L
	+\left(B_{(1)}+\frac{1}{2}E_{(1)}+\frac{1}{2}D_{(1)}\right)L
	\nonumber \\
	&+\left(\frac{3}{2}B_{(1)}+\frac{3}{4}E_{(1)}\right)\frac{\partial_{0}^{2}}{\Delta}\Theta
	+\left(-2B_{(1)}-E_{(1)}+\frac{1}{2}C_{(1)}\right)\Theta
	\nonumber \\
	&-\frac{1}{2}\left[
	\left(
	B_{(2)}-6C_{(2)}-2D_{(2)}+2E_{(2)}-2H_{(2)}+3I_{(2)}+6J_{(2)}+2K_{(2)}
	\right)\partial_{0}\bar{A}
	\right.
	\nonumber \\
	&+\left(3B_{(2)}+2F_{(2)}-H_{(2)}+4I_{(2)}+6J_{(2)}+2K_{(2)}\right)\partial_{0}A
	\nonumber \\
	&+2F_{(2)}\partial_{0}^{2}m
	+3I_{(2)}\Delta m
	+4J_{(2)}\Delta m
	+2K_{(2)}\Delta m
	+4D_{(2)}\partial_{0}^{2}n
	+2F_{(2)}\partial_{0}^{2}n
	\nonumber \\
	&-2D_{(2)}\Delta n
	-2E_{(2)}\Delta n
	-2F_{(2)}\Delta n
	+2H_{(2)}\Delta n
	-2I_{(2)}\Delta n
	-4J_{(2)}\Delta n
	-2K_{(2)}\Delta n
	\nonumber \\
	&-I_{(2)}\Delta \Pi
	-2J_{(2)}\Delta \Pi
	+2C_{(2)}\left(\partial_{0}^{2}m+4\partial_{0}^{2}n-\partial_{0}\Delta\bar{\Pi}-\partial_{0}\Psi\right)
	\nonumber \\
	&-H_{(2)}\partial_{0}\Psi
	+H_{(2)}\Delta\bar{\Psi}
	-I_{(2)}\Delta\bar{\Psi}
	-2J_{(2)}\Delta\bar{\Psi}
	\nonumber \\
	&\left.
	-B_{(2)}
	\left(
	-3\partial_{0}^{2}m
	-2\partial_{0}^{2}n
	+\partial_{0}\Delta\bar{\Pi}
	+2\Delta n
	+\Delta\bar{\Psi}
	\right)
	\right]=0.
\end{eqnarray}			

\section{The relationship between $\mathcal{L}\left[g_{\mu\nu}, R^{\mu}_{~\nu\rho\sigma}, Q_{\lambda\mu\nu}\right]$ and parameters}
\label{app: F}

Due to the unique geometric significance of the curvature tensor $R^{\mu}_{~\nu\rho\sigma} \coloneqq \partial_{\rho}\Gamma^{\mu}_{~\nu\sigma}-\partial_{\sigma}\Gamma^{\mu}_{~\nu\rho}+\Gamma^{\mu}_{~\tau\rho}\Gamma^{\tau}_{~\nu\sigma}-\Gamma^{\mu}_{~\tau\sigma}\Gamma^{\tau}_{~\nu\rho}$ and the non-metricity $Q_{\lambda\mu\nu}\coloneqq \nabla_{\lambda}g_{\mu\nu}$, algebraic combinations of $R^{\mu}_{~\nu\rho\sigma}$ and $Q_{\lambda\mu\nu}$ (sometimes with their covariant derivatives) are commonly employed to construct Lagrangian densities $\mathcal{L}$ ($S=\int d^{4}x \sqrt{-g} \mathcal{L}$), where the indices are contracted using the metric $g_{\mu\nu}$.

Now, we list all the possible terms in $\mathcal{L}$ that contribute to Eq. (\ref{the most general second-order perturbation action in general Palatini theory}), following the method outlined in the previous paragraph, and describe their relationship with the parameters in Eq. (\ref{the most general second-order perturbation action in general Palatini theory}).

Terms in the form of $QQ$ (the indices are contracted using the metric):
\begin{eqnarray}
	\label{QQ}
    Q^{\lambda\mu}_{~~~\mu}Q_{\lambda~\nu}^{~\nu} & \rightarrow & D_{(1)}=4,
    \nonumber \\
    Q^{\mu}_{~\mu\lambda}Q_{\nu}^{~\nu\lambda} & \rightarrow & E_{(1)}=2,
    \nonumber \\
    Q^{\lambda\mu}_{~~~\mu}Q_{\rho\lambda}^{~~\rho} & \rightarrow & B_{(1)}=1,~D_{(1)}=1,
    \nonumber \\
    Q^{\lambda\mu\nu}Q_{\lambda\mu\nu}& \rightarrow & C_{(1)}=4,~A_{(2)}=-1,
    \nonumber \\
    Q^{\lambda\mu\nu}Q_{\nu\lambda\mu}& \rightarrow & A_{(1)}=2,~C_{(1)}=2.
\end{eqnarray}	
Here, the contribution of the terms shown on the left side of the arrow (with a coefficient of 1) to the parameter is displayed on the right side of the arrow.

Terms in the form of $\nabla Q$ (these types of terms are not independent of terms like $QQ$):
\begin{eqnarray}
	\label{nabla Q}
	g^{\mu\nu}g^{\rho\lambda}\nabla_{\rho}Q_{\lambda\mu\nu}
	& \rightarrow & 
	C_{(1)}=4,~D_{(1)}=-1,~B_{(1)}=1,~A_{(2)}=-1,
	\nonumber \\
	g^{\lambda\nu}g^{\rho\mu}\nabla_{\rho}Q_{\lambda\mu\nu}
	& \rightarrow & 
	A_{(1)}=2,~B_{(1)}=-\frac{1}{2},~C_{(1)}=2,~D_{(1)}=-\frac{1}{2},~E_{(1)}=2.
\end{eqnarray}	

Terms in the form of $R$:
\begin{eqnarray}
	\label{R}
	R
	& \rightarrow & 
	A_{(1)}=-1,~B_{(1)}=\frac{1}{2},~D_{(1)}=-\frac{1}{2}.
\end{eqnarray}	
Here, $R \coloneqq g^{\nu\sigma}R^{\mu}_{~\nu\mu\sigma}$.

Terms in the form of $RR$:
\begin{eqnarray}
	\label{RR}
	R^{2} & \rightarrow & J_{(2)}=-1,
	\nonumber \\
    R^{\mu\nu}R_{\mu\nu} & \rightarrow & C_{(2)}=-1,~H_{(2)}=2,
    \nonumber \\
    R^{\mu\nu}R_{\nu\mu} & \rightarrow & H_{(2)}=2,
    \nonumber \\
    R^{\mu\nu}\tilde{R}_{\mu\nu}
    & \rightarrow & 
    I_{(2)}=-1,~B_{(2)}=1,
    \nonumber \\
    R^{\mu\nu}\tilde{R}_{\nu\mu}
    & \rightarrow & 
    I_{(2)}=-1,~H_{(2)}=-1,
    \nonumber \\
    R^{\mu\nu}\bar{R}_{\mu\nu}
    & \rightarrow & 
    C_{(2)}=-1,
    \nonumber \\
    \tilde{R}^{\mu\nu}\tilde{R}_{\mu\nu}
    & \rightarrow & 
    E_{(2)}=-1,
    \nonumber \\
    \tilde{R}^{\mu\nu}\tilde{R}_{\nu\mu}
    & \rightarrow & 
    J_{(2)}=-1,~K_{(2)}=-1,~I_{(2)}=2,
    \nonumber \\
    \tilde{R}^{\mu\nu}\bar{R}_{\mu\nu}
    & \rightarrow & 
    B_{(2)}=1,~H_{(2)}=1,
    \nonumber \\
    \bar{R}^{\mu\nu}\bar{R}_{\mu\nu}
    & \rightarrow & 
    C_{(2)}=-2,
    \nonumber \\
    R^{\mu}_{~\nu\rho\sigma}R_{\mu}^{~\nu\rho\sigma}
    & \rightarrow & 
    D_{(2)}=-2,
    \nonumber \\
    R^{\mu}_{~\nu\rho\sigma}R_{\mu}^{~\rho\nu\sigma}
    & \rightarrow & 
    D_{(2)}=-1,
    \nonumber \\
    R^{\mu}_{~\lambda\rho\sigma}R^{\lambda~~\rho\sigma}_{~\mu}
    & \rightarrow & 
    F_{(2)}=-2,~K_{(2)}=2,
    \nonumber \\
    R^{\mu}_{~\nu\lambda\sigma}R^{\lambda~~\nu\sigma}_{~\mu}
    & \rightarrow & 
    F_{(2)}=-1,~K_{(2)}=1,
    \nonumber \\
    R^{\mu}_{~\nu\lambda\sigma}R^{\lambda\nu~\sigma}_{~~\mu}
    & \rightarrow & 
    F_{(2)}=-1,
    \nonumber \\
    R^{\mu}_{~\nu\lambda\sigma}R^{\lambda\sigma~\nu}_{~~\mu}
    & \rightarrow & 
    K_{(2)}=-1.
\end{eqnarray}	
Here, $R_{\mu\nu} \coloneqq R^{\lambda}_{~\mu\lambda\nu}$, $\bar{R}_{\mu\nu} \coloneqq R^{\lambda}_{~\lambda\mu\nu}$ and $\tilde{R}^{\mu}_{~\nu} \coloneqq g^{\alpha\beta}R^{\mu}_{~\alpha\nu\beta}$.

It can be observed that for any set of parameters in Eq. (\ref{the most general second-order perturbation action in general Palatini theory}), there always exists a theory in the form of $\mathcal{L}\left[g_{\mu\nu}, R^{\mu}_{~\nu\rho\sigma}, Q_{\lambda\mu\nu}\right]$ that corresponds to this set of parameters. It should also be noted that, the theory corresponding to a given set of parameters is generally not unique.

\end{document}